\newcommand{\MSC}[2]{{\it #2}} 
\documentclass[final]{siamltex}

\usepackage{oldlfont} %uses LaTeX 2.09 fonts, e.g., \cal%
\usepackage{amsmath, amsfonts}
\usepackage{graphicx,color}
\input{epsf}
\input amssym.def
\input amssym

\title{Near-Resonant, Steady Mode Interaction: Periodic,
 Quasi-Periodic and Localized Patterns\thanks{The work of MH was
 supported by the Spanish Ministerio de Ciencia y Tecnolog\'ia (BFM2001-2363). 
The work of HR was supported by NASA grant NAG3-2113, the Department
 of Energy (DE-FG02-92ER14303) and NSF (DMS-9804673). The work of MS was
 supported by NASA grant NAG3-2364, and by NSF under DMS-9972059 
and the MRSEC Program under DMR-0213745.}}

\author{Mar\'ia Higuera\thanks{E. T. S. Ingenieros Aeron\'auticos, 
Universidad Polit\'ecnica de Madrid, Plaza Cardenal Cisneros 3, 28040 Madrid, Spain.
({\tt maria@fmetsia.upm.es}).}
        \and  Hermann Riecke\thanks{Department of Engineering Sciences and
       Applied Mathematics, Northwestern University, Evanston, IL 602028, USA.}
       \and Mary Silber\thanks{Department of Engineering Sciences and
       Applied Mathematics, Northwestern University, Evanston, IL 602028, USA.}}

\begin{document}

\maketitle

\begin{abstract}

Motivated by the rich variety of complex periodic and quasi-periodic
patterns found in systems  such as two-frequency forced Faraday waves, 
we study the interaction of two spatially periodic modes that are  
{\it nearly resonant}. Within 
the framework of two coupled one-dimensional Ginzburg-Landau equations
we investigate analytically the stability of the periodic solutions to general
perturbations, including perturbations that do not respect the
periodicity of the pattern, and which may lead to quasi-periodic solutions. We study the
impact of the deviation from exact resonance on the destabilizing
modes and on the final states. In regimes in which the mode
interaction leads to traveling waves our numerical simulations reveal
localized waves in which the wavenumbers are resonant and which
drift through a steady background pattern that has an off-resonant wavenumber
ratio.

\end{abstract}

% BEGIN MACRO DEFINITIONS
\def\ep{\epsilon}
\def\bs{\boldsymbol}
\def\O{\hbox{\cal O}}
\def\pref#1{(\ref{#1})}
\newcommand{\cierto}[1]{\label{#1}}
\newcommand\cc{\text{c.c.}}

\section{Introduction} 

Pattern-forming instabilities lead to an astonishing variety of
spatial and spatio-temporal structures, ranging from simple, periodic
stripes (rolls) to spatially localized structures and
spatio-temporally chaotic patterns. Even within the restricted class
of steady, spatially ordered patterns a wide range of
patterns \MSC{has}{have} been identified and investigated beyond
simple square or hexagonal planforms including patterns exhibiting
multiple length scales: superlattice patterns, in which the length
scales involved are rationally related rendering the pattern periodic
albeit on an unexpectedly large length scale, and quasipatterns, which
are characterized by incommensurate length scales and which are
therefore not periodic in space.  These more complex two-dimensional
patterns have been observed in particular in the form of Faraday
waves on a fluid 
layer that is vertically shaken with a two-frequency periodic
acceleration function \cite{EdFa94,ArFi98,KuPi98,ArFi02}, and to some
extent also in vertically vibrated Rayleigh-B\'enard
convection \cite{RoSc00a} and in nonlinear optical systems \cite{PaRa95}. The
Faraday system is especially suitable for experimental investigation
of pattern formation in systems with two competing spatial modes of
instability since such codimension-two points are easily accessible by
simply adjusting the frequency content of the periodic forcing
function~\cite{EdFa94,BeEd96}.  The observation of both
superlattices and quasipatterns in this physical system raises an
intriguing question concerning the selection of these kinds of
patterns: what determines whether, for given physical parameters, a
periodic or a quasi-periodic pattern is obtained?  This provides the
main motivation for the present paper in which we address certain
aspects of the selection problem within the somewhat simple framework
of mode competition in one spatial dimension.

To capture the competition between commensurate/incommensurate length
scales we focus on the interaction between two modes with a
wavelength ratio that is close to, but not necessarily equal to, the
ratio of two small integers. We are thus led to consider a {\it
near-resonant} mode interaction. We focus on the case in which both
modes arise in a steady bifurcation. While at first sight it may seem
that an analysis of steady-state modes would not be applicable to
two-frequency forced Faraday waves, for which most of the patterns of 
interest in the present context have been observed, it should be noted 
that very close to onset the amplitude equations for these waves can be reduced
to the standing-wave subspace, in which the waves satisfy equations
that have the same form as those for modes arising from a steady
bifurcation (see, for example,~\cite{PoSi02}). 

The competition between commensurate and incommensurate steady
structures has been addressed previously in the context of a
spatially-periodic forcing of patterns
\cite{LoGo83,Co86,CoHu86,CuEl87,OgZi96}. Using a
one-dimensional, {\it external} forcing of two-dimensional
patterns in electroconvection of nematic liquid crystals,
localized domain walls in the local phase of the patterns were
observed if the forcing wavenumber was 
sufficiently incommensurate with the preferred wavenumber of the
pattern \cite{LoGo83}. Theoretically, the domain walls were
described using a one-dimensional Ginzburg-Landau equation that
included a near-resonant forcing term, which reflects the small
mismatch between the forcing wavenumber and the wavenumber of
the spontaneously forming pattern \cite{Co86,CoEl87}.  The
situation we have in mind in the present paper is similar to
these studies in so far as the competing modes we investigate
also provide a periodic
forcing for each other.  The case of external forcing is
recovered if one of the two modes is much stronger than the
other and consequently the feedback from the forced mode on the
forcing mode can be ignored. We do not restrict ourselves to
this case, however, and thus both modes are active
degrees of freedom and the interaction between them is mutual. 

The analysis of the interaction of two exactly resonating modes near
a co-dimension-two point at which both modes bifurcate off the basic,
unpatterned state has revealed a wide variety of patterns and
dynamical phenomena.  Rich behavior has been found in small systems
in which the interacting modes are determined by the symmetry and
shape of the physical domain (e.g. \cite{SiGo89,MePr86,CrKn90,FeSe89,
Um91,MoKn98,HiVe02}). More relevant for
our goal are the studies of mode-interaction in the presence of
translation symmetry, since they allow the extension to systems with
large aspect ratio, which are required to address the difference
between commensurate and incommensurate structures. Therefore our
investigation builds on a comprehensive analysis, performed by Dangelmayr \cite{Da86}, 
of the interaction between two resonant spatial modes in $O(2)$-symmetric systems
with wavenumbers in the ratio $m:n$, $m<n$ (Here the $O(2)$-symmetry is a 
consequence of restricting to spatially periodic patterns in a translation-invariant
system.).  For $m>1$ there are two primary bifurcations off the
trivial state to pure modes followed by secondary bifurcations to
mixed modes. These mixed modes can in turn undergo a Hopf bifurcation
to generate standing wave solutions and a parity-breaking bifurcation
that produces traveling waves.  For $m=1$ similar results are found
except that there is only one pure-mode state, the one with the higher
wavenumber; the other primary bifurcation leads directly to a branch
of mixed modes.  Further analyses of this system have revealed
structurally and asymptotically stable heteroclinic cycles near the
mode interaction point when $m:n=1:2$
\cite{JoPr87,PrJo88,ArGu88}. More recently, complex dynamics organized
around a sequence of transitions between distinct heteroclinic cycles
has been discovered in resonances of the form $1:n$
\cite{PoKn00,PoKn01}, in particular, in the cases $n=2$ and $n=3$.

Although previous work on resonant mode interactions considered
only strictly periodic solutions it provided insight into
various phenomena that were observed experimentally in
large-aspect ratio systems involving large-scale modulations of
the patterns. For example, in steady Taylor vortex flow it was
found experimentally that  not too far from threshold the band
of experimentally accessible wavenumbers is
substantially reduced compared to the stability limit obtained
by the standard analysis of side-band instabilities in the
weakly nonlinear regime \cite{DoCa86}. The origin of this strong
deviation was identified to be a saddle-node bifurcation
associated with the $1:2$ mode interaction \cite{RiPa86}. In
directional solidification \cite{SiBe88} localized drift waves
have been observed, which arise from the parity-breaking bifurcation 
\cite{MaTr84,CoGo89,GoGu91} that is associated with the resonant mode interaction
with wavenumber ratio $1:2$ \cite{LeRa91}. Subsequently such
waves have also been obtained in a variety of other systems
including directional viscous fingering \cite{RaMi90}, Taylor
vortex flow \cite{WiMc92,RiPa92}, and premixed flames
\cite{BaMa94}. In our treatment of the near-resonant case the
mode amplitudes are allowed to vary slowly in space. It
therefore naturally incorporates phenomena like the localized
drift waves and the modification of side-band instabilities by
the resonance.  

In this paper we study the interaction of two nearly-resonant modes
in a spatially extended, driven, dissipative system. Near onset we
model the slow dynamics of such systems by two amplitude equations of
Ginzburg-Landau type, one for each mode. We focus on the weak
resonances ($m+n\ge 5$) in order to avoid some of the specific
features of the strong-resonance cases (for  example, the
structurally stable heteroclinic cycles in the case $m:n = 1:2$). Our
primary goal is to investigate the transitions between periodic and
quasi-periodic states that take place as the result of side-band
instabilities, with an eye on how the detuning from exact spatial 
resonance influences this process. We find that the detuning can
play  an important role in the selection of the final wavenumbers of
the modes involved. For example, it can  favor a periodic to
quasi-periodic transition that would otherwise (i.e., in the  case of
exact resonance) result in a second periodic state. Among the various
quasi-periodic states that we find in numerical simulations
are several that consist of drifting localized structures with alternating
locked (periodic) and unlocked (quasi-periodic) domains. 

It should be noted that at present there is no rigorous justification
for the description of quasi-periodic patterns using low-order
amplitude expansions. In fact, the lack of straightforward
convergence of such an expansion has been investigated recently for
two-dimensional quasi-patterns \cite{RuRu03}. We will not discuss
these issues; instead we use the coupled Ginzburg-Landau equations as
model equations that are known to be the appropriate equations for
periodic patterns and at the same time also allow quasi-periodic
solutions.

The organization of the paper is as follows. In Section 2 we set
up the coupled Ginzburg-Landau equations that are based on the
truncated normal form  equations for the $m:n$-resonance. In
Section 3 we utilize and build upon on the detailed results of 
Dangelmayr \cite{Da86} to describe the stability properties of
steady spatially  periodic states (i.e, pure and mixed modes)
with respect to perturbations that preserve the periodicity of
the pattern. In Section 4 we turn to the question of stability
of the steady periodic solutions with respect to side-band
instabilities. Here we also determine how these instabilities
are affected by the detuning from perfect  resonance.  Numerical
simulations of the system are carried out in Section 5 to 
investigate the nonlinear evolution of the system subsequent to
the side-band  instability discussed in the previous Section.
Our concluding remarks are  given in section 6. It
should be pointed out that a very recent preprint by Dawes {\em
et al.} presents a complementary investigation near the $1:2$
resonance \cite{DaPo03}.

\section{The amplitude equations}\cierto{sec:amplitudequations}
 
We consider driven, dissipative systems in one
spatial dimension that are
invariant under spatial translations
and reflections.  We further assume
that in the system of interest there are two distinct spatial
modes that destabilize the basic homogeneous state
nearly simultaneously.  The wavenumbers of these two modes,
$q_1$ and $q_2$, correspond to minima of the neutral curves and
are assumed to be in approximate spatial resonance, i.e., 
\begin{equation}  
n(q_1+\varepsilon\hat\gamma)= m q_2, \qquad |\varepsilon| \ll 1,
\cierto{1b}   
\end{equation} 
where $m$ and $n > m$ are positive
co-prime integers and the term  $\varepsilon\hat \gamma$
measures the deviation from perfect resonance.  Physical
fields,  near onset, may then be expanded in terms of these two
spatial modes:  
\begin{eqnarray} 
u(x,t)=\varepsilon[A_1(X,T)e^{i(q_1+\varepsilon\hat{\gamma} )x}+
A_2(X,T)e^{iq_2x}+\cc]+\cdots \cierto{1a}  
\end{eqnarray} 
The two modes are allowed to vary on slow spatial and temporal
scales $X=\varepsilon x$ and $T=\varepsilon^2t$, respectively. Note
that in the expansion~(\ref{1a}) we do not expand about about the
minima $q_{1,2}$ of the neutral stability curves, which for
$\hat\gamma \neq 0$ are not in spatial resonance,  but take
instead a mode $A_1$ which {\it is} in exact spatial resonance
with $A_2$.  This choice of $A_1$ and $A_2$  simplifies the
equations, avoiding any explicit dependence on the spatial
variable in the resulting amplitude equations. Furthermore, we
allow for a small offset between  the critical values of the forcing
amplitudes $F_{ic}$ of the two modes (see Fig.~\ref{fig:critic-modes}).

 %%%%%%% FIG00%%%%%%%%%%%%%
\begin{figure}[htb]
\centerline{\includegraphics[width=7cm]{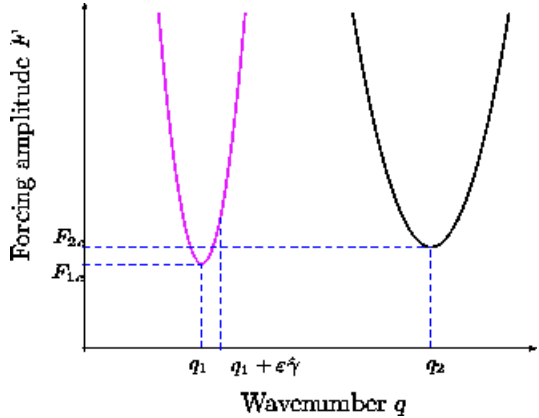}}
\caption{Sketch of the neutral stability curve for the two modes.} 
\cierto{fig:critic-modes}
\end{figure} 
%%%%%%%%%%%%%%%%%%%%%%%%%%%%%%%%%%%

The equations 
governing the evolution of $A_1$ and $A_2$ must be equivariant under the 
symmetry operations generated by spatial translations ($T_\varphi$) and 
spatial reflections  ($R$), which act on $(A_1,A_2)$ as follows: 
\begin{eqnarray} 
T_\varphi :(A_1, A_2)  \rightarrow &(e^{im\varphi} A_1, e^{in\varphi} A_2), 
\quad \text{for} \quad \varphi \in \text{$[$}0, 2\pi),\\   \nonumber
R: (A_1, A_2)  \rightarrow & (\bar A_1,\bar A_2),
\cierto{symmetries} 
\end{eqnarray}       
where the bar denotes the complex  conjugate. Consistent with this 
equivariance requirement the slow evolution of $A_1$ and $A_2$ can 
be approximated, after rescaling, by the Ginzburg-Landau equations

\begin{eqnarray} 
    A_{1 T}&=&\mu A_1+\delta A_{1XX}-i\gamma A_{1X}-(s|A_1|^2+\rho|A_2|^2)A_1
            +\nu \bar A_1^{n-1}A_2^m, \cierto{2}\\ 
    A_{2 T}&=&(\mu+\Delta\mu) A_2+\delta' A_{2XX}-(s'|A_2|^2+\rho'|A_1|^2)A_2 
               +\nu' \bar A_2^{m-1}A_1^n.\cierto{3}
\end{eqnarray} 
The subscripts indicate partial derivatives with respect to $X$
and $T$.  The main control parameter $\mu\propto
(F-F_{1c})/\varepsilon^2$ measures the  magnitude of the overall
forcing.  In addition, we keep track of the offset in  the two
critical forcing amplitudes with $\Delta \mu 
\equiv(F_{1c}-F_{2c})/\varepsilon^2$ (see
Fig.~\ref{fig:critic-modes}), and capture the detuning between $q_1$
and $m q_2/n$ with $\gamma\equiv 2\delta\hat \gamma$.  The local 
curvature of the neutral stability curves near $q_1$ and $q_2$ is measured by $\delta$
and $\delta'$, respectively.  We further assume that the nonlinear
self- and cross- interaction coefficients satisfy the non-degeneracy
conditions $ss' \neq 0$ and $ss'-\rho\rho'\neq 0$ and perform a simple
rescaling such that $s=\pm 1$, $s'=\pm 1$.

One goal of this paper is to gain insight into the difference
between periodic and quasi-periodic patterns in systems with two
unstable wavenumbers. If the two wavenumbers are not rationally
related and their irrational ratio is kept fixed as the onset
for the two modes is approached, then only terms of the form
$A_i|A_j|^{2l}$, $l=1,2,3\ldots$, appear in the equation for
$A_i$. For a rational ratio, however, additional nonlinear terms
arise, which couple the otherwise uncoupled phases of the two
modes $A_i$. For the $m:n$-resonance the leading-order resonance
terms are given by $\bar{A}_1^{n-1}A_2^m$ and
$\bar{A}_1^{m-1}A_1^n$ in the equations for $A_1$ and $A_2$,
respectively. In order to explore the connection between the
rational and the irrational case we consider a wavenumber ratio
which may be irrational, but its deviation from the ratio $m:n$
is of ${\mathcal O}(\varepsilon)$. This allows the
mismatch between the two wavenumbers to be captured by the slow
spatial variable $X$ and to describe periodic and quasi-periodic
patterns with the same set of equations (\ref{2},\ref{3}).
Equivalently, we could have expanded in the irrationally related 
wavenumbers $q_1$ and $q_2$ associated with the minima of the neutral 
curves in Fig.~\ref{fig:critic-modes}. Then the resonance terms would 
introduce space-periodic coefficients with a period that is related to 
the mismatch of the wavenumbers. Our choice of the expansion wavenumbers
(cf. (\ref{1b})) removes this space-dependence and introduces the first-order
 derivative $-i\gamma \partial_X A_1$ in its place.

We focus on the weak resonances, $m+n\geq 5$, in which the 
resonant terms are of higher order. We neglect, however, non-resonant terms 
of the form $A_{i} |A_{j}|^p$ ($i, j=1,\, 2$ and $4\leq p\in\Bbb N$) which may 
arise at lower order.  This is motivated by the observation that such terms
do not contribute any qualitatively new effects for small
amplitudes.  The resonant terms, in contrast, remove the unphysical
degeneracy that arises when the phases are left uncoupled   and can
therefore influence dynamics in a significant way despite appearing
at   higher order. Note, however, that near onset the resonant terms
are typically small and the phase coupling between $A_1$ and $A_2$
occurs on a very slow time scale. The coupling becomes stronger
further above onset where the weakly nonlinear analysis may no longer
be valid.

It is often useful to recast Eqs.~(\ref{2},\ref{3}) in terms of
real amplitudes $R_j \ge 0$ and phases $\phi_j$ by writing $A_j=R_je^{i\phi_j}$.  
This leads to
\begin{eqnarray}
  R_{1T} &=&\mu R_1-(sR_1^2+\rho R_2^2)R_1+\nu R_1^{n-1}R_2^m
              \cos(n\phi_1-m\phi_2) \nonumber\\
              &&+\delta R_{1XX}-\delta\phi_{1X}^2R_1
              -2\gamma\phi_{1X}R_1,\cierto{6a}\\
  R_{2T} &=&(\mu+\Delta\mu) R_2-(s'R_2^2+\rho' R_1^2)R_2+\nu' R_2^{m-1}R_1^n
                \cos(n\phi_1-m\phi_2)\nonumber\\
                &&+\delta' R_{2XX}-\delta'\phi_{2X}^2R_2 ,\cierto{7a}\\
  R_1\phi_{1T} &=&-\nu R_1^{n-1}R_2^{m} \sin(n\phi_1-m\phi_2)+
        \delta\phi_{1XX}R_1+ 2\delta\phi_{1X}R_{1X}+2\gamma R_{1X},\cierto{8aa}\\
 R_2\phi_{2T} &=&\nu' R_2^{m-1}R_1^{n} \sin(n\phi_1-m\phi_2)+
        \delta'\phi_{2XX}R_2+ 2\delta'\phi_{2X}R_{2X}.\cierto{9a}
 \end{eqnarray}
If the spatial dependence in Eqs.~(\ref{6a}-\ref{9a}) is ignored, the system reduces
to a set of ordinary differential equations  (ODEs), equivalent to the one analyzed
by Dangelmayr \cite{Da86}.  In this simplified problem, the translational symmetry
($T_\varphi$) causes the overall phase to decouple and leaves only the three real
variables $R_1$, $R_2$, and the mixed phase \begin{equation}
\phi=n\phi_1-m\phi_2, \cierto{9b}
\end{equation}
with dynamically important roles.  Dangelmayr's bifurcation analysis 
produced expressions for the location of  primary bifurcations to pure mode solutions,
secondary bifurcations to mixed mode solutions, and, in some instances, tertiary 
bifurcations to standing-wave and traveling-wave solutions.  These results apply 
to a general $m:n$ resonance, and prove useful in what follows.

\section{Steady Spatially-Periodic Solutions}
\cierto{sec:basicsolutions}

In this section we analyze steady solutions of
Eqs.~(\ref{2},\ref{3}) of the form 
\begin{equation}
A_1=R_1e^{i(kX+\hat\phi_1)},\quad A_2=R_2e^{i((nk/m)X+\hat\phi_2)}, \cierto{4}
\end{equation} 
where $R_{1,2}\geq 0$, and $\hat\phi_{1,2}$ and $k$ are real. 
Such states  represent spatially periodic solutions of the
original problem with wavenumbers $\tilde
q_1=q_1+\varepsilon\gamma +\varepsilon k$ and $\tilde
q_2=q_2+\varepsilon n k/m$ so that $\tilde{q}_1n=\tilde{q}_2m$.
These solutions break the continuous translational
symmetry ($T_\varphi$) but remain invariant under discrete
translations. 

Within this family of steady states there are generically only
two types of nontrivial solutions, pure modes and mixed modes, 
which we describe below. 

\noindent
{\bf I. Pure modes} ($S_{1,2}$).  These are single-mode states, which  take one
of two forms:
 \begin{align}
 S_1 &: (A_1, A_2) =(\sqrt{\alpha/s}\,\,e^{i (kX+\hat\phi_1)}, 0),
\quad\text{for $m>1$},\nonumber\\
 S_2 &: (A_1, A_2) =(0, \sqrt{\beta/s'}\,\,e^{i ((nk/m)X+\hat\phi_2)}),
\cierto{5}\end{align}
where $\hat\phi_1, \hat\phi_2\in [0, 2\pi)$ and
\begin{align}
&\alpha=\mu-\delta k^2+\gamma k,\nonumber\\
&\beta=\mu+\Delta\mu-\delta'(n k/m)^2.\cierto{5b}
\end{align}
Note that pure modes of type $S_1$ are not present if $m=1$ (see
Eqs.~(\ref{2},\ref{3})). Moreover, the pure modes $S_1$ and
$S_2$ are not isolated, but emerge as circles  of equivalent
solutions (parametrized by $\hat\phi_1$ or $\hat\phi_2$).
Hereafter we consider resonances $m:n$ where $m\geq 2$, in which
case both pure modes $S_1$ and $S_2$ are present.

\noindent
{\bf II. Mixed modes} ($S_\pm$). There are two types of mixed modes,

\begin{equation}
S_\pm: (A_1, A_2) =(R_1\,\,e^{i (kX+\hat\phi_1)}, 
R_2\,\,e^{i ((nk/m)X+\hat\phi_2)}),\cierto{mixmod}
\end{equation}
satisfying
\begin{align}
   S_{\pm} :\quad &\cos(\phi)=\pm 1, \nonumber\\
                  & \begin{array}{ll}
 &(s'\alpha-\rho\beta)=(ss'-\rho\rho')R_1^2\pm (s'
\nu R_2^2-\nu'\rho R_1^2)R_2^{m-2}R_1^{n-2},\\
&(s\beta-\rho'\alpha)=(ss'-\rho\rho')R_2^2\pm (s
\nu' R_1^2-\nu \rho' R_2^2)R_2^{m-2}R_1^{n-2}.\end{array}\cierto{6}
\end{align}
\noindent
Here $\phi$ is the mixed phase given by Eq.~(\ref{9b}),  and $\alpha$ and
$\beta$ are defined by Eq.~(\ref{5b}).   As in the case of the pure modes
$S_{1,2}$, translational symmetry implies that there are circles of equivalent
mixed-modes states (parametrized by $\hat\phi_1$, say). Like the
pure modes, the mixed modes are invariant under reflections $(R)$ through an appropriate origin. 
 
\subsection{Stability under homogeneous perturbations}

The stability of $S_{1,2}$ and $S_{\pm}$ under homogeneous
perturbations can be obtained from  Dangelmayr's analysis
\cite{Da86}, which we review here in some detail to provide the
background necessary for our analysis. The stability regions in
the $(\alpha, \beta)$-unfolding plane simply need to be mapped
to the $(k,\mu)$-plane with the (nonlinear) transformation
(\ref{5b}). Each intersection of the curves $\alpha=0$ and
$\beta=0$ corresponds to the codimension-two point of
\cite{Da86}; there are generically zero or two such
intersections. Since the nonlinear coefficients are identical in
the vicinity of both intersections, the response of the system
to homogeneous perturbations in  corresponding neighborhoods of
the (two) intersections  is identical.  In  particular, any
bifurcation set arising from one intersection arises from the 
other as well.  Several examples are given in
Fig.~\ref{fig:dgrgamma05}, which  shows the various bifurcation
sets in the $(k, \mu)$-plane in four  representative cases.
These bifurcations are described below. Note that for mode 
$A_1$ the deviation of the wavenumber from the critical
wavenumber is given by  $\varepsilon k$, whereas for mode $A_2$
it is given by $\varepsilon nk/m$.  Additional details, valid in
sufficiently small neighborhoods of the  intersections, are
available in \cite{Da86}.

%%%%%%% FIG1%%%%%%%%%%%%%
\begin{figure}
\centerline{\includegraphics[width=6cm]{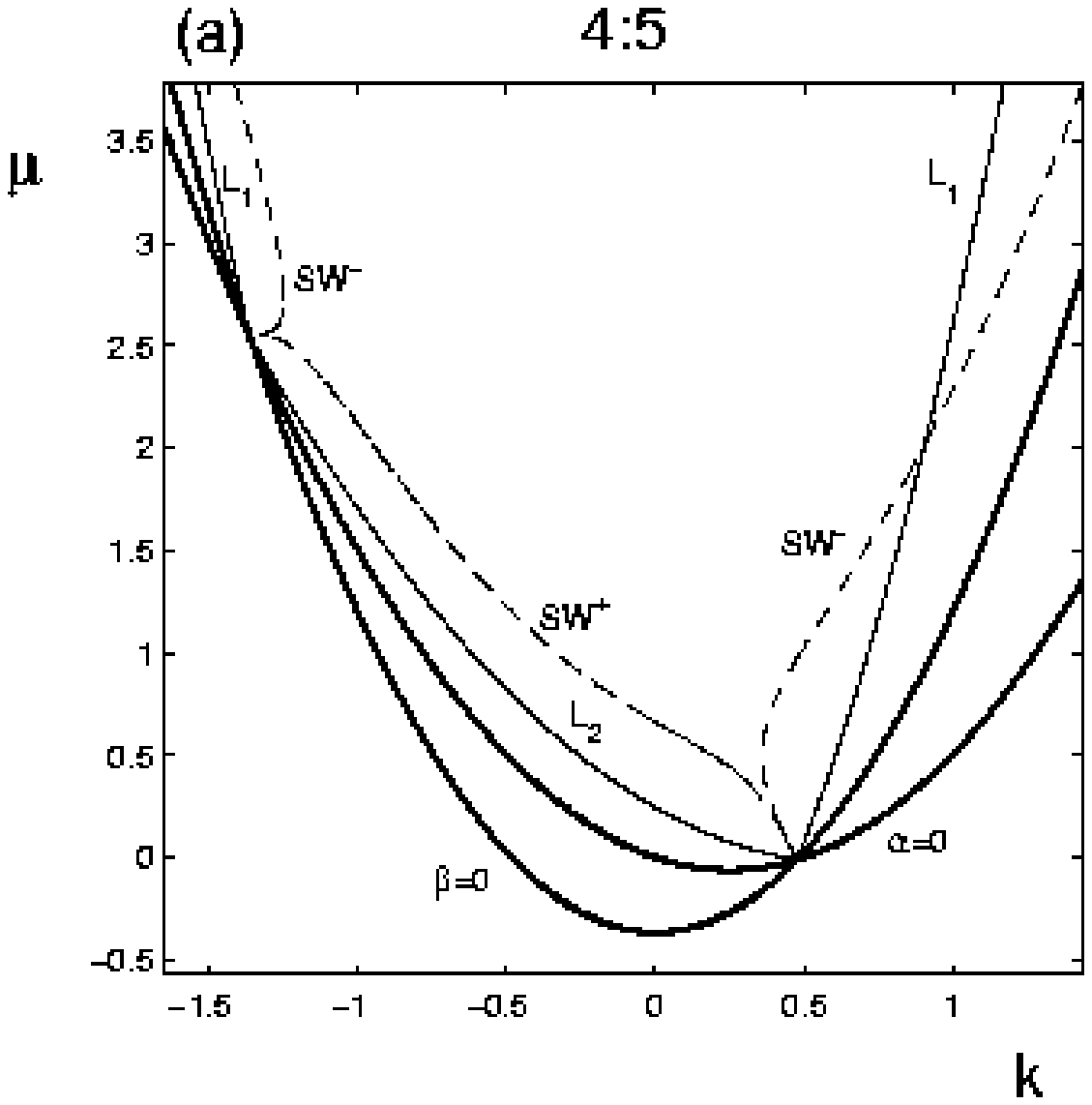}
\quad \includegraphics [width=6cm]{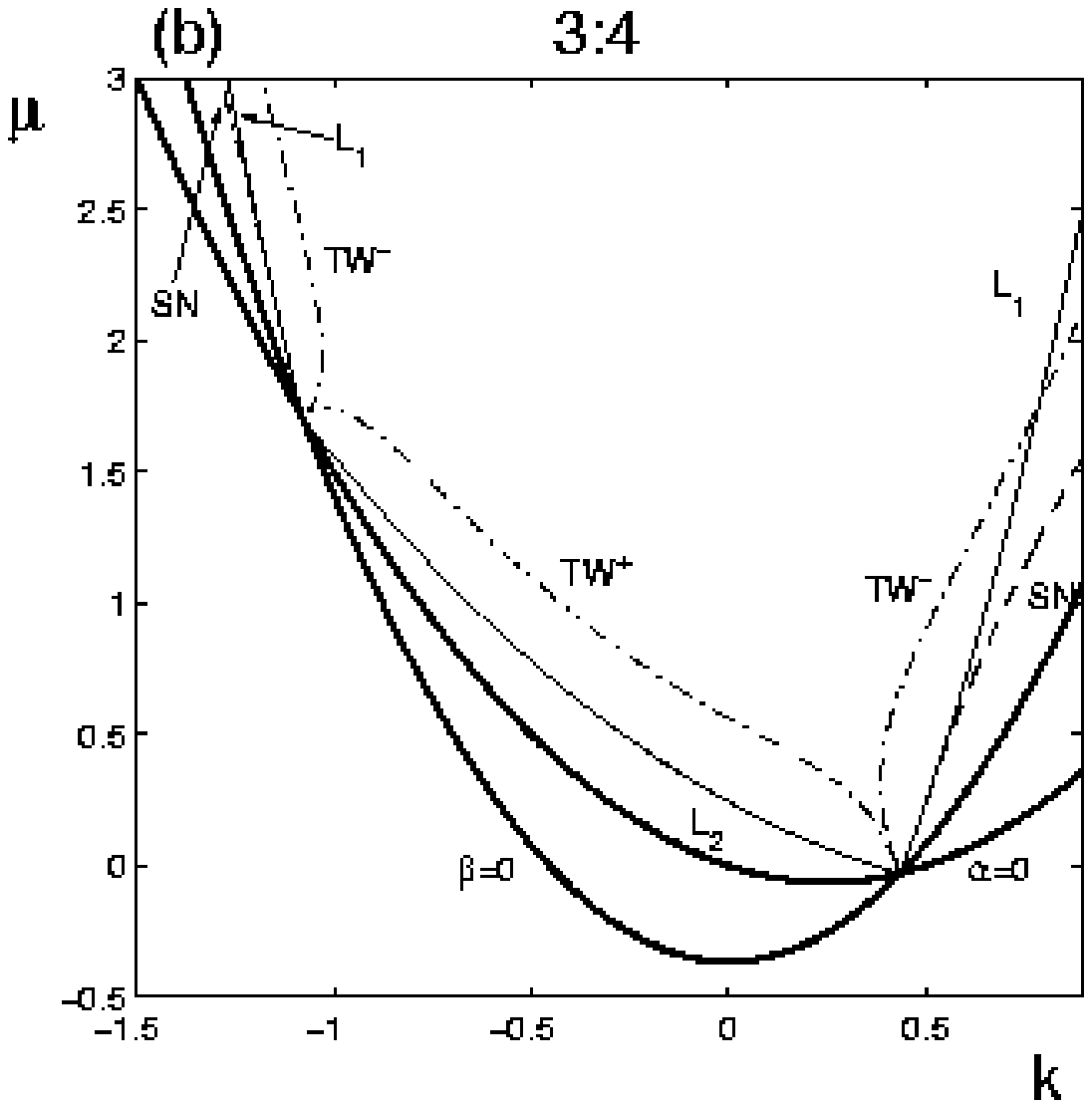}}
\centerline{\includegraphics [width=6cm]{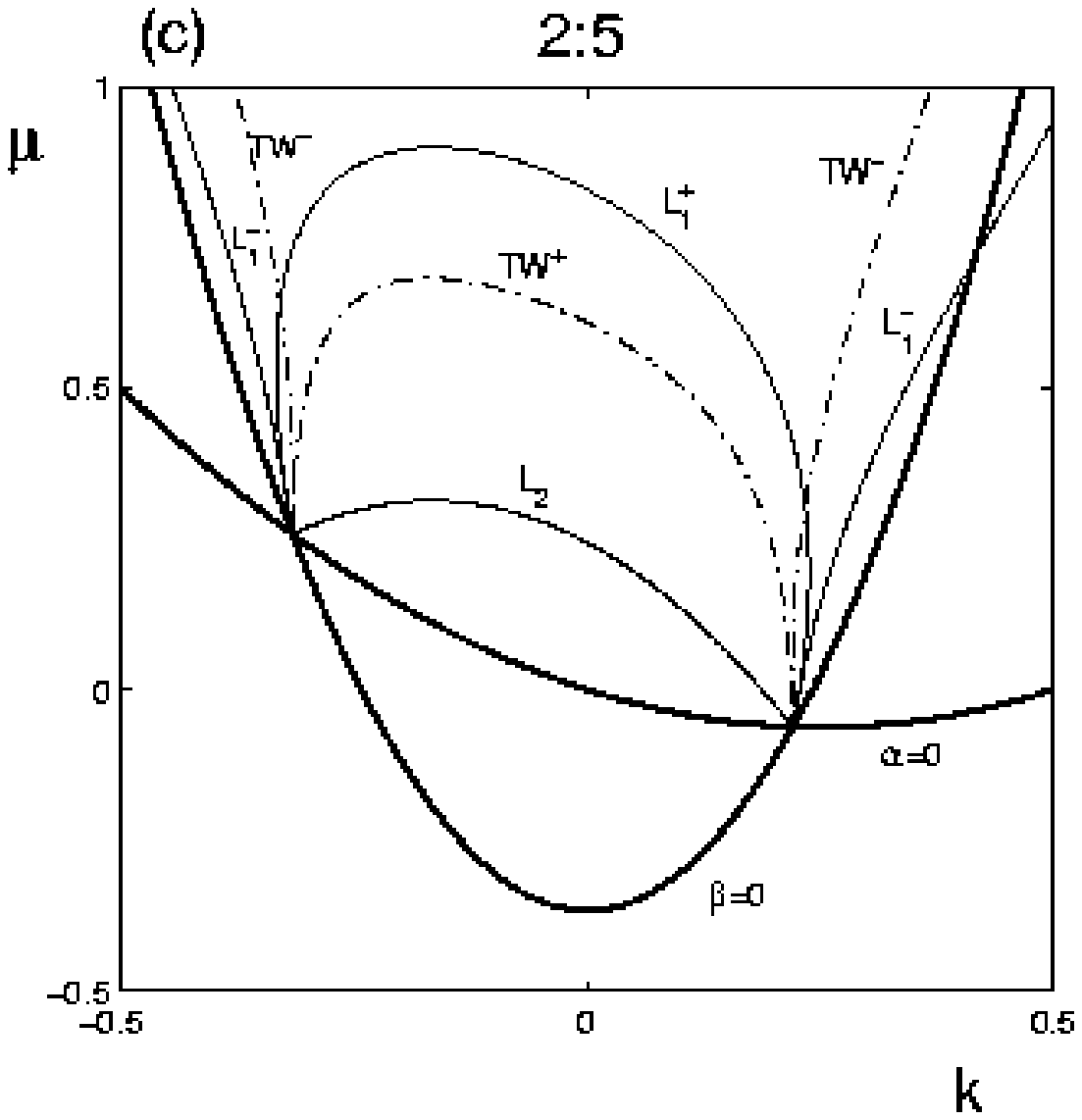}
\quad \includegraphics[width=6cm] {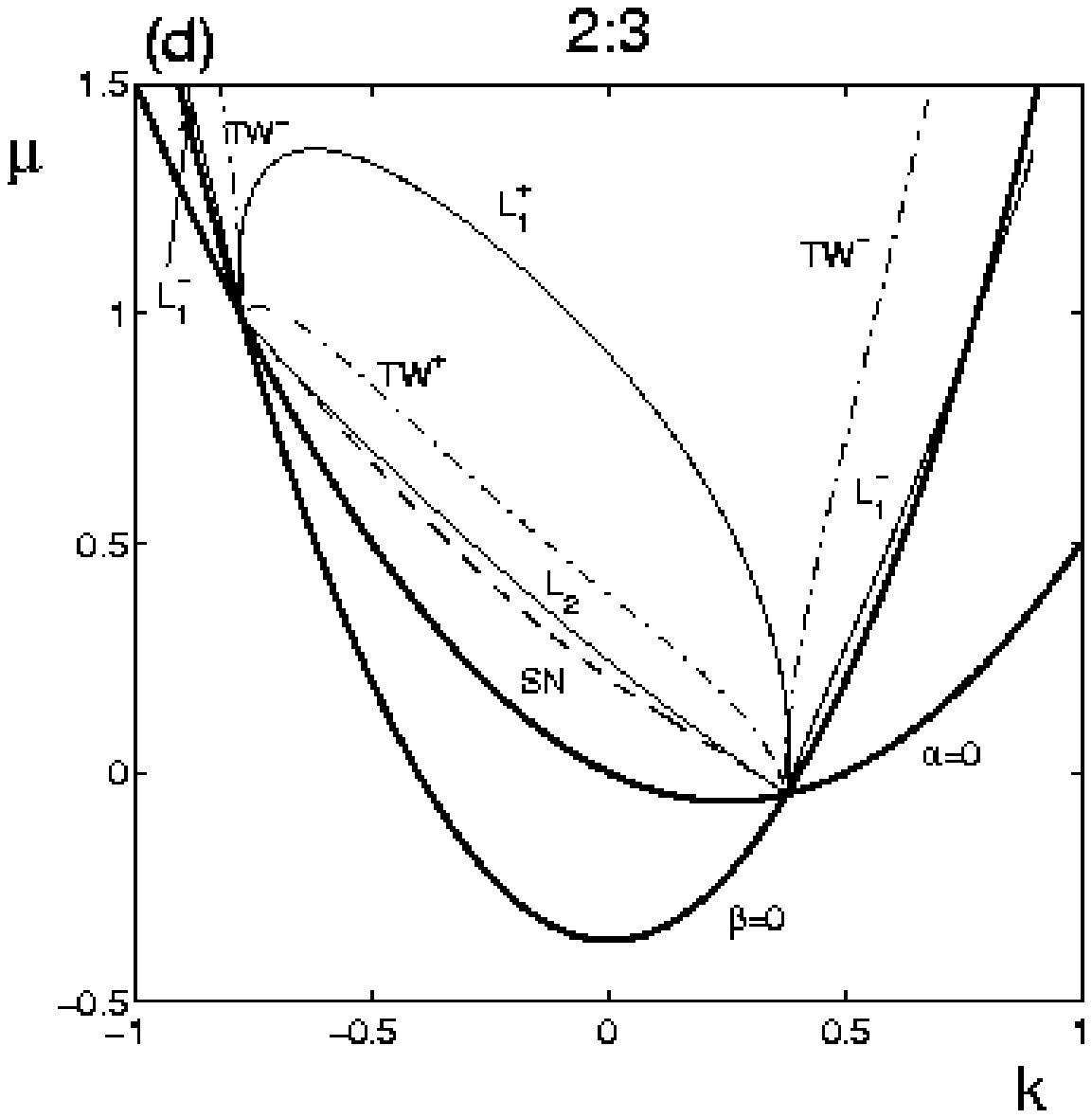}}
\caption{Bifurcation sets indicating the creation of pure modes (thick solid
lines), mixed modes (thin solid lines $L_1^\pm$, $L_1$ and $L_2$),  standing
waves (dashed lines $SW^\pm$) and traveling waves (dash-dotted lines
$TW^\pm$), as well as saddle-node bifurcations (dashed lines $SN$).  The specific
resonance $m:n$ is indicated above the plots.  In each case $\Delta\mu=0.5$,
$\gamma=0.5$, $\delta=\delta'=1$, $\rho=0.4$, and $\rho'=0.67$.  In (a)
$s=-s'=1$,
$\nu=0.62$ and $\nu'=1.02$, but in (b), (c) and (d)  $s=s'=-1$, $\nu= 0.62$ and
$\nu'=-1.02$.}
\cierto{fig:dgrgamma05}
\end{figure}
%%%%%%%%%%%%%%%%%%%%%%%

In this paper we study (\ref{2},\ref{3}) with the coefficients $\nu$
and $\nu'$ of the higher-order resonance terms taken
to be of order 1.  The resulting stability and bifurcation diagrams
therefore contain certain features that do not remain local to the
bifurcation when $\nu,\nu' \rightarrow 0$, i.e. in this limit these
features disappear at infinity, $\mu \rightarrow \infty$, and do not
represent robust aspects of the mode-interaction problem. We
return to this issue briefly at the end of this section, and indicate
which features of our sample bifurcation sets are not robust in the
limit $\nu,\nu'\rightarrow 0$.

The pure modes $S_1$ and $S_2$ bifurcate from the trivial state when
$\alpha=0$ and $\beta=0$, respectively: the bifurcation to $S_1$
($S_2$) is supercritical if $s=1$ ($s'=1$).
Their stability is determined by four eigenvalues, one of which is forced to be zero by 
translation symmetry. In the case of $S_1$ the remaining three eigenvalues are 
\begin{eqnarray}
&&\,\,\,\lambda_0^{(1)}=-2\alpha/s,\nonumber\\
&L_1:&\,\,\lambda_\pm^{(1)}=(\beta-\rho'\alpha)/s\pm\left\{
\nu'\left |\alpha/s\right|^{n/2}\right\}_{m=2},  \cierto{7}
\end{eqnarray}
where the  bracketed term with subscript $m=2$ is present only if $m=2$.
For $S_2$ the eigenvalues are given by
\begin{eqnarray}
&&\lambda_0^{(2)}=\beta/s',\nonumber\\
&L_2:&\,\,\lambda_+^{(2)}=\lambda_-^{(2)}=(s'\alpha-\rho\beta)/s'.
\cierto{8}\end{eqnarray}

When one of the eigenvalues $\lambda_\pm^{(1)}$
($\lambda_\pm^{(2)}$) changes sign the pure modes $S_1$ ($S_2$)
become unstable to mixed modes, respectively.
For $S_2$ the two eigenvalues coincide and the bifurcation occurs
along a single line denoted by $L_2$ in Fig.~\ref{fig:dgrgamma05}. 
Similarly, for $m>2$ the transition from $S_1$ to the mixed modes $S_\pm$ occurs along
the single curve $L_1$ (Fig.~\ref{fig:dgrgamma05}a,b). For $m=2$
the eigenvalues $\lambda_\pm^{(1)}$ are not degenerate and the
line $L_1$ splits into two curves, $L_1^+$ and  $L_1^-$; the
mixed mode $S_+$ bifurcates at $L_1^+$ and $S_-$ bifurcates at 
$L_1^-$ ($L_1^\pm$ in
Figs.~\ref{fig:dgrgamma05}c-d).  The splitting  is due to the
presence of the resonant terms which are linear in $A_2$ if
$m=2$   but not otherwise.   Both curves, $L_1^+$ and $L_1^-$,
become tangent to each  other at the intersection
$\alpha=\beta=0$.  

The response of the mixed modes to amplitude perturbations is
decoupled (due to reflection symmetry) from the effect of phase 
perturbations. The amplitude stability is determined by the 
eigenvalues of a $2 \times 2$-matrix $M_{\pm}$, whose determinant 
and trace can be written as (recall we consider $m>1$)
\begin{eqnarray}
\text{det}(M_\pm)
&=&-4(\rho\rho'-ss')R_1^2R_2^2\pm R_1^{n-2}R_2^{m-2}
H(R_1,R_2), \cierto{10}\\
\text{Tr}(M_\pm) & = &-2
s(R_1^2+ss'R_2^2)\pm R_1^{n-2}
R_2^{m-2}(\nu(n-2)R_2^2+\nu'(m-2)R_1^2).\cierto{11}
\end{eqnarray}
with
\begin{equation}
H(R_1,R_2)=-2\left[\nu's (m-2) R_1^4+
\nu s' (n-2)R_2^4-R_1^2R_2^2(\rho'\nu m+\rho\nu'n)\right]
\end{equation}
Here $R_1$ and $R_2$ are solutions of Eqs.~(\ref{6}).
Since in general $\nu$ and $\nu'$ are of ${\mathcal
O}(\varepsilon^{m+n-4})$,  the contribution $H(R_1,R_2)$ from the
resonance term affects the steady bifurcation determined by
(\ref{10}) only for $m\leq 3$. 
In particular, in the cases $m:n=3:n$ and $m:n=2:3$ the function $H$ can balance the 
first term along curves through the codimension-2 point along which
$R_2 \ll R_1$ and $R_1 \ll R_2$, respectively, and one of the mixed states experiences 
a saddle-node bifurcation (curves labeled $SN$ in Figs.~\ref{fig:dgrgamma05}b,d) 
\cite{Da86}. For $2:n$ resonances with $n\ge 5$ the 
term $R_1^4$ drops out in $H(R_1,R_2)$. Consequently, the $H$-term cannot balance 
the first term in (\ref{10}) and no saddle-node bifurcations occur.

For $m \ge 4$ the sign of det($M_\pm$) 
does not depend on $R_{1,2}$. Then the $S_\pm$ solutions are always 
unstable for $(\rho\rho'-ss')>0$, while for $(\rho\rho'-ss')<0$  their
stability must be deduced from the sign of Tr($M_\pm$). If
$ss'=1$ it follows from Eq.~(\ref {11}) that for $\varepsilon
\ll 1$ sign(Tr($M_\pm))=-s$; the mixed modes $S_\pm$ are then stable to
amplitude  perturbations if $s=+1$.  On the other hand, if
$ss'=-1$, the trace of $M_\pm$  can change sign, indicating the
possibility of a Hopf bifurcation. The resulting  time-periodic
solutions inherit the reflection symmetry of $S_\pm$ (so
$\dot\phi_1=\dot\phi_2=0$) and therefore correspond to standing
waves ($SW$  curves in  Fig.~\ref{fig:dgrgamma05}a).

Instabilities associated with perturbations of the mixed phase
(\ref{9b}) lead to bifurcations breaking the reflection
symmetry. The relevant eigenvalues are given by 
 \begin{equation}
TW:\,\,   e_\pm=\mp(\nu n R_2^2+\nu' mR_1^2) R_1^{n-2}
R_2^{m-2}, \cierto{12}
\end{equation} 
and may pass through zero only if sign$(\nu\nu')=-1$.  In this
case the $S_\pm$ states undergo a pitchfork bifurcation,
reflection symmetry is broken, and traveling waves appear ($TW$
curves in Figs.~\ref{fig:dgrgamma05}b-d).  These
traveling-wave solutions manifest themselves as
fixed points of the three-dimensional
ODE system involving $R_1$, $R_2$ and $\phi$, but are seen to be
traveling waves by the fact that the individual phase velocities
are nonzero: $\dot\phi_1/\dot\phi_2=n/m$. Since the phase
velocity of these waves goes to 0 at the bifurcation, they are
often called drift waves. We do not consider 
the stability properties of the
traveling-wave solutions in this paper.

The stability results for $S_{1,2}$ and $S_{\pm}$ described
above are illustrated in
Figs.~\ref{fig:dgrm2n5-gam0deltamu0}-\ref{fig:dgrm2n5Delta-s-}
for $m:n=2:5$ and the indicated parameter values.
They all satisfy  
\begin{equation}  s=s'=1,\qquad
ss'-\rho\rho'>0,\qquad  \nu\nu'<0, 
\cierto{paramset} 
\end{equation}  
so that the pure modes $S_{1,2}$ bifurcate supercritically in
all cases. Next to these plots we sketch the type of bifurcation
diagram one obtains when increasing $\mu$ at constant $k$
along the thin dashed vertical lines. Since in
all cases $ss'-\rho\rho'>0$, both mixed modes $S_{\pm}$ are
stable to amplitude perturbations (see the explanation following
Eqs.~(\ref{10},\ref{11})).  The stability of $S_\pm$ with regard
to phase perturbations depends,
however, on the eigenvalue $e_\pm$ given by (\ref{12}). Because
$\nu\nu'<0$, this eigenvalue may change sign, causing the mixed
modes $S_\pm$ to undergo a symmetry-breaking bifurcation to
traveling waves.
Figs.~\ref{fig:dgrm2n5-gam0deltamu0}-\ref{fig:dgrm2n5Delta-s-}
present six different cases characterized by the following
quantities:
\begin{equation}
{\chi}=\gamma^2-4\Delta\mu\left(\delta-(\frac{n}{m})^2\delta'\right) \quad
\text{and} \quad {\Lambda}= \frac{\gamma^2}{4 \delta}-\Delta\mu.
\cierto{deltas} 
\end{equation} 
The parameter $\chi$ controls the intersection of the parabolas
$\alpha=0$ and $\beta=0$; they intersect if $\chi \geq 0$ and
not otherwise.  The quantity $\Lambda$ determines the relative
position ($\mu$-value) of the minima of the curves $\alpha=0$
and $\beta=0$. It thus indicates which of the two modes, $S_1$
or $S_2$,  is excited first.  $S_2$ appears first when
$\Lambda>0$, while $S_1$ takes priority for $\Lambda<0$; if
$\Lambda=0$  both pure modes onset simultaneously.

%%%%%%%%%%%% FIG%%%%%%%%%%%%%%%%%%%%%%%
\begin{figure}
{\includegraphics[width=14cm]{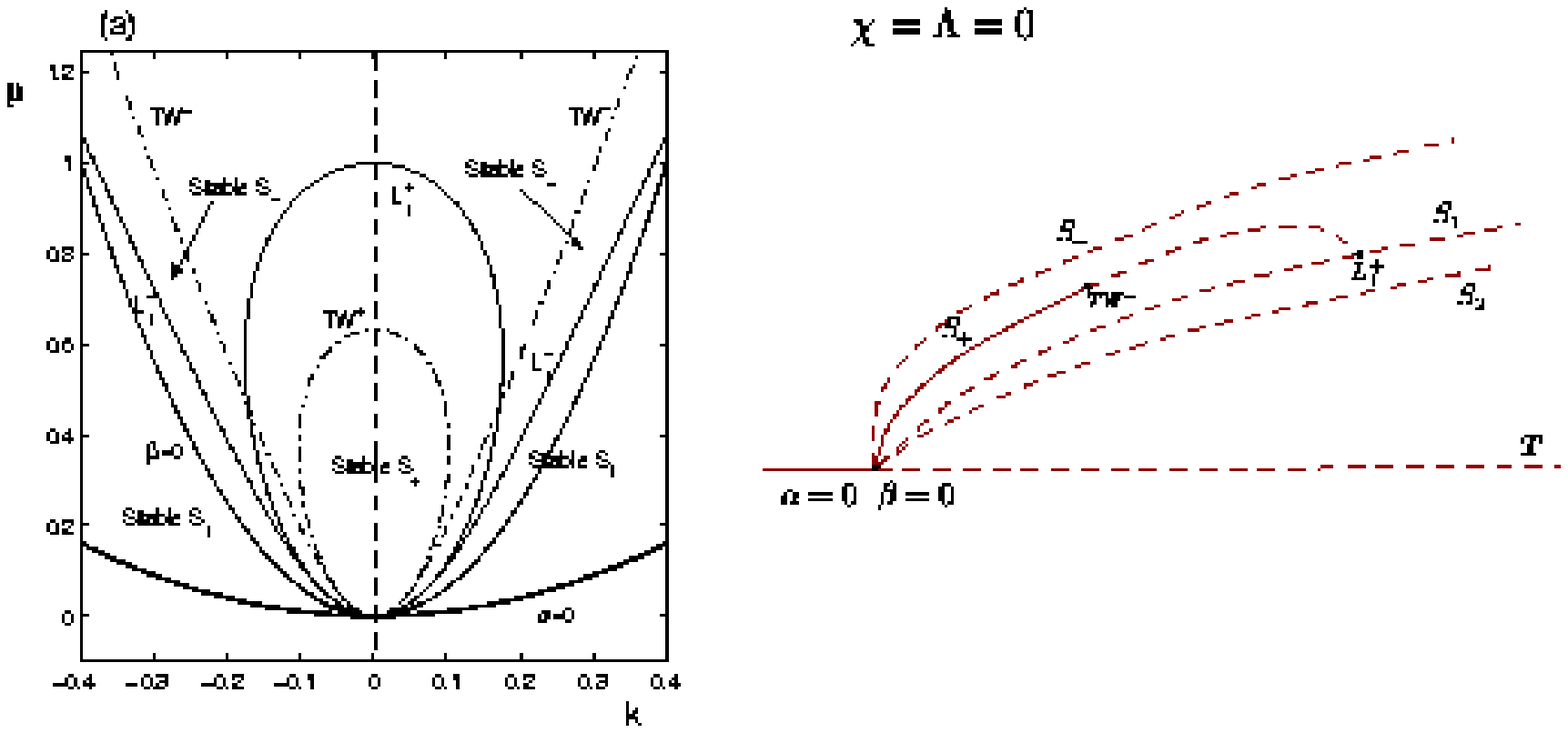}}
{\includegraphics[width=14cm]{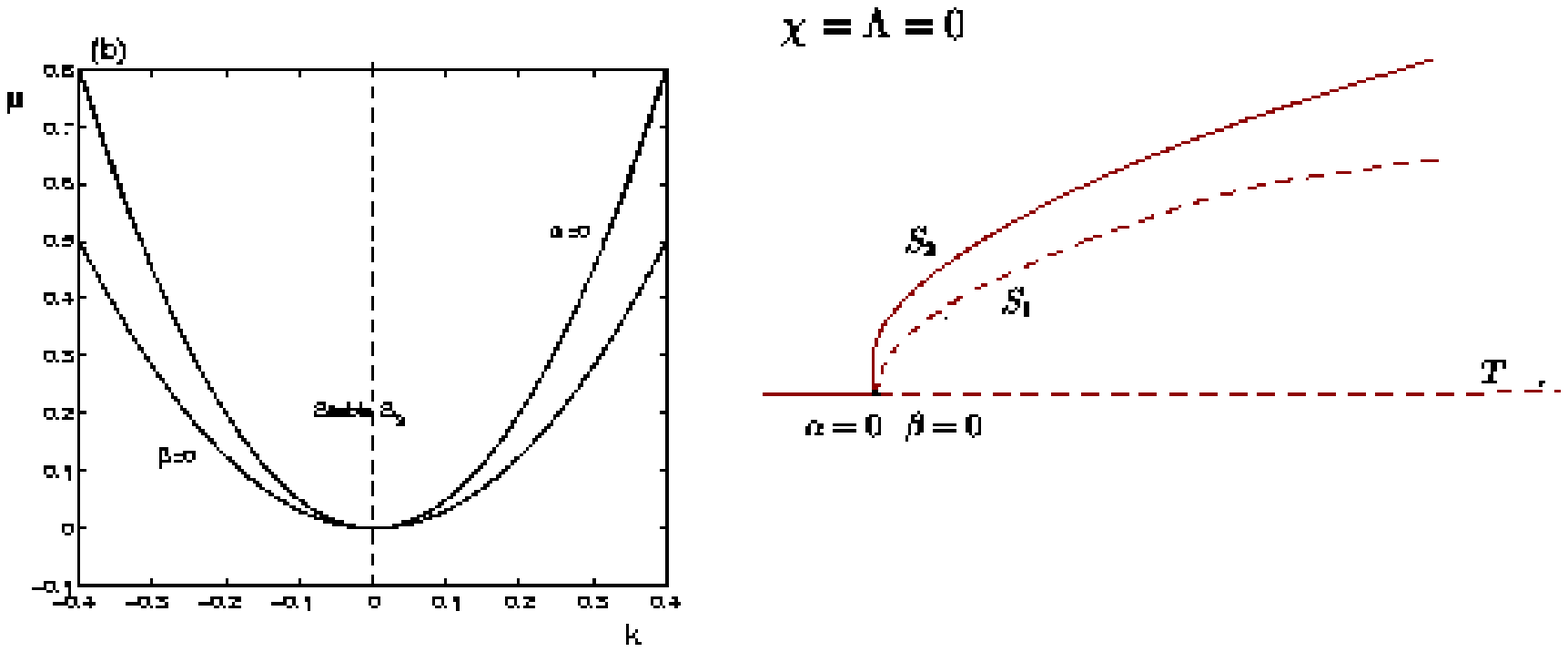}}
\caption{Stability regions of the pure modes and mixed modes for
the resonance $2:5$ for $\Delta\mu=0$ and
$\gamma=0$. Figures on the right are
sketches of bifurcation diagrams associated with the vertical paths
(dashed lines) in the stability regions on the left. Solid lines correspond 
to stable states and dashed lines to unstable states. $T$ stands for trivial 
state. (a) $\delta=\delta'=1$, $s=s'=1$, $\rho=0.4$, $\rho'=0.67$
$\nu=0.62$ and $\nu'=-1.02$. (b) $\delta= 5$, $\delta'=0.5$, $s=s'=1$, $\rho=1.5$,
$\rho'=0.5$ and  $\nu=-\nu'=0.05$. In (a) the traveling-wave branch
that arises at $TW^+$ is not shown since we do not consider stability properties 
of traveling solutions.} \cierto{fig:dgrm2n5-gam0deltamu0}
\end{figure}
%%%%%%%%%%%%%%%%%%%%%%%%%%%%%%%%%%%

The degenerate case $\chi=\Lambda=0$, i.e $\gamma=\Delta\mu=0$,
is illustrated in Fig.~\ref{fig:dgrm2n5-gam0deltamu0}. In
this case the curves $\alpha=0$ and $\beta=0$ intersect only once
 and their minima coincide. While in
Fig.\ref{fig:dgrm2n5-gam0deltamu0}a the neutral curve for mode
$A_1$ is wider than that for $A_2$, it is  the other way around
in Fig.\ref{fig:dgrm2n5-gam0deltamu0}b. Note that the wavenumber
$nk/m$ of $A_2$ is larger than that of $A_1$; therefore to make
the neutral curve of $A_2$ wider than that of $A_1$ requires  a
large ratio of $\delta/\delta'$. Depending on the nonlinear
coefficients, all four branches of pure modes and mixed modes
($S_1$, $S_2$, $S_\pm$), or just the pure modes ($S_1$,$S_2$)
may arise at the intersection point of the neutral
curves. Because we are considering $s=s'=1$, $S_1$ and $S_2$
bifurcate supercritically.
For $k=0$ the eigenvalues  $\lambda_\pm^{(1,2)}$  (cf. 
Eqs.~(\ref{7},\ref{8})) determining the  stability of $S_1$ and
$S_2$  take the simpler form: 
\begin{eqnarray}
\mbox{L}_1: 
\lambda_\pm^{(1)}&=&\mu(1-\rho')\pm\left\{\nu'\left|\mu\right|^{n/2}\right\}_{m=2} 
\quad \text{ for $S_1$,}\cierto{eigPM}\\ 
\mbox{L}_2: \lambda_\pm^{(2)}&=& \mu(1-\rho)\qquad \text{  for
$S_2$.}\nonumber 
\end{eqnarray}

Near onset (i.e., $0<\mu \ll 1$) $S_1$ and $S_2$ are stable if $1-\rho'<0$ and 
$1-\rho<0$, respectively.  In the case $m=2$, the stability of $S_1$ can be modified 
at larger values of $\mu$ by the resonant terms. Near onset the
amplitudes of the mixed modes are approximated by 
 \begin{align}
S_{\pm}: \quad &
R_1^2=\frac{1}{1-\rho\rho'}\left(\mu(1-\rho)\mp(\nu-\nu'\rho)o(\mu^2)\right)
\cierto{aprxMM}\\
&R_2^2=\frac{1}{1-\rho\rho'}\left(\mu(1-\rho')\mp(
\nu-\nu'\rho)o(\mu^2)\right),\nonumber
\end{align} 
Note that 
small-amplitude mixed modes $S_\pm$ can
therefore exist only if
$(1-\rho)(1-\rho')>0$.  As discussed
above, when  $\rho\rho'<1$, the stability of $S_\pm$ is controlled 
by the phase eigenvalue $e_\pm$ of Eq.~(\ref{12}).  Upon substituting 
Eqs.~(\ref{aprxMM}) into Eq.~(\ref{12}) we find that
\begin{equation}
\text{sign}(e_\pm)=\text{sign}\left\{\mp\frac{\mu}{1-\rho\rho'}\left(n \nu 
(1-\rho')+ m\nu'(1-\rho)\right)\right\}.\cierto{eigMM}
\end{equation}

The results for $\Delta\mu=\gamma=0$ and $s=s'=1$ may now be summarized:
\begin{enumerate}
\item  If $1-\rho>0$ and $1-\rho'>0$ both pure and mixed
modes bifurcate from the trivial state as the forcing $\mu$ is increased. The two pure
modes are both unstable, while one of the mixed modes, either $S_+$ or $S_-$, is stable.
This case is illustrated in the diagram of Fig.~\ref{fig:dgrm2n5-gam0deltamu0}a; in
this example sign$(e_\pm)=\mp 1$ (see Eq.~(\ref{eigMM})), implying that
$S_+$ is stable and $S_-$ unstable at onset.

\item  If $1-\rho<0$ or $1-\rho'<0$ only the
two pure modes are present at onset.When
$1-\rho\rho'>0$  $S_1$ is stable if $1-\rho' < 0$ and $S_2$ is
stable if $1-\rho < 0$.  In particular, a bistable situation is
permitted.  In the example shown in  
Fig.~\ref{fig:dgrm2n5-gam0deltamu0} only $S_2$ is stable because $1-\rho'>0$. 
\end{enumerate}

Fig.~\ref{fig:dgrm2n5Delta-s+} presents two possible unfoldings
of the degenerate diagrams shown in
Fig.~\ref{fig:dgrm2n5-gam0deltamu0}. In both cases the curves
$\alpha=0$ and $\beta=0$ intersect twice because $\chi>0$.   In
Fig.~\ref{fig:dgrm2n5Delta-s+}a the pure mode $S_2$ appears
first ($\Lambda>0$) and is stable.   With increasing $\mu$,
however, perturbations in the direction of the other mode,
$A_1$, become increasingly important, eventually destabilizing
$S_2$ at $L_2$ in favor of the mixed modes $S_\pm$.  Since $R_1
\ll R_2$ in the vicinity of $L_2$ and $\nu>0$,  $S_+$ is the
stable mixed mode (cf.~Eq.~(\ref{12})). In
Fig.~\ref{fig:dgrm2n5Delta-s+}b we have $\Lambda<0$, and the
roles of $S_1$ and $S_2$ are switched. The mode $S_1$, which is
now stable at onset, is destabilized by a perturbation in the
direction of $A_2$ at $L_1^-$, generating the mixed state $S_-$.
It undergoes a second bifurcation involving a  perturbation in
the direction of $A_2$ at $L_1^+$, leading to the mixed mode
$S_+$. In this case $S_-$ is stable, not
$S_+$.  In contrast with Fig.~\ref{fig:dgrm2n5Delta-s+}a,
however, the pure mode $S_2$ is ultimately stabilized at large
$\mu$ since $\rho>s'=1$. Therefore the cross-interaction
term
$\rho|A_2|^2\equiv\rho(\mu+\Delta \mu)/s'$ in
Eqs.(\ref{2},\ref{3}),  
dominates the linear growth rate $\mu$ of $A_1$
for large $\mu$ and suppresses the perturbations in the
direction of $A_1$.

%%%%%%%%%%%%%FIG2%%%%%%%%%%%%%
 \begin{figure}[htb]
\hskip0.2truecm{\includegraphics [width=14cm]{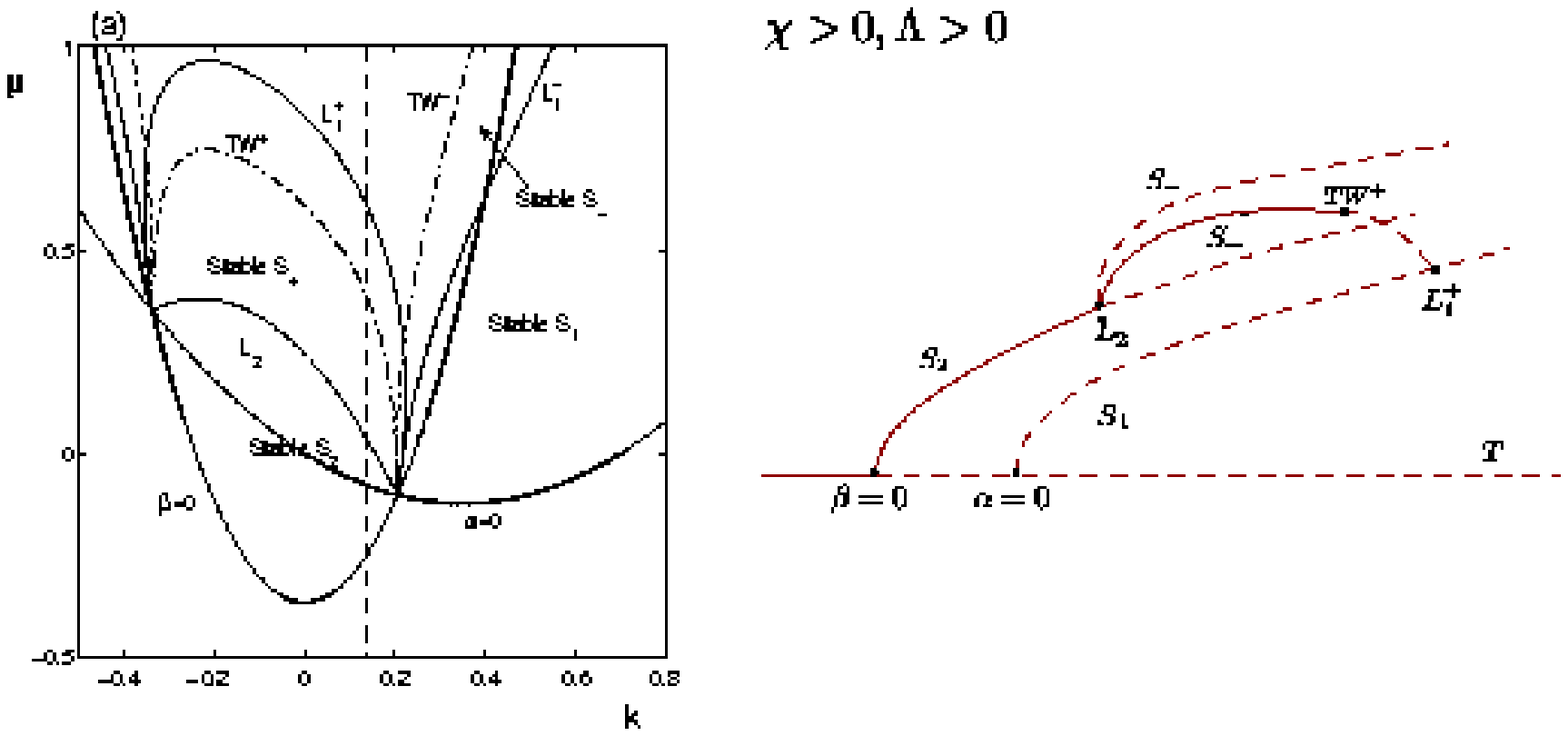}}
{\includegraphics[width=14.45cm]{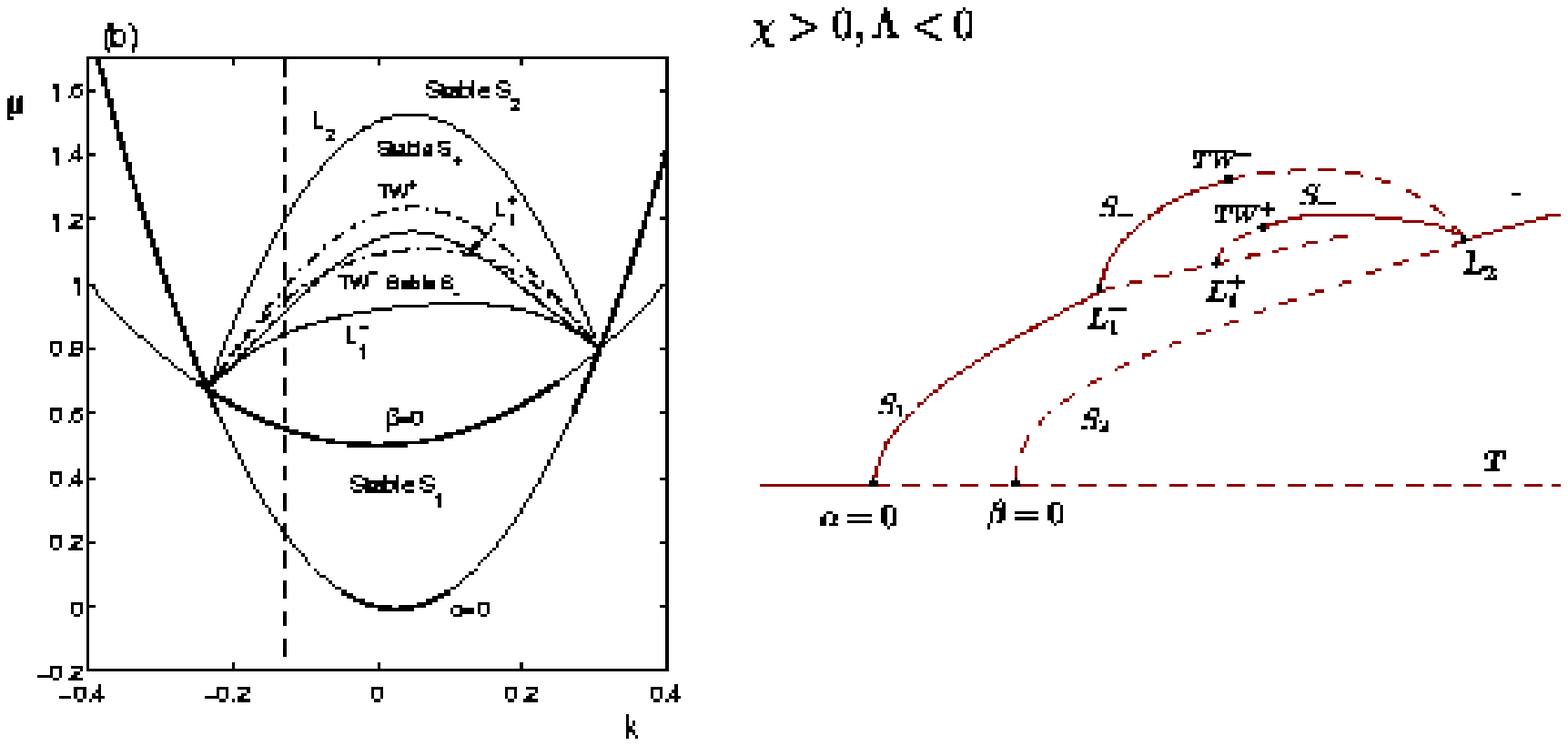}}
 \caption{Stability regions of the pure modes and mixed modes for the
 resonance $2:5$. Figures on the right are bifurcation diagrams
 associated with the vertical paths (dashed lines) in the
 stability diagrams on the left. $T$ stands for trivial state
 $A_1=A_2=0$. (a) Same parameters as in
 Fig.~\ref{fig:dgrm2n5-gam0deltamu0}a but $\Delta\mu=0.366$ and
 $\gamma=0.7$. (b) Same parameters as in
 Fig.~\ref{fig:dgrm2n5-gam0deltamu0}b but $\Delta\mu=-0.5$ and
 $\gamma=0.5$.  The branch of traveling waves that arises from
 $TW^-$ in (a), or that (possibly) connects the
 $S_+$ with the $S_-$ solutions in (b) is not shown.}
 \cierto{fig:dgrm2n5Delta-s+}
\end{figure}
%%%%%%%%%%%%%%%%%%%%%%%%%%%%%%%%%%%%

Fig.~\ref{fig:dgrm2n5Delta-s-} shows diagrams for the case 
where the curves $\alpha=0$ and $\beta=0$ do not
cross ($\chi<0$). In Fig.~\ref{fig:dgrm2n5Delta-s-}a the pure
mode $S_2$ bifurcates first, while $S_1$ does so in
Fig.~\ref{fig:dgrm2n5Delta-s-}b.  The subsequent bifurcations
that these states undergo are basically of the same type as
those in
Fig.~\ref{fig:dgrm2n5Delta-s+}. For example, in
Fig.~\ref{fig:dgrm2n5Delta-s-}a the mode $S_2$ becomes unstable
to the mixed modes at $L_2$, while $S_1$ eventually
gains stability at large $\mu$ because $\rho > s'$.  
A feature of Fig.~\ref{fig:dgrm2n5Delta-s-}b
that is not present in the previous diagrams is the saddle-node
bifurcation on the $S_+$ branch. The appearance of this
bifurcation is not specific to the case $\chi<0$ but
depends on the nonlinear coefficients, in
particular on the resonance coefficients $\nu$
and $\nu'$ (see Eq.~(\ref{6})).

A comment regarding the validity of the phase diagrams is in
order. For the weak resonances we are considering in this
paper the resonant terms proportional to $\nu$ and $\nu'$ are of
higher order in the amplitudes. Consequently, they have to be
considered as perturbation terms. Since they are responsible for the
splitting of the lines $L_1$ and $TW$ into $L_1^\pm$ and $TW^\pm$,
only those aspects of the results presented in Figures
\ref{fig:dgrm2n5-gam0deltamu0}-\ref{fig:dgrm2n5Delta-s-}
that persist as this splitting becomes small are expected to hold
systematically. Thus, the transition to traveling waves at $k=0$
in Fig.\ref{fig:dgrm2n5Delta-s+}b is robust, while that at $k=0$
in Fig.\ref{fig:dgrm2n5Delta-s+}a is shifted to ever larger
values of $\mu$ as the resonant terms become weaker. Similarly,
the saddle-node bifurcation of $S_+$ in
Fig.\ref{fig:dgrm2n5Delta-s-}b is not robust. 
It disappears through a sequence of
bifurcations that involves a merging of the branch $S_+$ with
$S_1$. For very small $|\nu|$ and $|\nu'|$ the branch $S_+$
merges with $S_1$ close to $L_1^-$.

 %%%%%%%%%%%%%FIG%%%%%%%%%%%%%%%%%%
 \begin{figure}[htb]
\hskip-0.5truecm{\includegraphics[width=14.5cm]{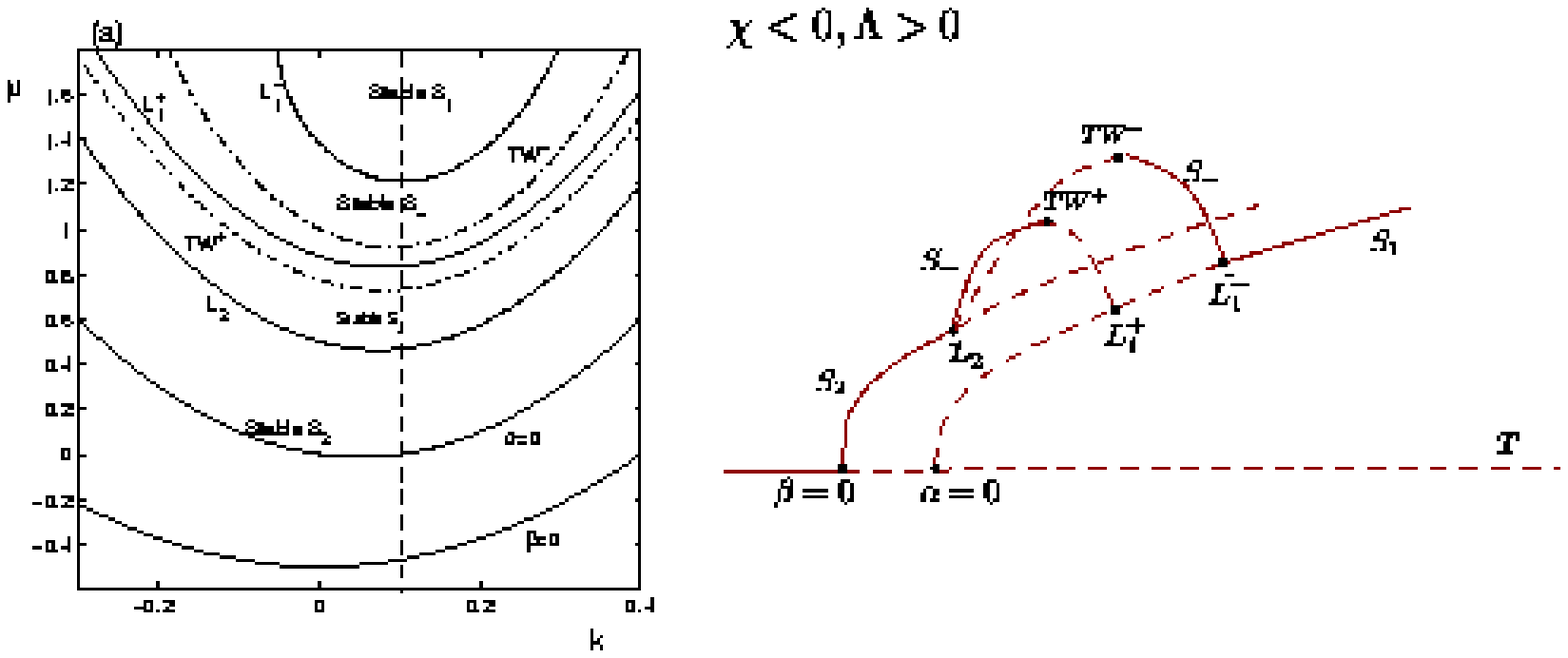}}
{\includegraphics[width=13cm]{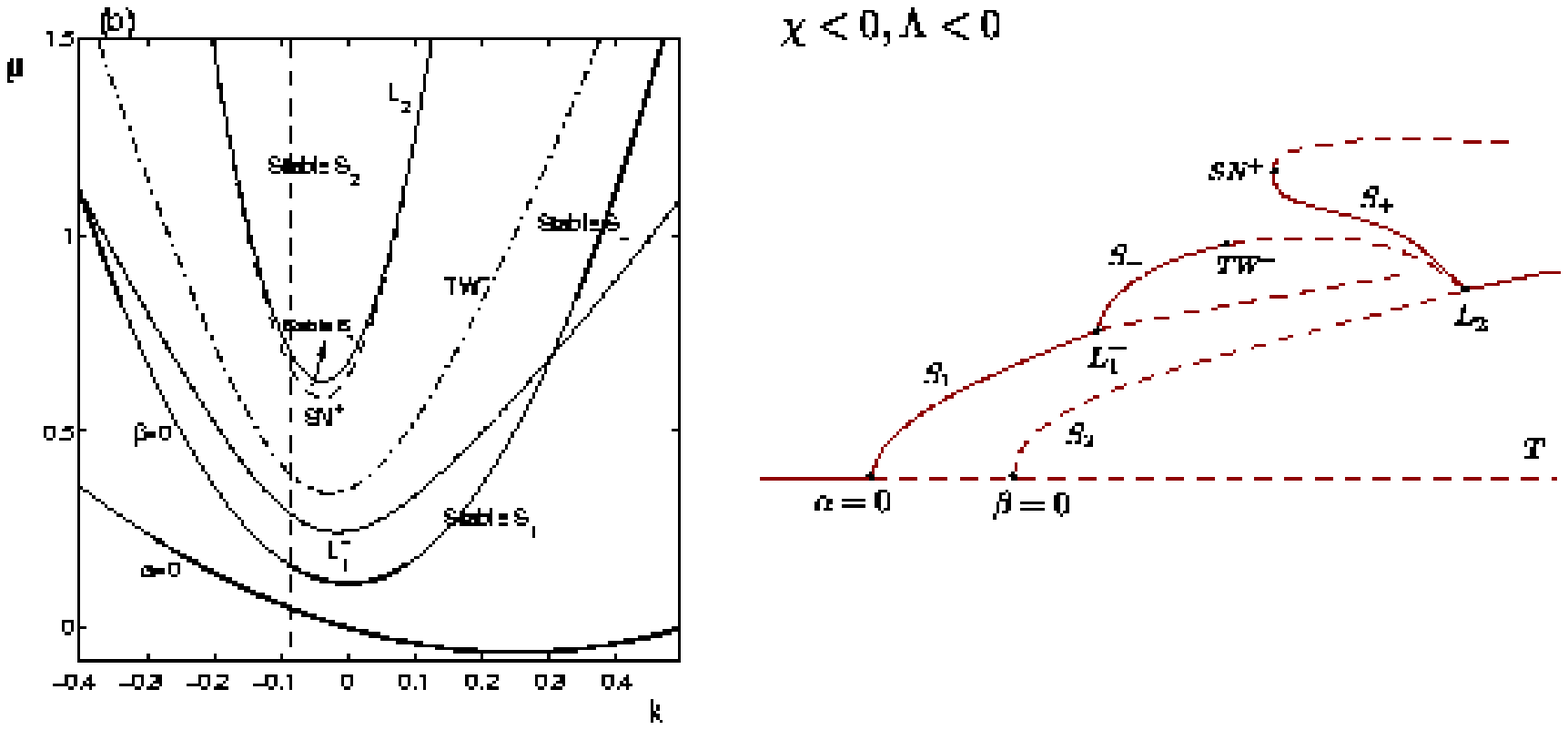}}
\caption{Stability regions of the pure modes and mixed modes for
the resonance $2:5$. Figures on the right are sketches of 
bifurcation diagrams associated with the vertical paths
(dashed lines) in the
stability diagram on the left. $T$ stands for trivial state
$A_1=A_2=0$. 
(a) $\Delta\mu=0.5$, $\gamma=0.5$, $\delta=5$, $\delta'=0.5$, $s=s'=1$, $\rho=0.5$,
$\rho'=1.5$  and
$\nu=-\nu'=0.085$
(b) $\Delta\mu=-0.1125$, $\gamma=0.5$, $\delta=\delta'=1$, $s=s'=1$, $\rho=1.5$,
$\rho'=0.5$ and  $\nu=0.5$ and $\nu'=-1.01$. The branch of
traveling waves that (possibly) connects the $S_+$
with the $S_-$ solutions in (a), or that arises from $TW^-$  in (b) is not shown.}
\cierto{fig:dgrm2n5Delta-s-}
\end{figure}
%%%%%%%%%%%%%%%%%%%%%%%%%%%%%%%%%%%%%%%%%% 

 \clearpage
\section{Side-Band Instabilities}
\cierto{sec:side-band}

We now turn to the analytical core of this paper, the stability
of the spatially periodic solutions with respect to side-band
instabilities. Such instabilities are expected to  destroy the 
periodicity of the solutions and provide a possible connection
with  quasi-periodic patterns. We therefore consider
perturbations of the periodic  solutions in the form 
\begin{equation}  
A_1=(1+a_1(X,T))R_1 e^{i(kX+\hat\phi_1)},\quad
A_2=(1+a_2(X,T)) R_2e^{i((nk/m)X+\hat\phi_2)},\cierto{8a}
\end{equation}
where $a_j(X,T)= (a_j^+(T) e^{i Q X}+a_j^-(T)e^{-i QX})$ with $Q\neq
0$ and $j=1, 2$. Note that the  perturbation wavenumber  $Q$ is measured 
relative to the deviation wavenumbers $k$ and $nk/m$. The linear stability of 
the pure modes $S_{1,2}$ and the mixed modes $S_{\pm}$ is calculated by 
inserting the ansatz (\ref{8a}) into Eqs.~(\ref{2},\ref{3}) and linearizing in 
$a_j^{\pm}$. The details of these calculations are given in Appendix A.

\subsection{Pure modes $\mathbf{S_{1,2}}$}

The linearized system for the perturbations associated with the
pure modes $S_{1,2}$ separates into two uncoupled
$2\times2$-blocks, which allows the stability of each pure mode
to be calculated analytically.  These two blocks can be
associated with longwave and shortwave instabilities,
respectively.  The block corresponding to the longwave
instability contains the eigenvalue related to spatial
translations (i.e. phase modulations).  The other (shortwave)
block describes the evolution of amplitude perturbations in a
direction tranverse to the relevant pure-mode subspace.   We
find that in addition to the instabilities discussed in Section
3 the destabilization of the pure modes can occur by longwave
(Eckhaus) or shortwave instabilities. 

In the case of $S_1$, a straightforward calculation yields
\begin{eqnarray} E_1 & : & \mu-\delta k^2-\gamma k-(\gamma+2
k\delta)^2/2\delta=0,\cierto{7e1}\\[0.5cm] \Gamma_1 & : &
(s\beta-\rho' \alpha)/s+ \delta'(nk/m)^2
+\left\{\nu'^2|\alpha/s|^n/4\delta'(nk/m)^2\right\}_{m=2}=0,\cierto{7g1}
\end{eqnarray} where $E_1$ is the Eckhaus curve and $\Gamma_1$ is the
stability limit associated with shortwave perturbations.  If the
curve $\Gamma_1$ is crossed, $S_1$ undergoes a steady-state
bifurcation with a perturbation wavenumber $Q$ given by
\begin{eqnarray} 
&Q^2 &= \left(\frac{n}{m}\right)^2
k^2-\left\{{\nu'}^2\left|\frac{\alpha}{s}\right|^n
\left(\frac{m}{n}\frac{1}{2\delta'k}\right)^2
\right\}_{m=2}.\cierto{7wn1} 
\end{eqnarray} 
When $m>2$ the bracketed terms  in Eqs.~(\ref{7g1})  and
(\ref{7wn1}) are absent. The destabilizing  eigenfunction then takes
the simple form $(a_1, a_2) \propto (0,e^{-i k\frac{n}{m} X})$
(see Eqs.~(\ref{8a}) and (\ref{eqa2S1}) of Appendix A) and
allows a correspondingly simple physical interpretation: the
wavenumber of  the destabilizing mode is the wavenumber at the
band center of  the other mode $S_2$ (see Eqs.~(\ref{8a})). This
is no longer the case for $m=2$  because the
resonance terms (which are linear in $A_2$)
affect the stability of  $S_1$.  One consequence of this is that
the destabilizing mode is composed of  two wavenumbers: $
k\frac{n}{m}\pm Q$ with $Q$ given by Eq.~(\ref{7wn1}). 
Moreover, it is possible to have one or more non-zero wavenumbers
$k^*$, say, for which the
perturbation wavenumber $Q$ vanishes.   At  these points
$(k,\mu)=(k^*,\mu^*)$ the curve $\Gamma_1$ merges with the 
curves $L_1^+$ (or $L_1^-$) describing stability under
homogeneous  perturbations.

In the case of $S_2$ we find
\begin{eqnarray}
 E_2 & : & \mu+\Delta\mu-3\delta'(n k/m)^2=0,\cierto{7e2} \\
 \Gamma_2 & : & (s'\alpha-\rho\beta)/s'+(\gamma+2\delta k)^2/4 \delta=0,
\cierto{7g2}
\end{eqnarray}
where $E_2$ is the Eckhaus curve and $\Gamma_2$ is the stability
limit for shortwave instabilities.  As $\Gamma_2$ is crossed the pure mode
$S_2$ undergoes a steady-state bifurcation with perturbation wavenumber $Q$
given by
\begin{equation}
Q=\pm|k+\gamma/2\delta|.\cierto{7wn2}
\end{equation}
The associated eigenfunction is of the form $(a_1,a_2) \propto
(e^{-i (k+\frac{\gamma}{2\delta})X},0)$ (see Eqs.~(\ref{8a}) and (\ref{eqa1S2})) and, as in the
case of $S_1$, points to a simple interpretation: the mode that destabilizes
$S_2$ lies at the band center of $S_1$, i.e., its wavenumber is 
$\gamma/(2\delta)=\hat\gamma$ (cf.~Eqs.~(\ref{8a}), and Appendix A; see also 
Fig.~\ref{fig:critic-modes}). This result holds for all resonances except 
$m:n=1:2$, in which case the linear stability of the pure mode $S_2$ is affected 
by the resonance terms (as $S_1$ was when $m=2$).

Stability results for the pure modes
$S_{1,2}$ are shown in Figs.~\ref{fig:dgrm2n5withtw(a)} and
\ref{fig:dgrm2n5withtw(b)}. In
Fig.~\ref{fig:dgrm2n5withtw(a)}b,d and Fig.~\ref{fig:dgrm2n5withtw(b)}b
the parameters are as in
Fig.~\ref{fig:dgrm2n5Delta-s+}a,b and  Fig.~\ref{fig:dgrm2n5Delta-s-}a, 
respectively, while in the diagrams of Fig.~\ref{fig:dgrm2n5withtw(a)}a,c 
and Fig.~\ref{fig:dgrm2n5withtw(b)}a we depart from
those parameters only in setting $\gamma=0$. The
inclusion of these last three cases allows us to gauge how much the
side-band instabilities are affected by the detuning from
perfect resonance. 
                                                           
 Note that due to the large ratio $\delta/\delta'$ the detuning
of $\gamma=0.5$ has only a small effect on the neutral curve of
$A_1$ ($\alpha=0$). Observe that in Fig.~\ref{fig:dgrm2n5withtw(a)}a,b
the pure mode $S_2$ is stable within the region bounded by the
curves $E_2$ and $\Gamma_2$ while $S_1$ is everywhere unstable
due to shortwave instabilities.  In contrast,
Fig.~\ref{fig:dgrm2n5withtw(a)}c,d depicts a situation with
stable regions for both pure modes.  Note that the pure mode
$S_2$ experiences only shortwave instabilities as the forcing is
decreased while the pure mode $S_1$ can lose stability to either
longwave or shortwave perturbations, provided the wavenumber $k$
is not within the interval defined by the merging points of
$\Gamma_1$ and $L_1^-$ (see Fig.~\ref{fig:dgrm2n5withtw(a)}c and
~\ref{fig:dgrm2n5withtw(a)}d); over that interval the
instability suffered by $S_1$ is to 
perturbations preserving the periodicity of $S_1$
and gives rise to (steady) mixed modes.  The parameters of
Fig.~\ref{fig:dgrm2n5withtw(b)} do not allow for codimension-two
points (intersection of the curves $\alpha=0$ and $\beta=0$). 
Despite this difference, the stability results for $S_2$ are
qualitatively similar to those of
Fig.~\ref{fig:dgrm2n5withtw(a)}. The effect of $\gamma$ on
$S_1$,  however, is more striking than in
Fig.~\ref{fig:dgrm2n5withtw(a)}c-d. In  particular, $S_1$ can
undergo a shortwave steady-state instability (along  $\Gamma_1$)
when $\gamma = 0.5$ but not when $\gamma = 0$.

%%%%%%%%%%%%%FIG3%%%%%%%%%%%%%%%

\begin{figure}[htb]
\includegraphics[width=15cm,height=8cm]{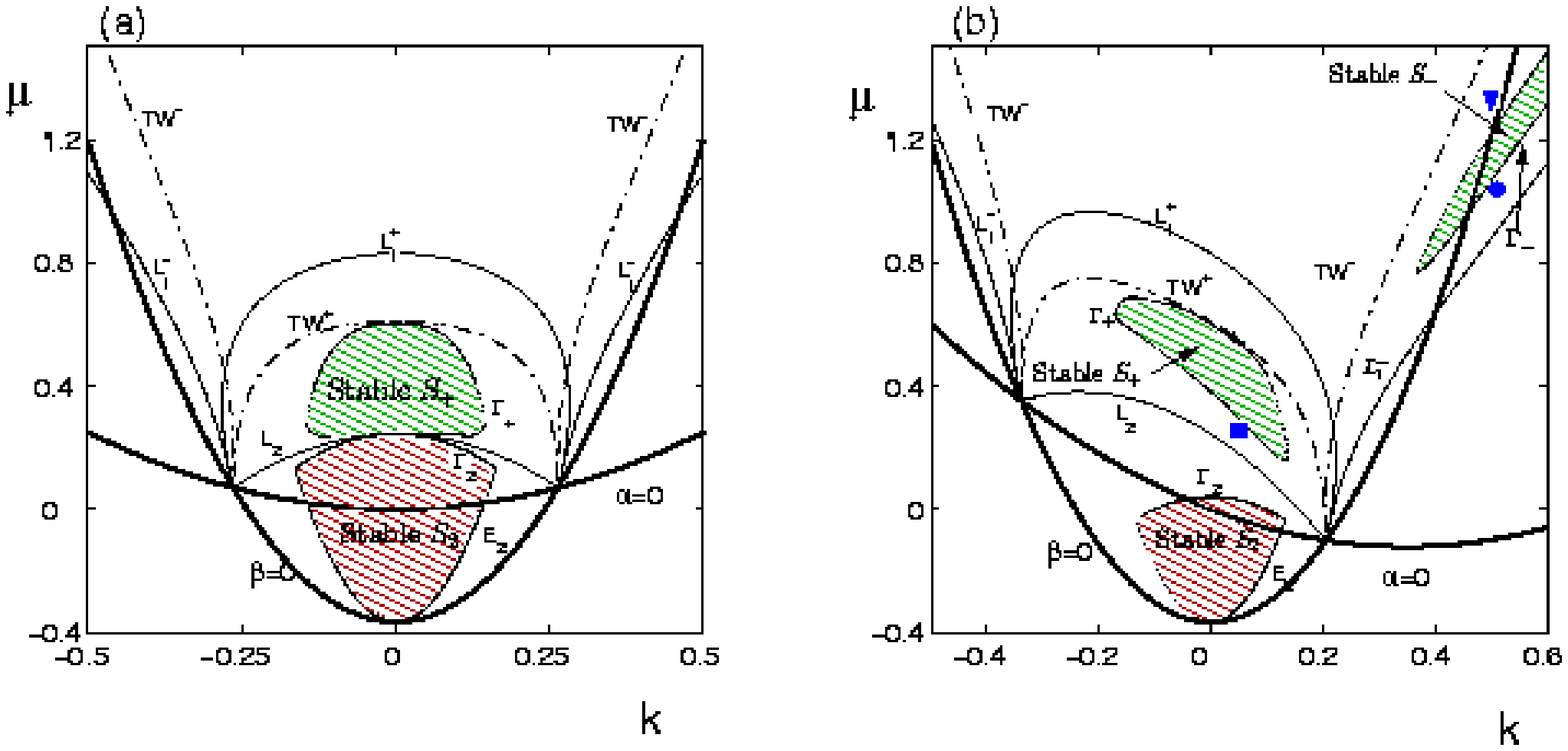}
\hskip-1.5truecm\includegraphics [width=15cm,height=8.2cm]{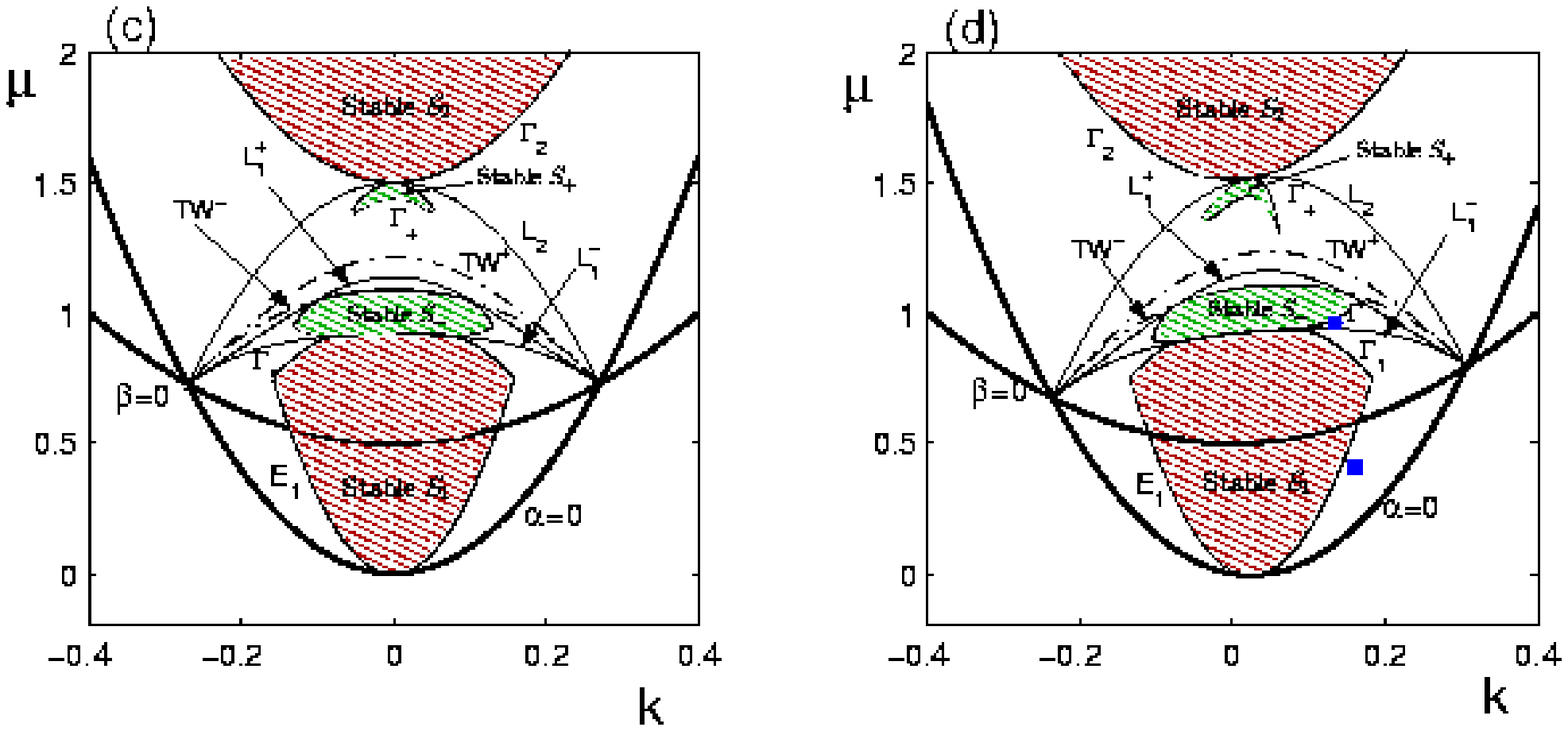}
\caption{Stability regions of the pure modes $S_{1,2}$ and the mixed modes
$S_\pm$ for the resonance $2:5$ and the following parameter sets: (a) $\Delta\mu=0.366$,
$\delta=\delta'=1$, $s=s'=1$, $\rho=0.4$, $\rho'=0.67$, $\nu=0.62$, $\nu'=-1.02$ and
$\gamma=0$; (b) as in (a) but $\gamma=0.7$ (cf.
Fig.\ref{fig:dgrm2n5Delta-s+}a); (c) $\Delta\mu=-0.5$,
$\delta=10 $, $\delta'=0.5$, $s=s'=1$, $\rho=1.5$, $\rho'=0.5$, $\nu=-\nu'=0.05$ and
$\gamma=0$; (d) as in (c) but $\gamma=0.5$
(cf. Fig.\ref{fig:dgrm2n5Delta-s+}b). The symbols in b) and d) indicate the
parameters used in the numerical simulations described in Sec.~\ref{sec:numericalresults}.}
\cierto{fig:dgrm2n5withtw(a)}
\end{figure}
 
%%%%%%%%%%%%%%%%%%%%%%%%%%%%%%%%%%%%%%%%

  %%%%%%%%%%%%%FIG4%%%%%%%%%%%%%%%
 \begin{figure}[htb]
\centerline{\includegraphics [width=15cm]{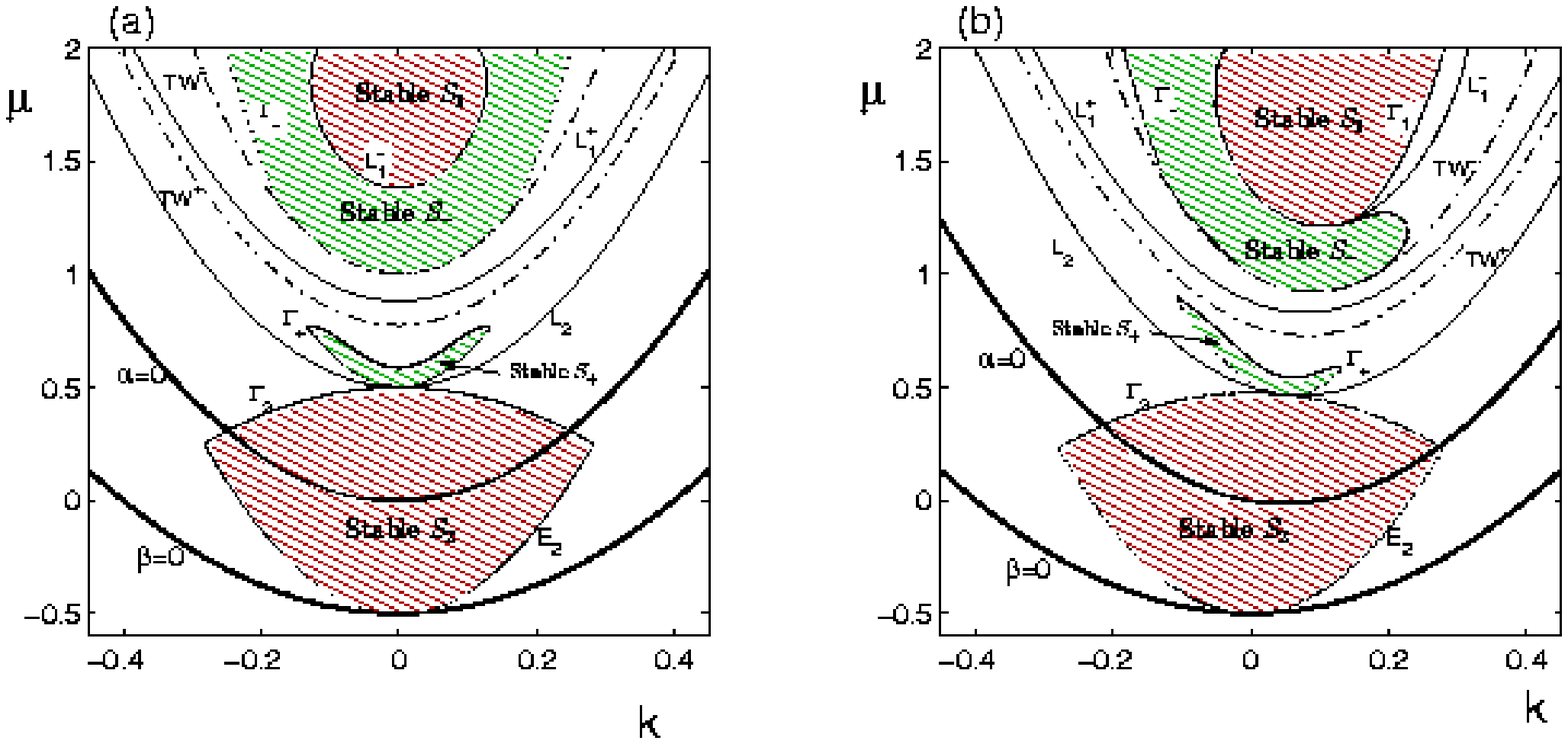}}
\caption{Stability regions of the pure modes $S_{1,2}$ and the mixed modes
$S_\pm$ for the resonance $2:5$ and the following parameter sets:
(a) $\Delta\mu=0.5$, $\gamma=0$, $\delta=5$, $\delta'=0.5$, $s=s'=1$,
$\rho=0.5$,
$\rho'=1.5$, $\nu=-\nu'=0.085$. (b) as in (a) but $\gamma=0.5$ (cf.
Fig.\ref{fig:dgrm2n5Delta-s-}a).}
\cierto{fig:dgrm2n5withtw(b)}
\end{figure}

%%%%%%%%%%%%%%%%%%%%%%%%%%%%%%%%%%%%%%%%

\subsection{Mixed modes $S_\pm$}

The stability of the mixed modes $S_\pm$ is governed by four
eigenvalues which must be determined numerically.  Two of these
are associated with amplitude modes and are always real. The
remaining two may be real or complex and are related,
respectively, with the translation mode and with the relative
phase between the two modes $A_1$ and $A_2$. Throughout this
stability analysis we focus on the influence of the detuning
parameter $\gamma$ on the (side-band)
instabilities of $S_\pm$.  Furthermore, because
of the invariance of Eqs.~(\ref{2},\ref{3}) under the
transformation
$$\gamma\rightarrow -\gamma,\quad X\rightarrow -X,$$ we may take
$\gamma\geq 0$ without any loss of generality.

The results of this stability analysis are shown in Fig.~\ref{fig:dgrm2n5withtw(a)}, which
depicts a situation with two codimension-two points, and Fig.~\ref{fig:dgrm2n5withtw(b)},
which presents a case with no codimension-two points.  Since this difference seems to have 
only a minor effect on the stability of $S_\pm$ we focus our discussion
on the case of Fig.~\ref{fig:dgrm2n5withtw(a)} (i.e. two codimension-two points).

\subsubsection{Mixed mode $\mathbf {S_+}$}

When $\gamma=0$ (Fig.~\ref{fig:dgrm2n5withtw(a)}a,c), the mixed
mode $S_+$ becomes  unstable to short-wavelength perturbations
on the  part of $\Gamma_{\!+}$   closer to $L_2$, while a
longwave stability analysis captures the part of  $\Gamma_{\!+}$
closer to $TW^+$ .  Both instabilities are of steady-state  type
with the one close to $TW^+$ breaking the reflection symmetry of
the pattern. This is expected to lead to a drift of the pattern;
additional calculations for other parameter sets (not shown)
suggest that this situation is characteristic of $S_+$ and
$\gamma=0$.

When $\gamma\neq 0$ the mixed mode $S_+$ displays both
steady-state and Hopf bifurcations.  The two relevant
eigenvalues are associated with the translation mode and the
symmetry-breaking  mode, respectively, with the latter eigenvalue
going to zero at $TW^+$.  The change
in character of  the instability along $\Gamma_{\!+}$ is
illustrated in Fig.~\ref{fig:growthgamma04}.
Figs.~\ref{fig:growthgamma04}b,c give the
growthrates of the dominant modes as $\Gamma^+$ is crossed at
specific points marked in Fig.~\ref{fig:growthgamma04}a. At  points 1-5 the
bifurcation is steady but as one continues in a counterclockwise
direction  (points 4-6) the eigenvalues merge and become complex
over an interval of $Q$-values.  This interval containing 
complex conjugate eigenvalues continues to grow, leading
eventually to an oscillatory instability superceding the steady one. 
A codimension-two point
therefore exists near $(k,\mu)=(-0.155, 0.581)$ at the transition from
steady-state to Hopf bifurcation; at this point there are two unstable
modes, at different $Q$, one steady and one oscillatory.  As one goes
still further in the counterclockwise direction, the $Q$-band over which 
the eigenvalues are complex
reaches $Q=0$ and the oscillatory instability becomes a long-wave
instability. This interaction was studied in the context of Taylor vortex flow
\cite{RiPa92} and has been found in many others problems
\cite{RaRi92,FoRa94}.

%%%%%%%%%%%FIG%%%%%%%%%%%%%%%
\begin{figure}
\centerline{\includegraphics{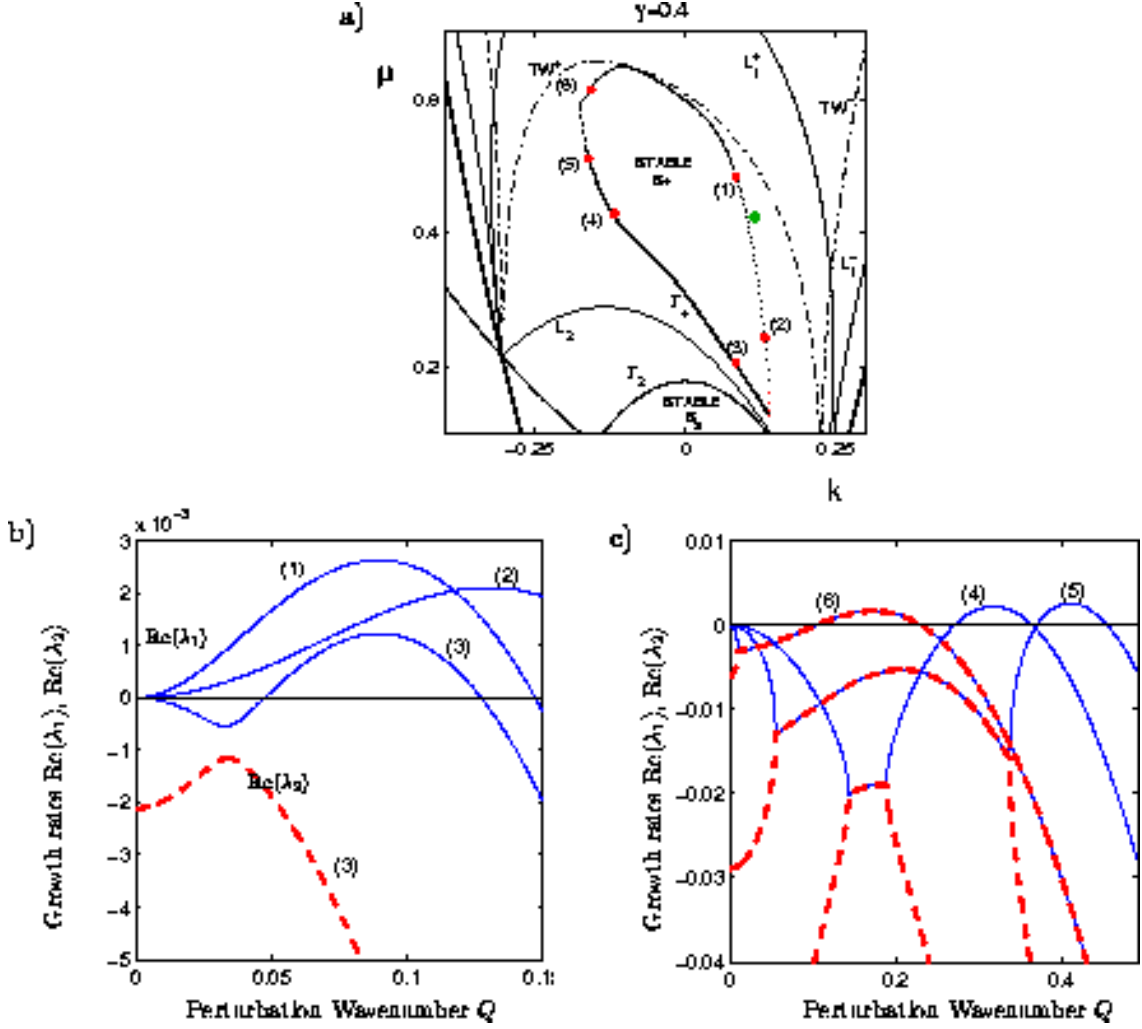}}
\caption{Growth rates of the two most unstable perturbations of $S_+$ when 
$\Gamma_{\!+}$ is crossed at the points indicated in the upper figure: 
$Re(\lambda_1)$ (solid-thin line) and $Re(\lambda_2)$ (dashed-thick);  
$\lambda_1$ and $\lambda_2$ are complex where there is a single curve, i.e., 
where $Re(\lambda_1)=Re(\lambda_2)$.  We use $\gamma=0.4$ with the remaining 
parameters as in Figs.~\ref{fig:dgrm2n5withtw(a)}b.} \cierto{fig:growthgamma04}
\end{figure}
 %%%%%%%%%%%%%%%%%%%%%

Similar transitions occur if $\mu$ and $\gamma$ are varied (rather
than $\mu$ and $k$). We illustrate this in
Fig.~\ref{fig:growthm2n5k-01} by plotting the growth rate of  the
most unstable perturbations after crossing, at fixed $k=-0.1$, the
upper part  of $\Gamma_{\!+}$ which is close to $TW^+$
(Fig.~\ref{fig:growthm2n5k-01}a),  and the lower part of
$\Gamma_{\!+}$ which is close to $L_2$
(Fig.~\ref{fig:growthm2n5k-01}b). Along the upper part of
$\Gamma_{\!+}$ the mixed mode $S_+$ typically sees a   shortwave
oscillatory instability; for small $\gamma$, $S_+$ can also lose 
stability to longwave perturbations.  The lower part of
$\Gamma_{\!+}$ is  characterized by either steady-state or
oscillatory instability of shortwave  type (see
Fig.~\ref{fig:growthm2n5k-01}b).  There is again a codimension-two
point,  near $(\gamma,\mu)\simeq(1.035, 0.47)$, where Hopf and
steady-state bifurcations  coalescence.  For $\gamma$ less (greater)
than this  critical value the  bifurcation is steady (oscillatory).
The route to the oscillatory instability  is as described above: with
increasing $\gamma$ the two modes associated with  translation and
reflection symmetry interact more and more,  ultimately leading to a
Hopf bifurcation. 

The steady-state shortwave instability of $S_+$ (see
Fig.~\ref{fig:growthm2n5k-01}b) always  appears in the vicinity
of $L_2$. This proximity implies that the amplitude
$A_1$ of the mixed mode is very small.   Consequently,
the behavior of $S_+$ in this region is similar to that of the 
pure mode $S_2$.  The growing perturbations are predominantly in
the direction  of $A_1$ and the associated wavenumber $Q$ is
given in  a first approximation by  Eq.~(\ref{7wn2}), i.e.,  
\begin{equation} Q\simeq \pm |k+\gamma/2\delta|. \cierto{13e} 
\end{equation}

%%%%%%%%%%%%%FIG6%%%%%%%%%%%%%
\begin{figure}
\centerline{\includegraphics[height=11cm]{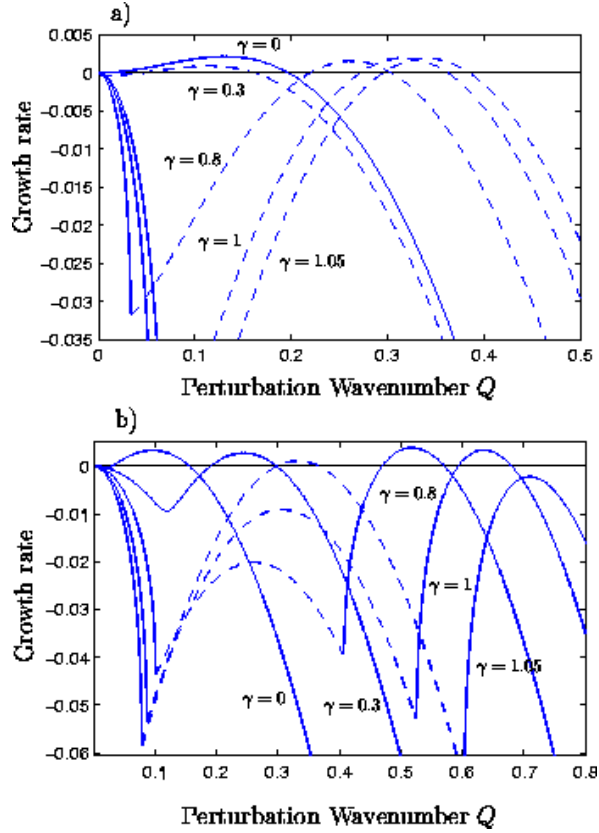}}
\caption{Growth rate of the most unstable perturbation of $S_+$ when
$\Gamma_{\!+}$ is crossed vertically at $k=-0.1$ for the indicated values of $\gamma$,
(a) on the upper part of $\Gamma_{\!+}$ (close to $TW^+$) and (b) on  the lower part
(close to $L_2$ ). The remaining parameters are as in Fig.~\ref{fig:dgrm2n5withtw(a)}a-b.
Solid (dashed) lines correspond to real (complex) eigenvalues. }
\cierto{fig:growthm2n5k-01}
\end{figure}
%%%%%%%%%%% %FIG%%%%%%%%%%%%%%

%%%%%%%%%%%%%FIG7%%%%%%%%%%%%%
\begin{figure}[htb]
\centerline{\includegraphics[width=12cm] {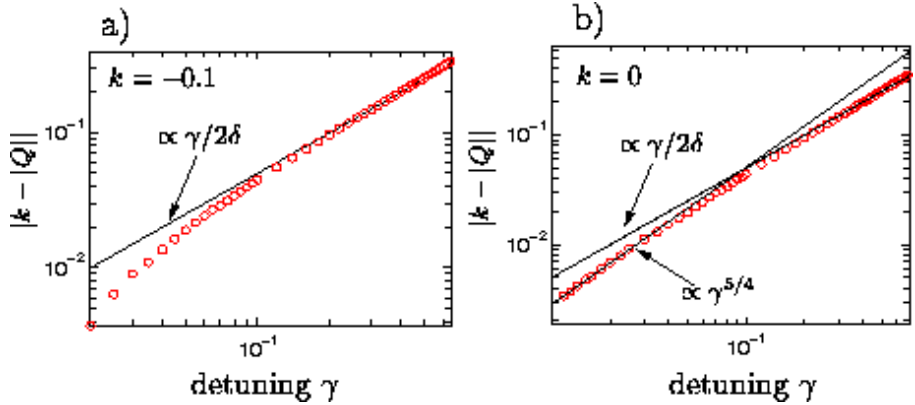}}
\caption{Dependence of the perturbation wavenumber $Q$ of the steady mode
destabilizing $S_+$ on the
detuning parameter $\gamma$ for the resonace case $m:n=2:5$.
Parameters as in Fig.\ref{fig:dgrm2n5withtw(a)}a,b 
with a) $k=-0.1$ and b) $k=0$.
Open circles  correspond to  the perturbation wavenumber 
$Q$ calculated numerically from Eq.~(\ref{14b},\ref{14c}). 
Solid lines correspond to the indicated approximations.}
\cierto{fig:maxinstvsgammam2n5k-01k0k01}
\end{figure}
%%%%%%%%%%%%%%%%%%%%%%%

As demonstrated in Fig.~\ref{fig:maxinstvsgammam2n5k-01k0k01},
Eq.~(\ref{13e}) provides a good approximation over the interval
$0.1\lesssim\gamma\lesssim 0.7$ when $|k|\leq 0.1$.  It is interesting
to note that, when $\gamma$ is not too small, Eq.~(\ref{13e}) applies
independently of the resonance $m:n$.
The robustness of this result is not unexpected since the resonance
terms do not influence at first order the stability properties of the
mixed mode $S_+$ (because $|A_1|$ is small).  However,
Eq.~(\ref{13e}) ceases to apply when $\gamma$ becomes very small
because resonance effects start to play a significant role in the
wavenumber selection of the linear response when crossing the lower
part of $\Gamma_{\!+}$.  In particular, for $k=0$ and $\gamma\ll 1$
one can show \begin{equation} Q\simeq(\gamma^n\nu)^{1/4}\left[ r_{20}
(2s'/(ss'-
\rho\rho'))^{3/4}\right]^{1/2}/2\delta+O(\gamma^{n+1}). \cierto{16a}
\end{equation}
where $r_{20}$ denotes the amplitude of $S_2$ (with $k=0$) on the
curve $L_2$. The derivation of Eq.~(\ref{16a}), given in Appendix B, relies on
the fact that for $\gamma=0$ the curve $\Gamma_{\!+}$ becomes tangent to the
curve $L_2$ at $k=0$.  
Fig.~\ref{fig:maxinstvsgamma} shows that (\ref{16a}) is
in excellent agreement with the numerical results obtained directly from the
characteristic equation~(\ref{14b}) and applies to other weak resonances with
$m:n\neq 2:5$. Note that for larger values of $n$ Eq.~(\ref{16a}) is valid over a
wider range of $\gamma$ values (since the error is $O(\gamma^{n+1})$)
and the cross-over to the behavior (\ref{13e}) is shifted to
larger values of $\gamma$.

%%%%%%%%%%%%%FIG8%%%%%%%%%%%%%
\begin{figure}[htb]
\centerline{\includegraphics[width=8cm]{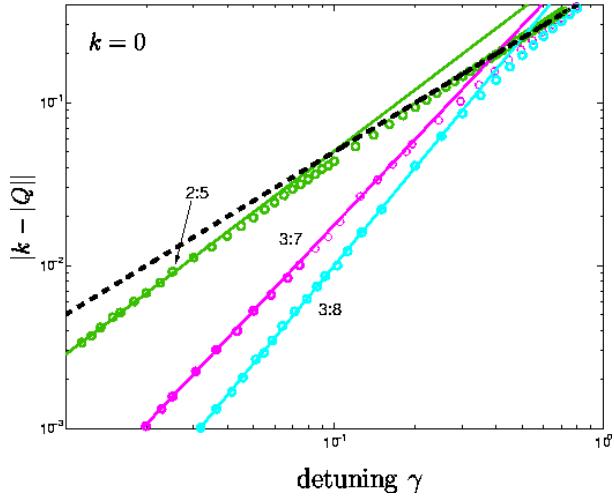}}
\caption{Dependence of $Q$ on the
detuning parameter $\gamma$ for the indicated resonances $m:n$ and $k=0$.
Remaining parameters are $\Delta\mu=0.366$, $\delta=\delta'=1$, $s=s'=1$, 
$\rho=0.4$, $\rho'=0.67$, $\nu=0.62$ $\nu'=-1.02$.
Open circles correspond to numerical results  of
Eq.~(\ref{14b},\ref{14c})
and   lines to the approximations (\ref{13e}) and (\ref{16a}),
$Q\propto\gamma/2 \delta $ (dashed)
and $Q\propto\gamma^{n/4}$  (solid), respectively.
}
\cierto{fig:maxinstvsgamma}
\end{figure}

%%%%%%%%%%%%%%%%%%%

\bigskip
\subsubsection{\bf Mixed mode $\bold {S_-}$}

As illustrated in Fig.~\ref{fig:dgrm2n5withtw(a)}a, the mixed mode $S_-$ can be
everywhere unstable when $\gamma=0$. For the parameters of
Fig.~\ref{fig:dgrm2n5withtw(a)}a, such a situation persists until $\gamma \approx 0.6$, when
a stability region for $S_-$ emerges.  This stability region widens as $\gamma$ is increased
(see Fig.~\ref{fig:dgrm2n5withtw(a)}b).   

As in the case of $S_+$, the two most important eigenvalues for
the stability of  $S_-$ are those associated with the
translation and parity-breaking modes.  The  oscillatory
instability results from an interaction between these two 
eigenvalues and, consequently, is found in the vicinity of the
bifurcation set  $TW^-$; the steady (shortwave) instability
occurs near the initial bifurcation  producing the mixed mode
$S_-$ from the pure mode (i.e., near $L_1^-$).   Thus,  the
mixed mode $S_-$ undergoes the same kinds of transitions
described above for  $S_+$.  Depending on where $\Gamma_{\!-}$
is crossed in the $(k,\mu)$- or the
$(\gamma,\mu)$-plane, the instability may be either oscillatory  or
steady, and either of long-wave or short-wave type. 

%%%%%%%%%%%%%FIG6%%%%%%%%%%%%%
\begin{figure}[htb] 
\centerline{\includegraphics[height=10cm]{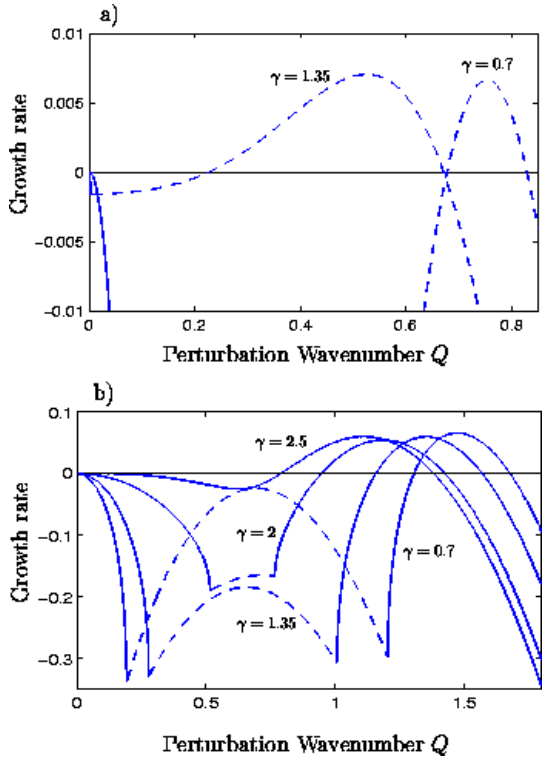}} 
\caption{Growth rate of the most unstable perturbation when  $\Gamma_{\!-}$ is 
crossed vertically at $k=0.5$ for the indicated values of
$\gamma$ and the remaining parametersas in 
Fig.~\ref{fig:dgrm2n5withtw(a)}a-b.  (a) on the 
upper part of $\Gamma_{\!-}$ (close to $TW^-$) and (b) on the lower part (close 
to $L_1^-$ ). A solid line is used when the associated eigenvalue is real and 
a dashed line when it is complex.} \cierto{fig:growthm2n5k05}
\end{figure}

 %%%%%%%%%%% %FIG%%%%%%%%%%%%%% 
 
Note that  on the upper part of $\Gamma_{\!-}$ in
Fig.\ref{fig:growthm2n5k05}a, as $\gamma$ is increased the two
eigenvalues merge and become complex at ever  smaller values of
$Q$.  In contrast, on the lower part of $\Gamma_{\!-}$ 
raising $\gamma$ increases
the damping of the parity-breaking mode, thus
discouraging its interaction with the translation mode, this 
behavior is opposite to what occurs with $S_+$.  It is also
interesting to note  that variations in the detuning parameter
$\gamma$ do not induce a significant  change in the perturbation
wavenumber $Q$ associated with the bifurcation.  This  is
because the instability of $S_-$ takes place close to $L_1^-$
where the  destabilizing  perturbations are (preferentially) in
the direction of $A_2$.   The perturbation wavenumber $Q$
associated with the instability is given, to  first
approximation, by Eq.~(\ref{7wn1}). Thus $Q$ depends mainly on
the  resonance $m:n$ and only implicitly on the parameter
$\gamma$ (through the  function $\alpha$ defined in 
Eq.~(\ref{5b})).

\section{Numerical Simulations}  

\cierto{sec:numericalresults}

The main objective in this section is to investigate the
nonlinear evolution  of the side-band instabilities of the pure
and mixed modes discussed in Section~\ref{sec:side-band}. The
results are obtained for a periodic domain of length $L$ using a
finite-difference code with Crank-Nicholson time-stepping.  
Each simulation of (\ref{2},\ref{3}) uses as an initial
condition an unstable periodic steady state (i.e., a pure or
mixed mode of the type discussed in
Section~\ref{sec:basicsolutions}) with a small random
perturbation. The final solutions obtained in this way are
typically `quasi-periodic' although periodic solutions are also
found.  Here we use the term quasi-periodic to refer to any
solution for which the wavenumbers of $A_1$ and $A_2$ are {\it
not} in the ratio $m/n$.  Therefore the full reconstructed
field (\ref{1a}) will not be composed of wavenumbers in the
ratio $m/n$; in general it will be quasi-periodic. 

First we describe the evolution of system~(\ref{2},\ref{3}) when
the pure modes become unstable. If the instability is longwave
it affects only the excited  amplitude and the solution remains
in the pure-mode subspace. For example, in the case of $S_1$  (see
Fig.~\ref{fig:spacetime1a}) the longwave instability involves
$A_1$  but not $A_2$, which remains zero. As expected, the
changes affect predominantly the phase of $A_1$ and evolve into 
a series of phase-slips.  The final state,  which is
time-independent, is shown on the right part of
Fig.~\ref{fig:spacetime1a}.  In the case of $S_2$ (not shown)
the longwave instability behaves analogously: it is now $A_2$
that  changes, while $A_1$ remains zero. 

%%%%%%%%%%%%%%%%%%%%%%%%%%%%%%%%%%%%%%%%%%%%%%%%%%%
\begin{figure}[htb]
\centerline{\includegraphics[width=11cm]{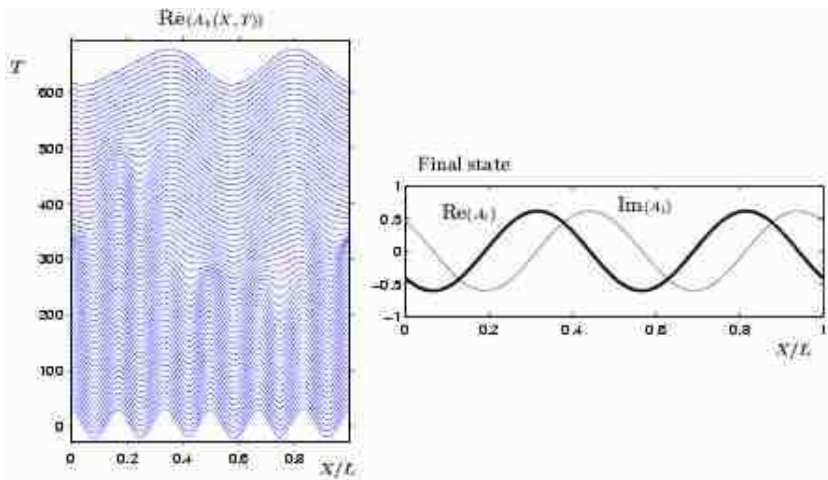}}
\caption{Evolution of the longwave instability of the pure mode
$S_1$. Space-time diagram of the real part of $A_1$ (left) and
final state  (right) with thick (thin) lines denoting the real (imaginary)
part of $A_1$. Parameters as in Fig.~\ref{fig:dgrm2n5withtw(a)}d with
$k_1=k=0.15$, $\mu=0.37$ (indicated by a square in
Fig.~\ref{fig:dgrm2n5withtw(a)}d) and $L=250$.}
\cierto{fig:spacetime1a}
\end{figure}
%%%%%%%%%%%%%%%%%%%%%%%%%%%%%%%%%%%%%%%%%%%%%%%%%%%

Depending on the detuning parameter $\gamma$ the transition from
the pure modes to the mixed modes can also involve side-band
instabilities. For the parameters corresponding to the stability
diagram shown in Fig.~\ref{fig:dgrm2n5withtw(b)} the pure mode
$S_1$ becomes unstable to the mixed mode $S_-$ $via$ a side-band
instability for $\gamma \ne 0$ and $k>0.15$ (cf.
Fig.~\ref{fig:dgrm2n5withtw(b)}b), while for $\gamma=0$, i.e.
when the two critical wavenumbers are resonant, the instability
preserves the periodicity of the solution for all $k$ (cf.
Fig.~\ref{fig:dgrm2n5withtw(b)}a). The final states that are
reached after a small random perturbation has been applied to a
pure-mode solution $S_1$ are depicted in
Fig.\ref{fig:puremixed}. The movies {\tt fig5.2a.avi} and {\tt
fig5.2b.avi} show the corresponding temporal evolution of $A_2$.
The sequence of phase slips that is evidenced in the sharp peaks of the local
wavenumber during the early stages of the evolution is a
consequence of the random initial perturbation. After this
transient the relevant perturbation mode starts
to dominate. In the case $\gamma=0.5$ that mode has a strongly
modulated wavenumber reflecting the fact that the wavenumbers of
$A_1$ and $A_2$ do not satisfy the resonance condition
$m:n=2:5$, resulting in a quasi-periodic reconstructed solution
$u(x)$ (see also discussion leading to (\ref{19}) below).

%%%%%%%%%%%%%FIG%%%%%%%%%%%%%
\begin{figure}[htb]
\centerline{\includegraphics[width=11cm]{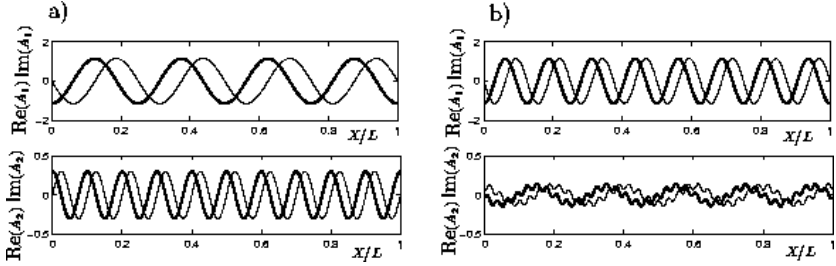}}
\caption{Final state after a small random perturbation of the
pure mode $S_1$.
Parameters in (a) and (b) as in
Fig.~\ref{fig:dgrm2n5withtw(b)}a,b, respectively
with $L=250$ and  $\gamma=0$, $k=0.1$, $\mu=1.4$
in a)
and $\gamma=0.5$, $k=0.2$, $\mu=1.4$ in b). The corresponding
movies [movie:fig5.2a.avi] and [movie:fig5.2b.avi] show the
temporal evolution of $A_2$. Red and yellow lines show the real
and imaginary part of $A_2$, respectively, white its magnitude $|A|$,
and green the local wavenumber.}
\cierto{fig:puremixed}
\end{figure}
%%%%%%%%%%%%%%%%%%%%%%%%%%%%%%%%%%%

The longwave instability of the mixed modes affects $A_1$ as
well as $A_2$ (see Fig.~\ref{fig:spacetime1b}). As with the
pure modes, the phase perturbations evolve into phase slips,
which change the wavenumber and lead eventually to a stationary
state. In the case shown in Fig.~\ref{fig:spacetime1b} phase
slips occur only in mode $A_2$. As in the mixed mode shown in
Fig.\ref{fig:puremixed}b, the strong deviations of $A_{1,2}(x)$
from purely sinusoidal behavior, which are due to the resonant
terms proportional to $\nu$ and $\nu'$, indicate that the
reconstructed solution $u(x)$ is not periodic.

%%%%%%%%%%%%%FIG%%%%%%%%%%%%%%%%%%%%

\begin{figure}[htb]
\centerline{\includegraphics[height=7.5cm]{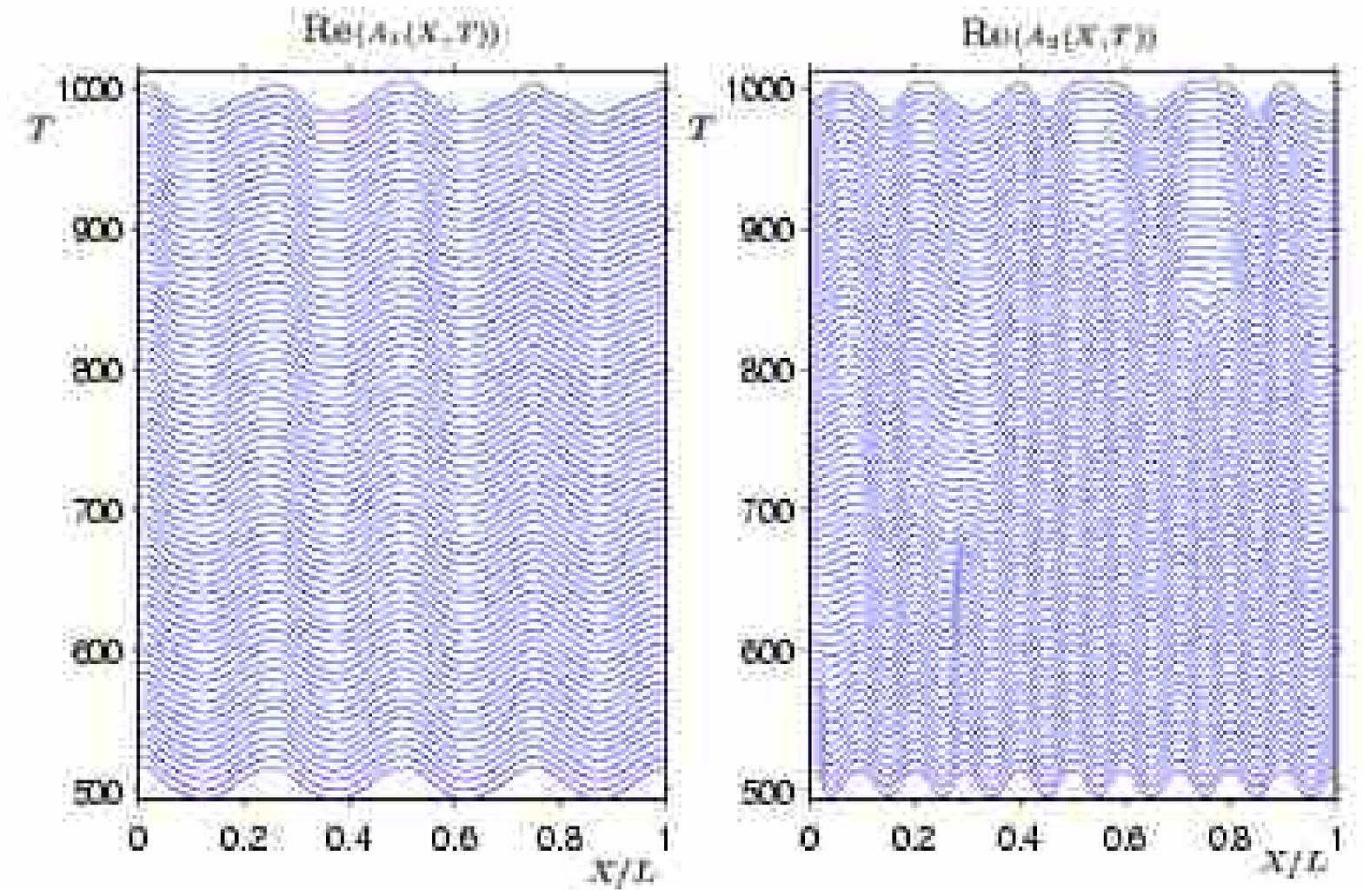}}
\centerline{\includegraphics[height=4.5cm]{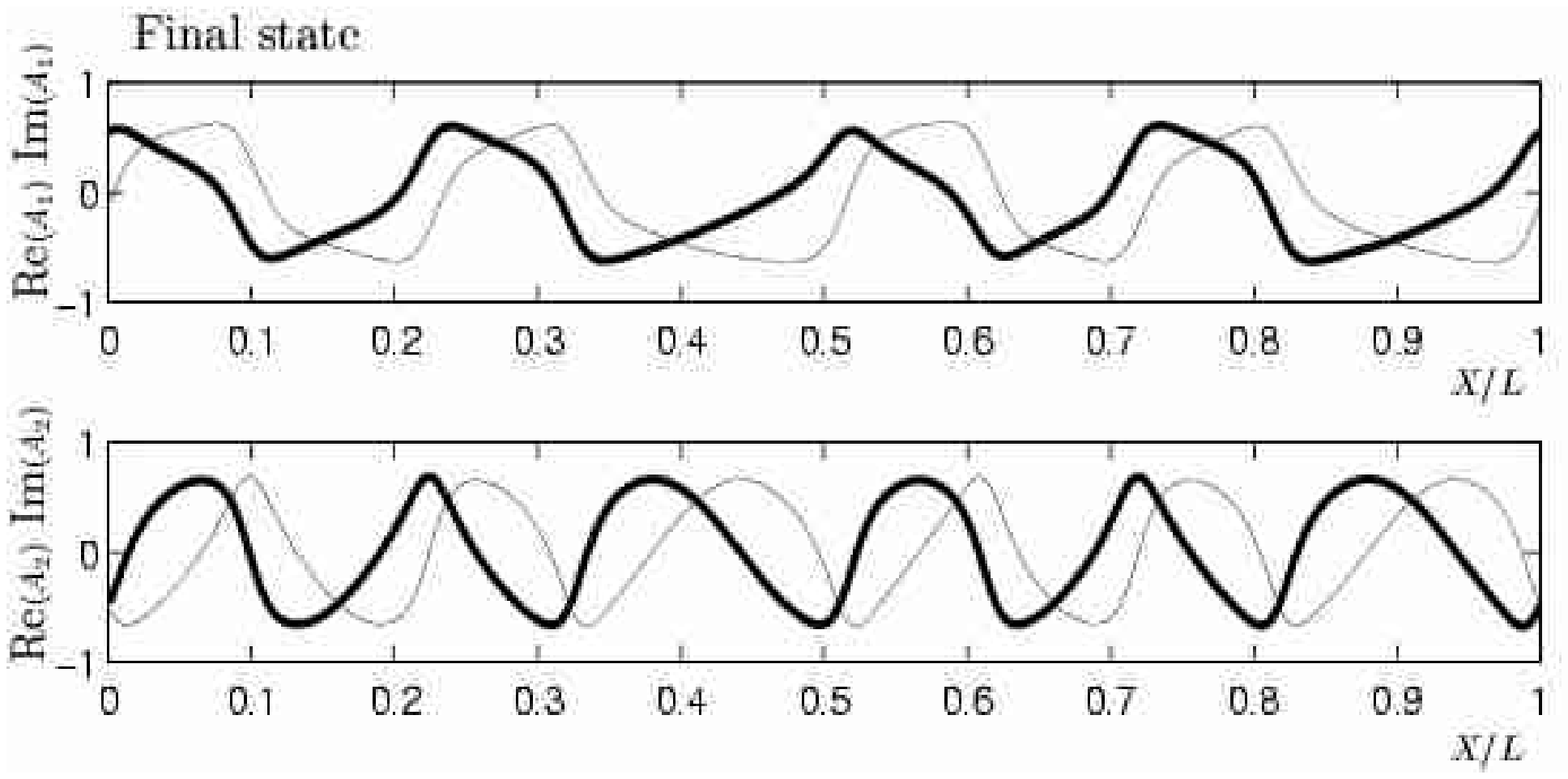}}
\caption{Evolution of longwave instability of the mixed mode $S_+$.
Shown are space-time diagrams of the
real part of the amplitudes and the final state
with thick (thin) lines denoting the real (imaginary) parts of  $A_1$ and $A_2$,
respectively. Parameters as in
Fig.~\ref{fig:growthgamma04}a with $k=0.1$, $\mu=0.423$
(indicated by rhomb) and
$L=300$.} \cierto{fig:spacetime1b}
\end{figure}
%%%%%%%%%%%%%FIG%%%%%%%%%%%%%

The mixed modes can undergo a shortwave steady instability as
well. As discussed in Section~\ref{sec:side-band} it occurs
when the
stability limit $\Gamma_+$ ($\Gamma_-$) of $S_+$ ($S_-$) is
close to the transition line $L_2$ ($L_1^-$). The instability
is primarily in the direction  of the mode with smaller
amplitude: $A_1$ for $\Gamma_+$ near $L_2$ and $A_2$ for
$\Gamma_-$ near $L_1^-$.  The resulting evolution of
system~(\ref{2},\ref{3})  is shown in Fig.~\ref{fig:spacetime2}
for parameters near $\Gamma_+$. It  confirms the expectation
that at least initially only $A_1$ changes, but not $A_2$.  The
space-time diagram for Re$(A_1(X,T))$ in
Fig.~\ref{fig:spacetime2} shows how the growing perturbation
determines the wavenumber of the final state; this change from
the initial wavenumber takes place $via$  phase-slips.

 The wavenumber $k_1$ of $A_1$ in the final state is controlled by the
 detuning parameter $\gamma$ in the manner predicted by the linear
 analysis: $k_1\simeq \gamma/(2\delta)$.  In other words, the number of maxima
observed in $A_1$ is $N_1  \simeq L \gamma/(4\pi\delta)$ (see lower figures of
Fig.~\ref{fig:spacetime2}).

%%%%%%%%%%%%%FIG%%%%%%%%%%%%%
\begin{figure}[htb]
\centerline{\includegraphics[height=6cm, width=10cm]{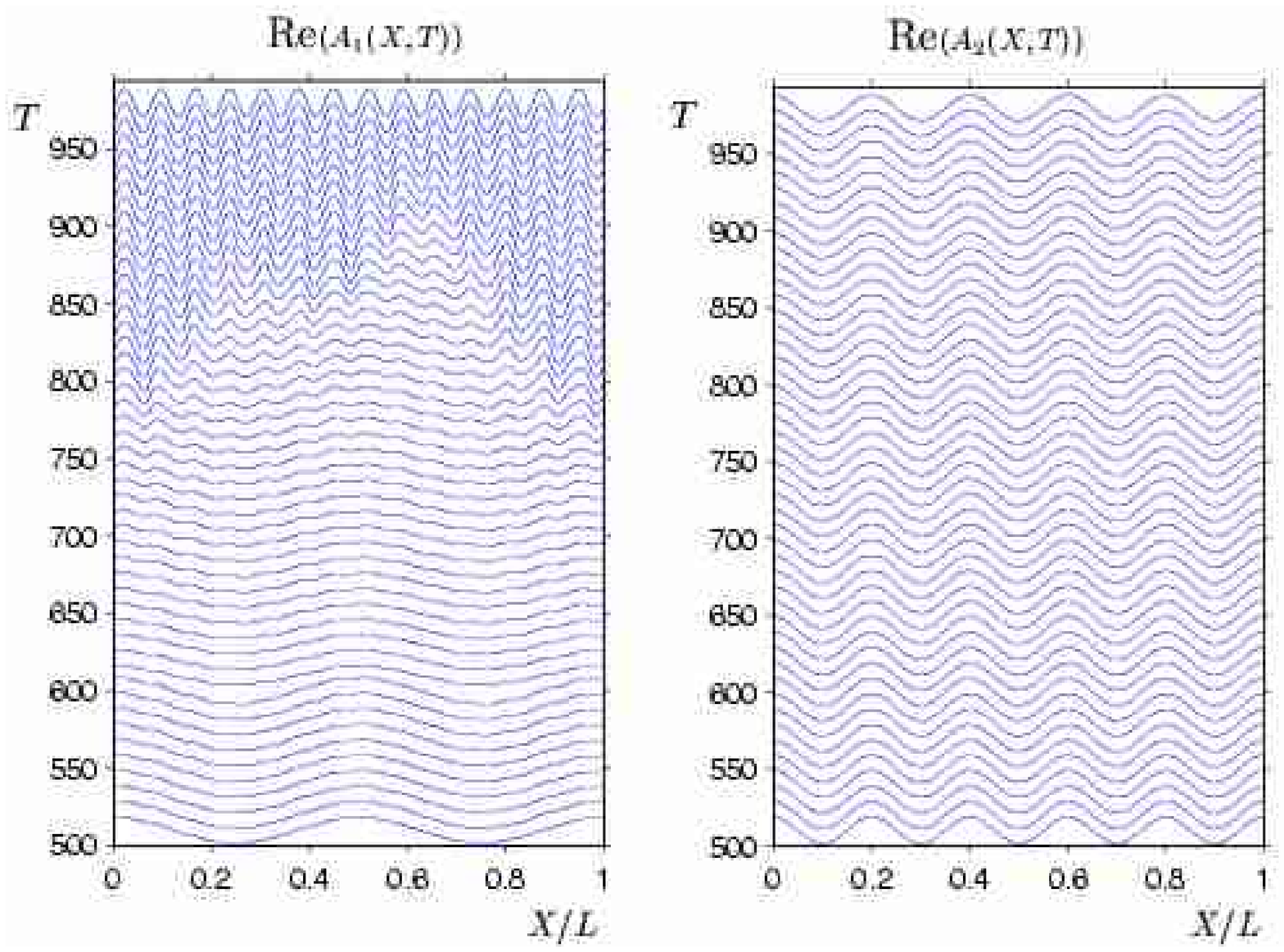}}
\centerline{\includegraphics[height=4cm, width=10cm]{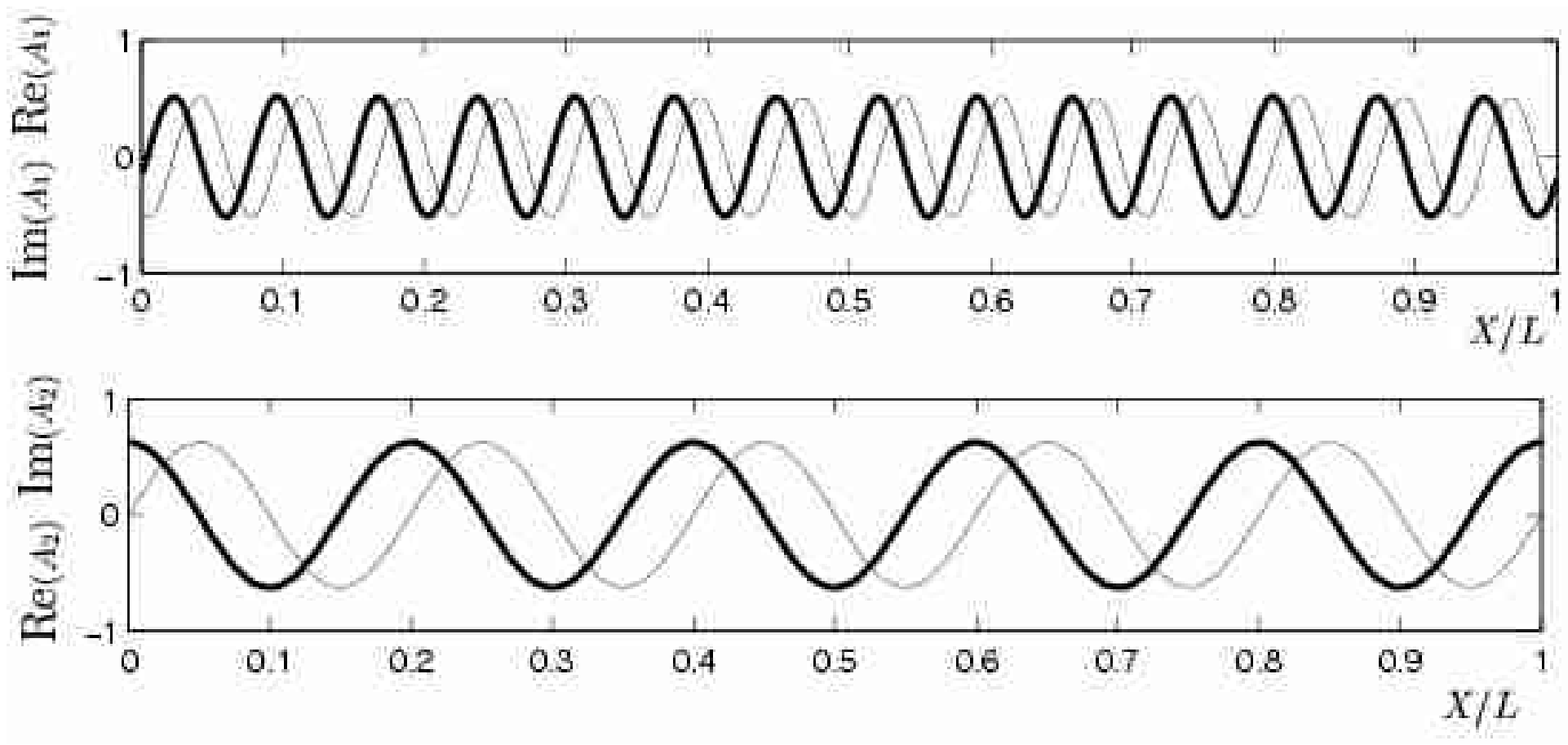}}
\caption{Space-time diagram showing the real part of the amplitudes $A_1$ and $A_2$.   The
lower figures show the real (thick line) and imaginary (thin line) parts of $A_1$ and $A_2$
corresponding to the final stable stationary state.  The parameters are as in
Fig.~\ref{fig:dgrm2n5withtw(a)}b ({\it square }) with $k=-0.05$, $\mu=0.27$  and $L=250$.}
\cierto{fig:spacetime2}
 \end{figure}
 %%%%%%%%%%%%%FIG%%%%%%%%%%%%%

The evolution of $A_1$ and $A_2$ after a shortwave steady
instability on $\Gamma_-$  near $L_1^-$ is shown in
Fig.~\ref{fig:spacetime3}.  This time the amplitude $A_1$
remains nearly constant while $A_2$ changes.   The destabilizing
mode was shown in Section~\ref{sec:side-band} to be of the form
$a_2^+(T)e^{i(k\frac{n}{m}+Q)X}+ a_2^-(T)e^{i(k\frac{n}{m}-Q)X}$
with $Q$ defined by Eq.~(\ref{7wn1}).  For the cases illustrated
in Fig.~\ref{fig:spacetime3} we have $|a_2^+(T)| \ll |a_2^-(T)|$
and the  wavenumber of the linearly unstable mode is
approximately given by $k_2\approx k \frac{n}{m}-|Q|$.  Note
that this wavenumber is essentially determined by the resonance
terms and would be 0  without them (see Eq.~\ref{7wn1}).  In
particular, for the example considered in
Fig.~\ref{fig:spacetime3}a, where $\nu=0.62$ and
$\nu'=-1.02$,
we have  $|k_2|\simeq 0.144$, while
in Fig.~\ref{fig:spacetime3}b $\nu=-\nu'=0.05$ and $|k_2|\simeq 0.018$;
note that with  regard to this effect both cases are
are analogous, except for the magnitude of the resonance coefficients.

In addition to the wavenumber $k_2$ determined by the linear
stability analysis, Fig.~\ref{fig:spacetime3} reveals a second
prominent wavenumber $k_2'$, which is the result of the
resonance term $\bar{A}_2^{m-1}A_1^n$. It acts as a driving term
for $A_2$ and generates a mode with wavenumber $k_2'$ that is
determined by
\begin{equation}  nk_1\pm(m-1)|k_2|=\pm |k_2'|,
\cierto{19}  \end{equation}
where $k_1$ is the wavenumber associated with amplitude $A_1$.
In the evolution shown in Fig.\ref{fig:spacetime3} $k_1$ remains
unchanged.  The same mechanism is at work for the short-wave
instability of $S_+$ (cf. Fig.\ref{fig:spacetime2}), which
occurs near $L_2$. There the wavenumber modulation of $A_1$ is,
however, negligible. It is driven by the resonance term
$\bar{A}_1^{n-1}A_2^m$, which is proportional to the fourth
power of the amplitude $A_1$ ($n=5$) which is very small where $S_+$
branches off the pure mode $S_2$. This is not the case for the
short-wave instability of $S_-$ (Fig.\ref{fig:spacetime3}),
which occurs near $L_1^-$.
There it is $A_2$ that is small; but
it enters the resonance term $\bar{A}_2^{m-1}A_1^{n}$ linearly
($m=2$) and therefore the resonance term
provides a relatively strong modulation of $A_2$.  Substituting the
final values of $k_1$ (same as the initial value) and $k_2$
(calculated above) into Eq.(\ref{19}) one obtains
 $|k_2'|=2.644$ in the case of Fig.~\ref{fig:spacetime3}a
and $|k_2'|=0.63$ for Fig.\ref{fig:spacetime3}b.
For the numbers ($N_1$,$N_2$,$N'_2$)
of wavelengths associated with $k_1$, $k_2$, and $k_2'$,
respectively this implies $(N_1,N_2,N_2')=(20,6,106)\simeq
L/(2\pi)(0.5,0.144,2.644)$ for
Fig.~\ref{fig:spacetime3}a (where $L=250$) and
$(N_1,N_2,N_2')=(6,1,31)\simeq L/(2\pi)(0.124,0.02,0.63)$ for
Fig.~\ref{fig:spacetime3}b (where $L=305$).  These results
compare quite well with the
results in Fig.~\ref{fig:spacetime3}
where  $(N_1,N_2,N_2')=(20,6,106)$  in
Fig.~\ref{fig:spacetime3}a and $(N_1,N_2,N_2')=(6,0,30)$  in
Fig.~\ref{fig:spacetime3}b.

%%%%%%%%%%%%%FIG%%%%%%%%%%%%%
\begin{figure}[htb]
\hskip1truecm{\includegraphics[height=5.5cm, width=11cm]{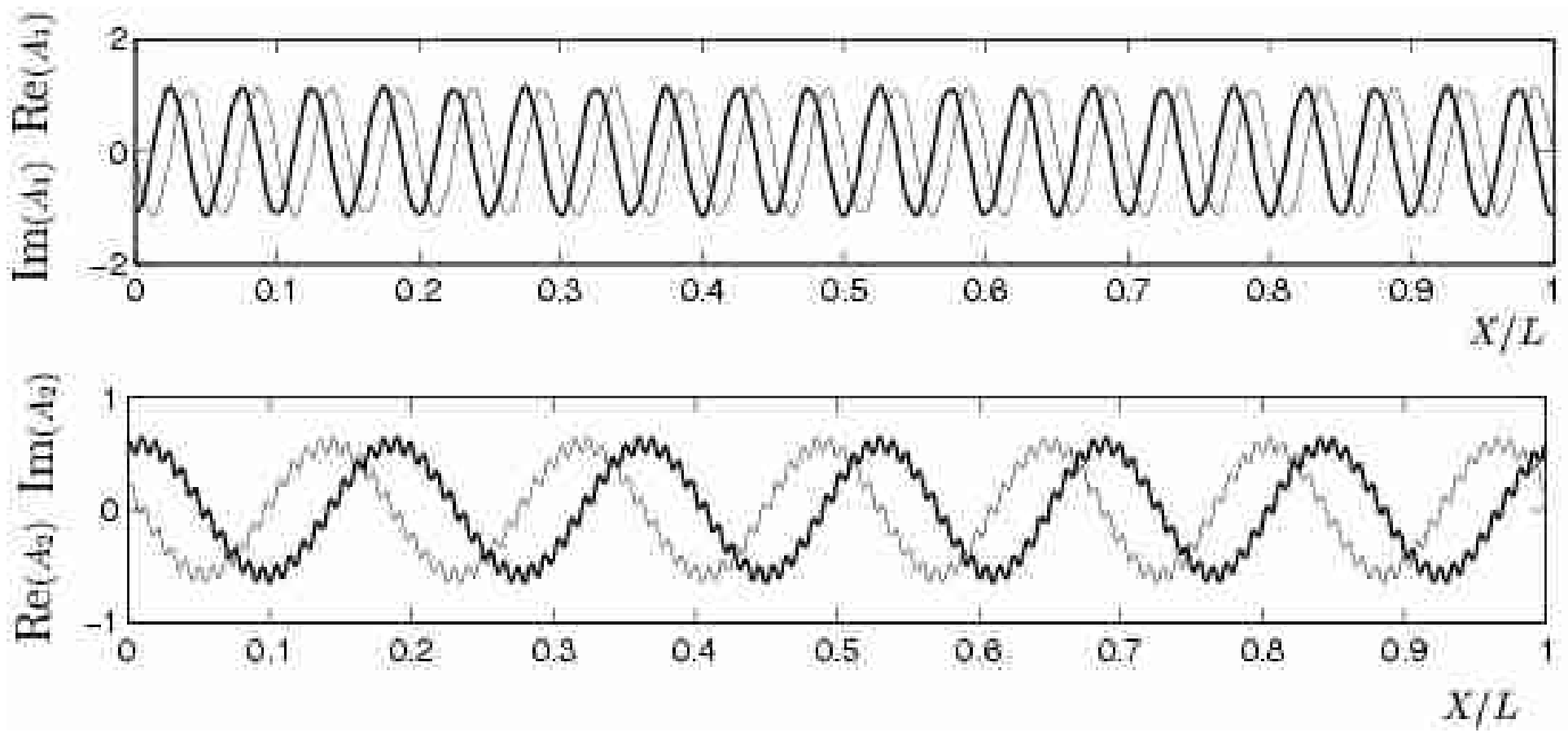}}
\centerline{\includegraphics[height=3.5cm, width=10cm]{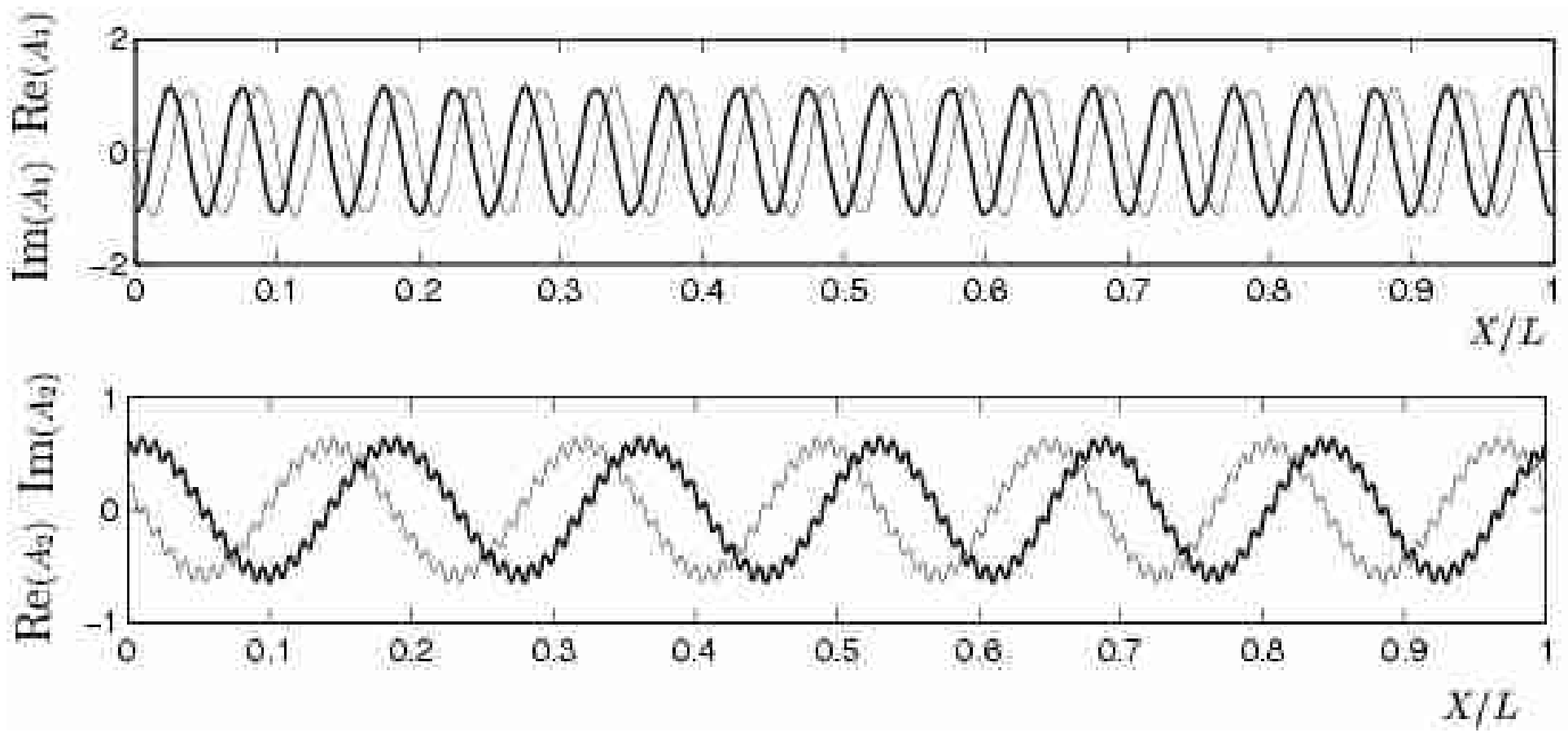}}
\vskip0.5truecm
\hskip1truecm{\includegraphics[height=5.5cm, width=11cm]{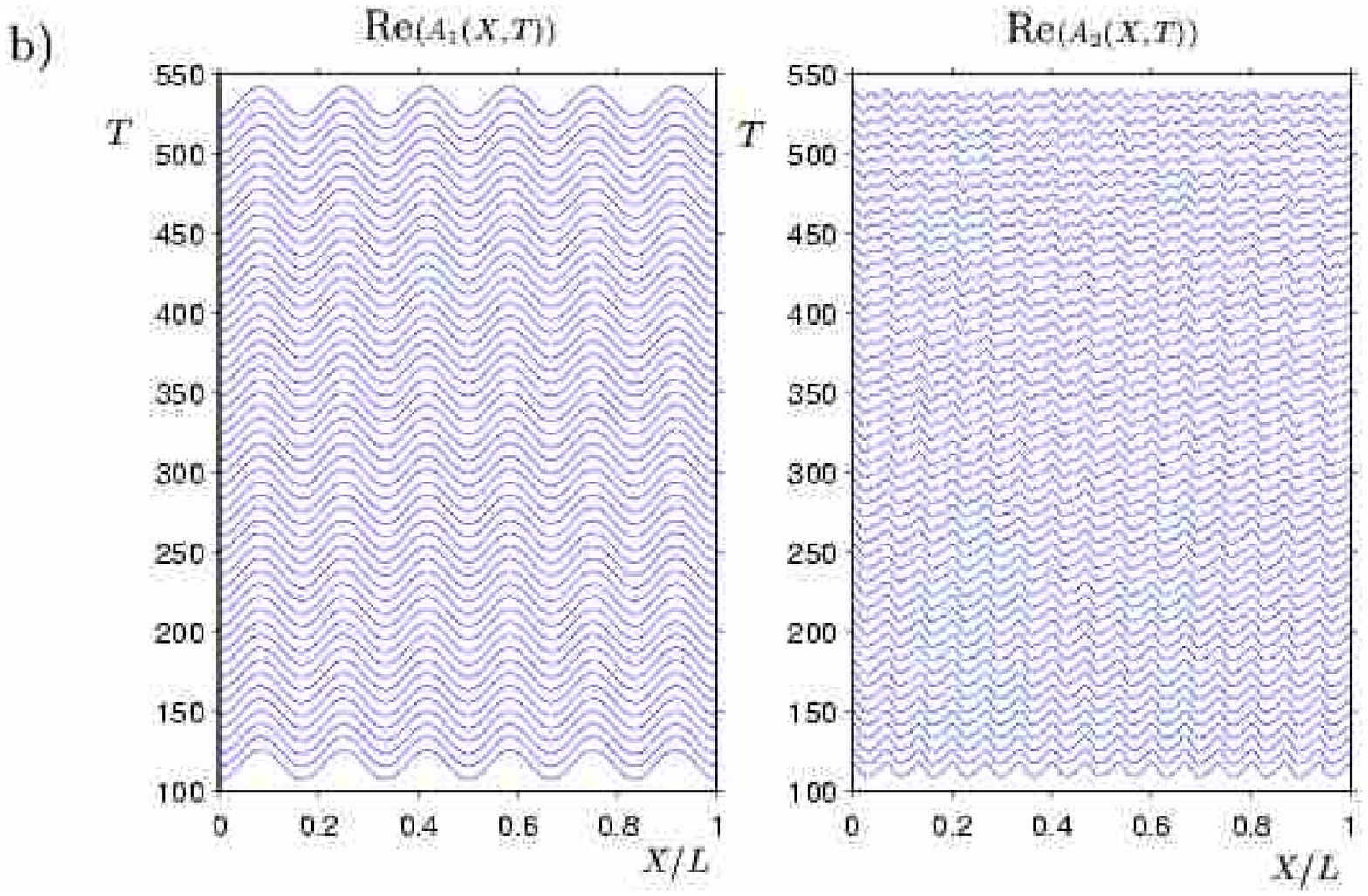}}
\centerline{\includegraphics[height=3.5cm, width=10cm]{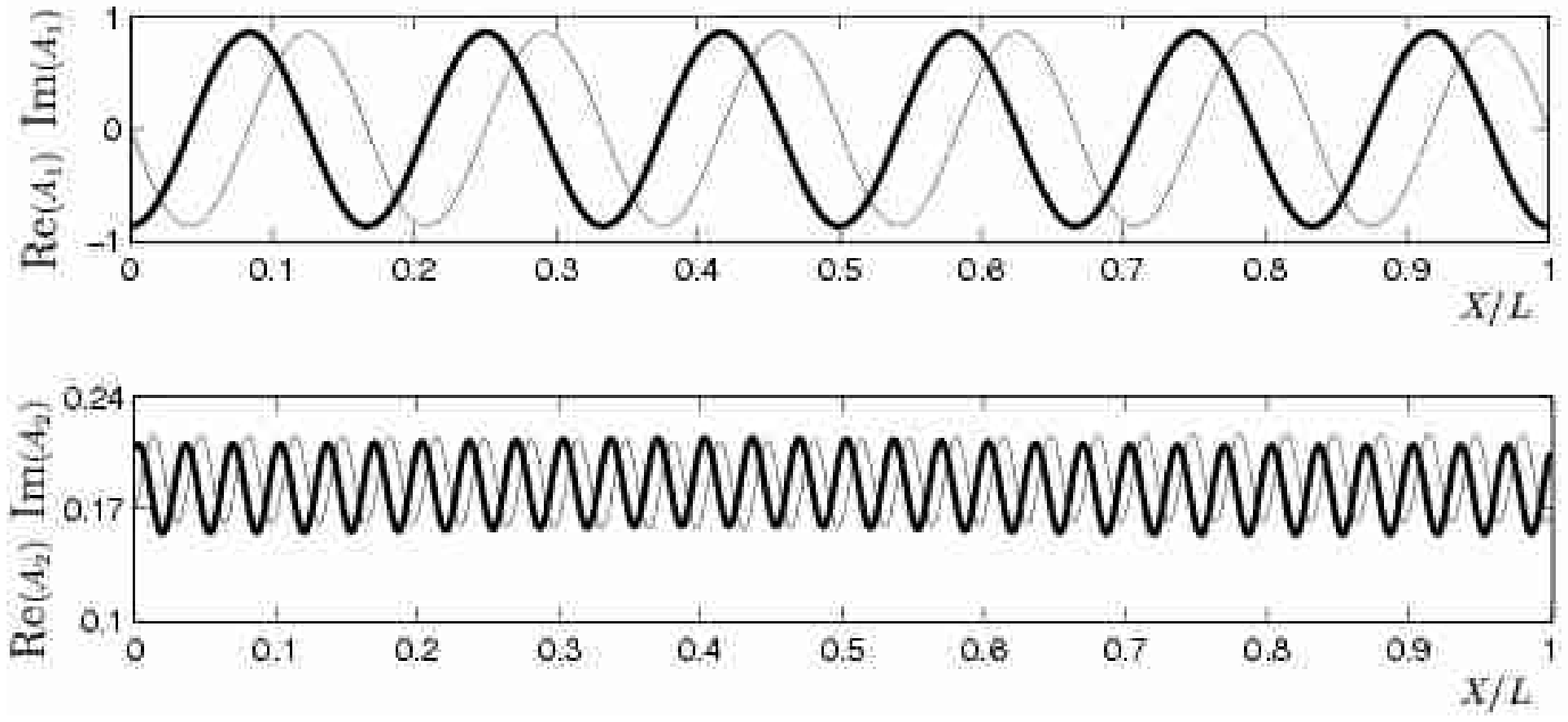}}
\caption{Space-time diagram showing the real parts of $A_1$ and
$A_2$ when mixed
mode $S_-$ undergoes a shortwave steady instability.  The parameters in (a) are those of
Fig.~\ref{fig:dgrm2n5withtw(a)}b ({\it circle}) with $k=0.5$, $\mu=1.05$  and $L=250$.  Parameters
in (b) are as in Fig.~\ref{fig:dgrm2n5withtw(a)}d ({\it circle})
with $k=0.125$, $\mu=0.95$  and $L=305$.}
\cierto{fig:spacetime3}
 \end{figure}
%%%%%%%%%%%%%%%%%%%%%%%%%%%%%%%%%%%%%%%%%%%%%%%%%

\clearpage
%%%%%%%%%%%%%FIG%%%%%%%%%%%%%
\begin{figure}[htb]
\centerline{\includegraphics[height=8cm]{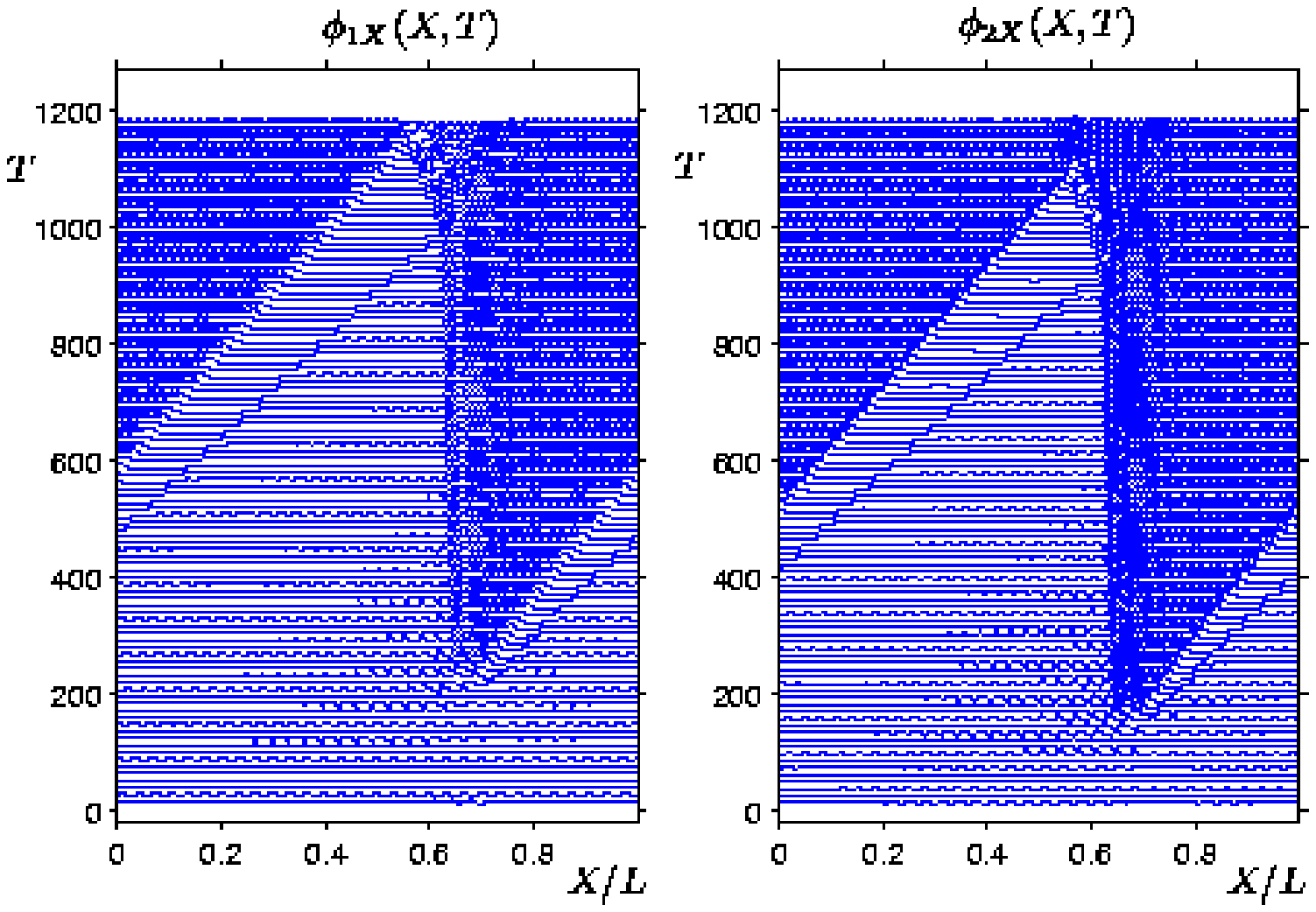}}
\centerline{\includegraphics[height=8cm]{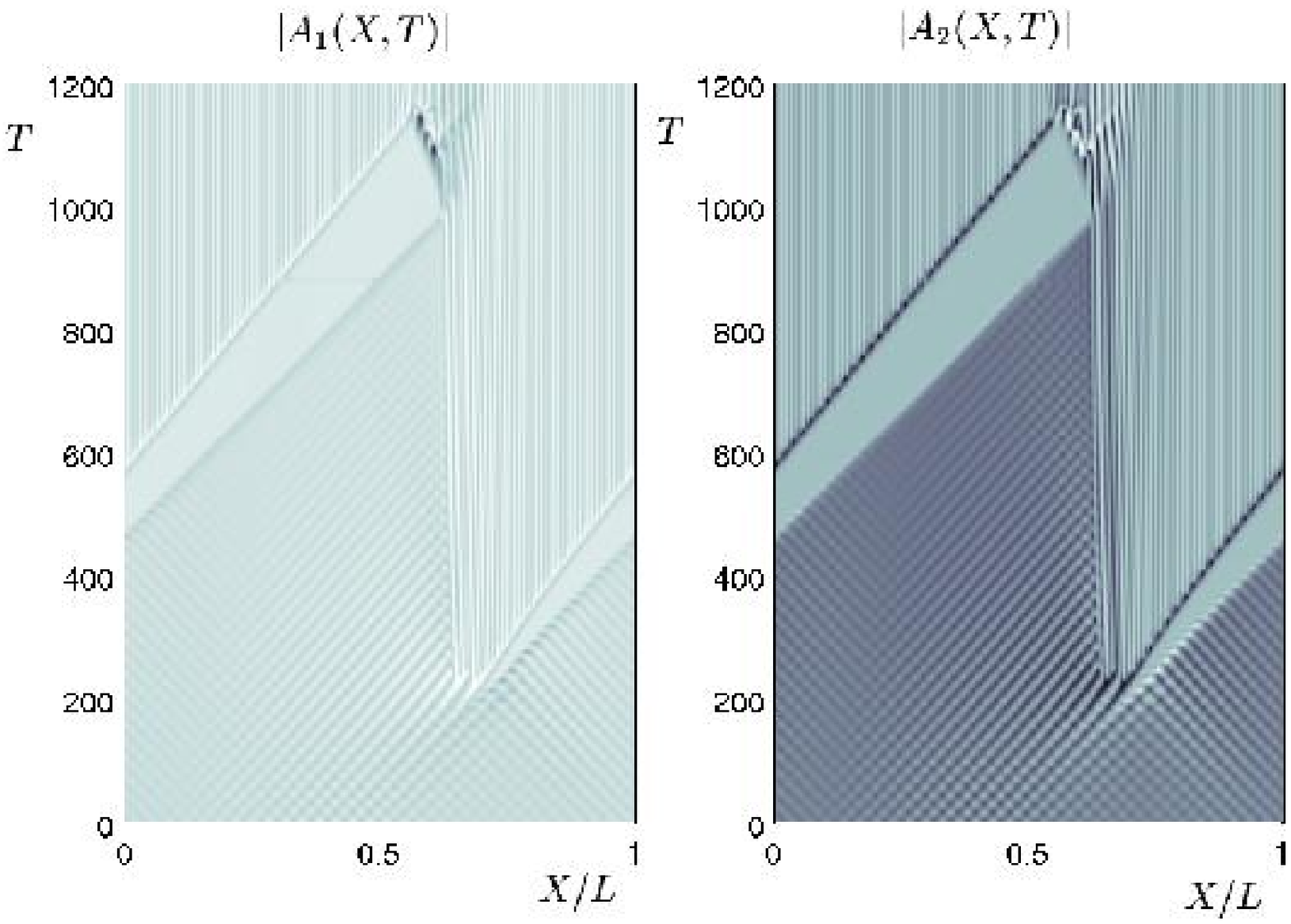}}
\centerline{\includegraphics[height=5.5cm,width=12cm]{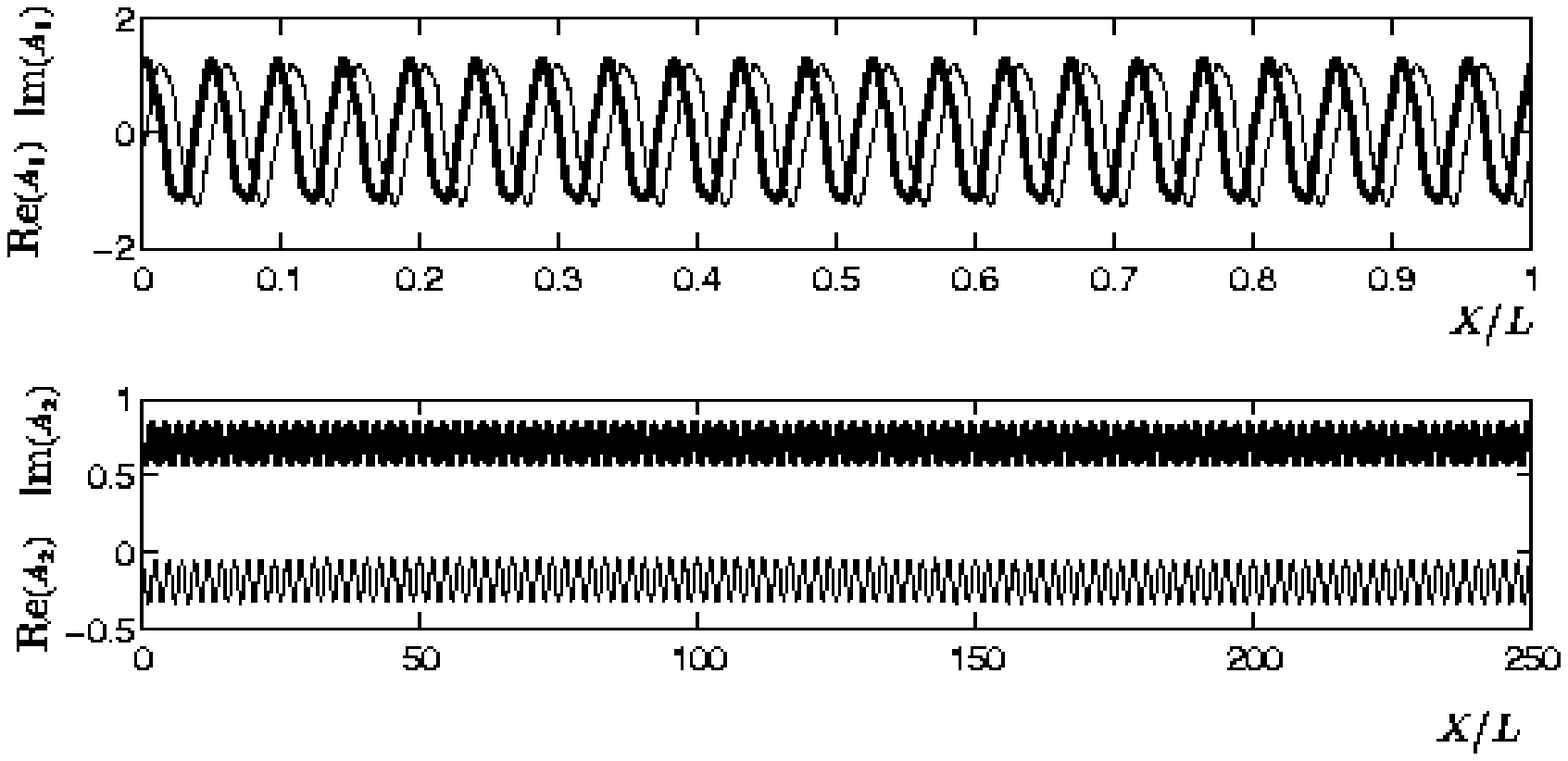}}
\caption{Space-time diagram of the phase-gradient and the magnitude
of the  amplitudes $A_1$  and $A_2$. Lower figures:
Real (thick) and imaginary (thin) part of  $A_1$  and $A_2$ corresponding
to the final stable state. Parameter as in
Fig.~\ref{fig:dgrm2n5withtw(a)}b ({\it triangle})
 with $k=0.5$, $\mu=1.308$  and $L=250$. In  the gray-scale plots dark
 (light) stands for low (high) amplitudes.}
\cierto{fig:spacetime4}
\end{figure}
 %%%%%%%%%%%%%FIG%%%%%%%%%%%%%

Qualitatively different is the evolution of the oscillatory
instability of the mixed mode, as illustrated in
Figs.~\ref{fig:spacetime4}-\ref{fig:spacetime8}. Small
perturbations initially evolve into standing-wave oscillations.
They do not last for long, however, and decompose into left- and
right-traveling disturbances, which then form localized
low-wavenumber domains of waves drifting to the right and to the
left, respectively. Such localized, propagating regions of
`drift waves' are typical for parity-breaking instabilities
because the extended drift waves are generically unstable at
onset \cite{RiPa92,CaCa92,FaDo90} and may become stable only for
larger amplitudes \cite{BaMa94}. The localized waves can be
described using equations for the amplitude of the
parity-breaking mode and the phase of the underlying pattern and
can arise stably  when the parity-breaking bifurcation is
subcritical \cite{CoGo89,GoGu91} as well as when it is
supercritical \cite{RiPa92,CaCa92}.  They are related to the
solitary modes observed in experiments in directional
solidification  \cite{SiBe88,FaCh89}  as well as in Taylor vortex
flow \cite{WiMc92}, and in simulations of premixed flames
\cite{BaMa94}. In these systems the parity-breaking bifurcation
arises from a 1:2 mode interaction \cite{LeRa91,RiPa92} rather
than the 2:5-resonance considered here.

%%%%%%%%%%%%%FIG%%%%%%%%%%%%%
\begin{figure}[htb]
\centerline{\includegraphics[height=9cm]{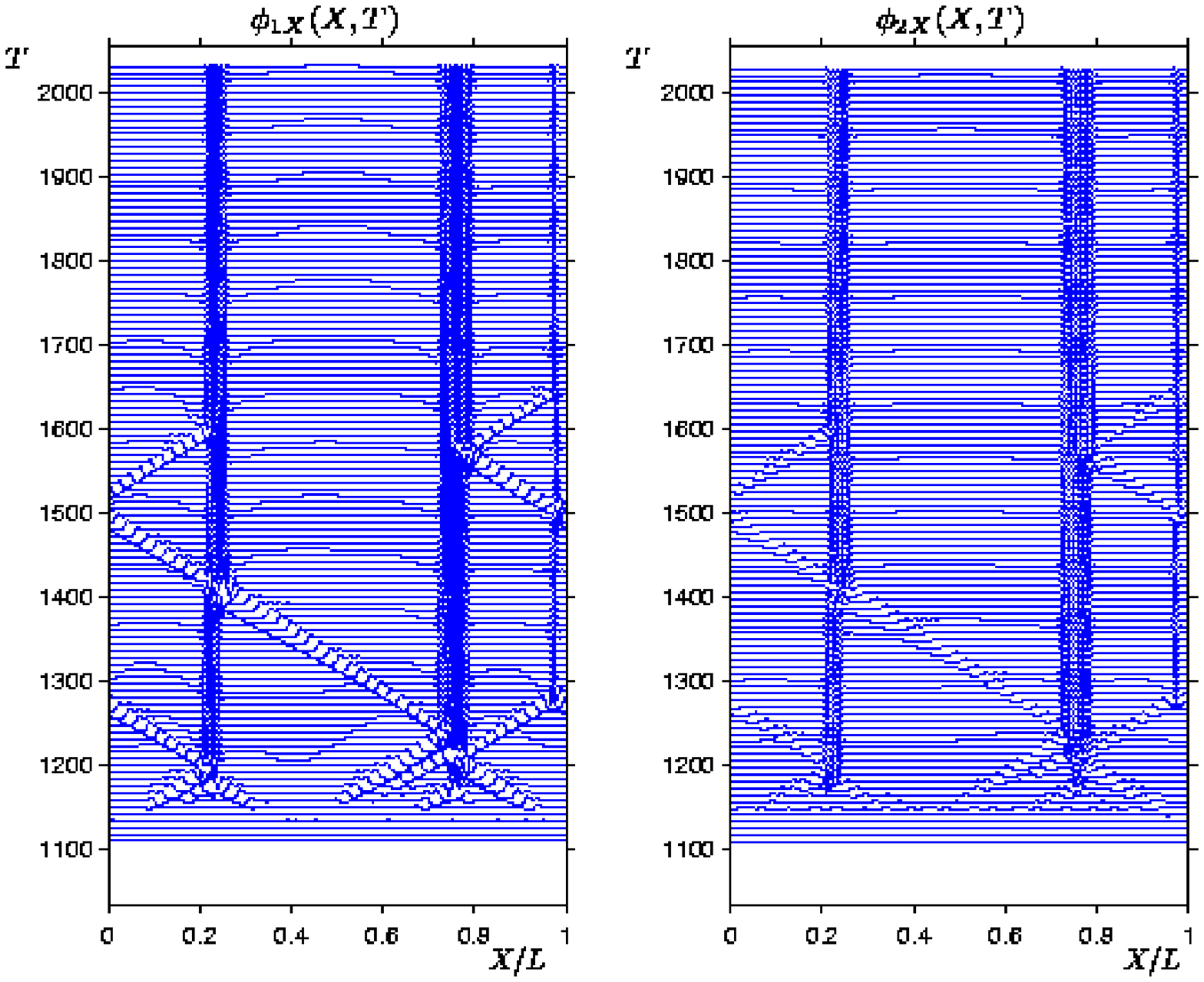}}
\centerline{\includegraphics[width=10cm]{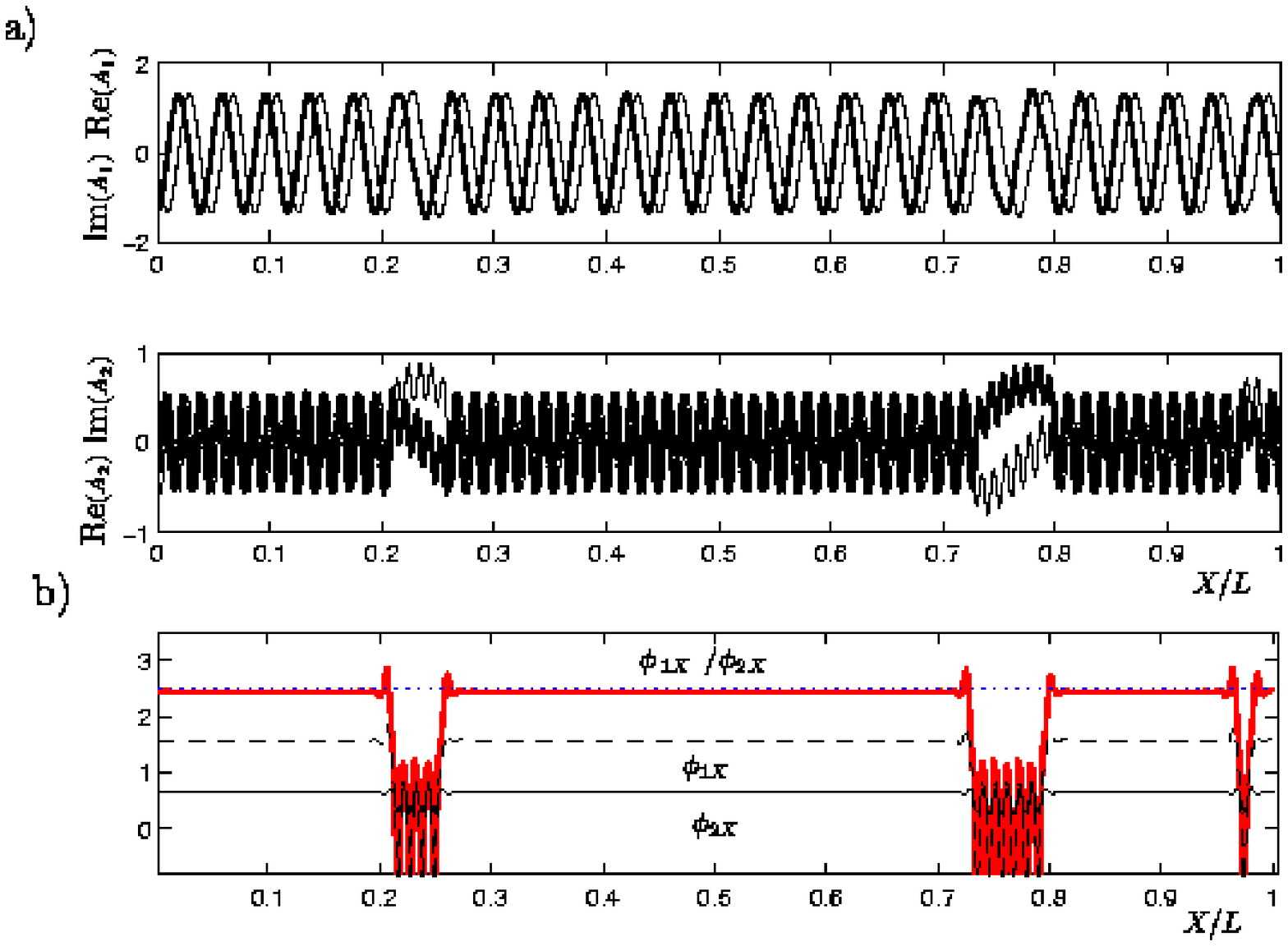}}
\caption{a) Space-time diagram of the
phase-gradients amplitudes $A_1$  and $A_2$, respectively.
Parameters as in Fig.~\ref{fig:dgrm2n5withtw(a)}b but $\gamma=1$
and with $k=0.5$, $\mu=1.42$ and $L=250$.
 Lower figures correspond
to the final state: a) Real and imaginary part of the amplitudes $A_1$ and $A_2$
b) phase gradients of $A_1$ and $A_2$. Dotted line indicates perfect $2:5$ resonance.
The temporal evolution of $A_2$ is also shown in movie
[movie:fig.5.7.avi]; red and yellow lines show the real
and imaginary part of $A_2$, respectively, white its magnitude $|A|$,
and green the local wavenumber.}
\cierto{fig:spacetime5}
\end{figure}
 %%%%%%%%%%%%%FIG%%%%%%%%%%%%%

The localized drift waves do not always persist. In the
simulation shown in  Fig.\ref{fig:spacetime4} a sequence of
phase slips occurs in mode $A_2$ at the trailing end of the
localized drift wave. The phase slips
substantially reduces the wavenumber in the growing domain
between the location where the localized drift wave first
is created (at $x \approx 0.6/L$) and its
trailing edge. Since no phase slips occur in mode $A_1$ the
wavenumbers of $A_1$ and $A_2$ are not resonant any more in this
growing domain, which results in a strong modulation of the
wavenumber. Due to the periodic boundary
conditions the growing, localized drift wave collides
eventually with this domain and is absorbed by it. After
that the pattern becomes stationary. The stationary domains with
strongly modulated wavenumber do not always absorb the localized
drift waves. Fig.\ref{fig:spacetime5} and the movie {\tt
fig5.7.avi} show a case in which the localized drift waves at
times are also reflected by the stationary domains or pass
through them. We have not investigated which
factors determine the outcome of the collisions. In the
simulation shown in Fig.\ref{fig:spacetime5} the system
eventually evolves into a stationary state. It consists of
relatively large domains in which the wavenumber ratio is very
close to 2:5 and small domains in which the wavenumbers are
strongly modulated and not in this rational ratio. One may have
expected that the localized stationary domains
attract each other and eventually merge; but this was not observed.
Presumably, the strong wavenumber oscillations, which are
particularly visible in $A_2$, lock the domains in place. Thus,
in the final, stationary state shown in
Fig.\ref{fig:spacetime5}b one has a periodic pattern that is
interrupted by small domains in which the pattern is not
periodic.

The transients and the resulting final state depend very
sensitively on the amplitude of the initial perturbations of the
periodic state. The relatively large perturbations used in
Fig.\ref{fig:spacetime5} lead to a large number of
localized drift waves, which upon their collision generate
localized stationary structures. Since the
localized stationary structures often absorb the  localized
drift waves the final state tends to be stationary in this case.
If the initial perturbations are much smaller, fewer drift waves
arise. If they travel in the same direction no stationary
localized structures are generated and the localized drift waves
persist. Such a case is illustrated in
Fig.\ref{fig:spacetime7},\ref{fig:spacetime8} and movie
{\tt fig5.8.avi}. Here, uniformly distributed
perturbations with maximal amplitude 0.001 were applied
independently to the real and imaginary part of both amplitudes.
In this case only a single domain forms. It
develops a defect in its interior, which eventually disappears
at its trailing end (cf. movie {\tt
fig5.8.avi}). Initially, the propagation velocities of the
leading and the trailing front of the domain are not equal and
the domain grows. As time goes on, however, the front velocities
converge to a common value and one sees a stable, localized,
propagating domain of fixed size containing traveling waves
(see Fig.~\ref{fig:spacetime8}). It is interesting to note that
the wavenumbers  selected in these traveling domains are in
$2:5$ resonance, while the stationary  regions have slightly
shifted wavenumbers that are not in resonance (see upper panel
of Fig.~\ref{fig:spacetime8}). The physical solution
(cf.(\ref{1a})) can therefore by described as consisting
of regions of localized periodic waves traveling through a
stationary  quasi-periodic pattern. This type of solution is
generally found when the  side-band oscillatory instability
occurs very close to the parity-breaking  bifurcation. To describe
the parity-breaking bifurcation in more detail the appropriate
amplitude-phase equations \cite{CoGo89,RiPa92,CaCa92,BaMa94} could be
derived from the coupled Ginzburg-Landau equations. They would, in particular,
allow a better understading of the localized drift waves.
Such an undertaking is, however, beyond the scope of this paper.

%%%%%%%%%%%%%FIG%%%%%%%%%%%%%
\begin{figure}[htb]
\centerline{\includegraphics[height=9cm]{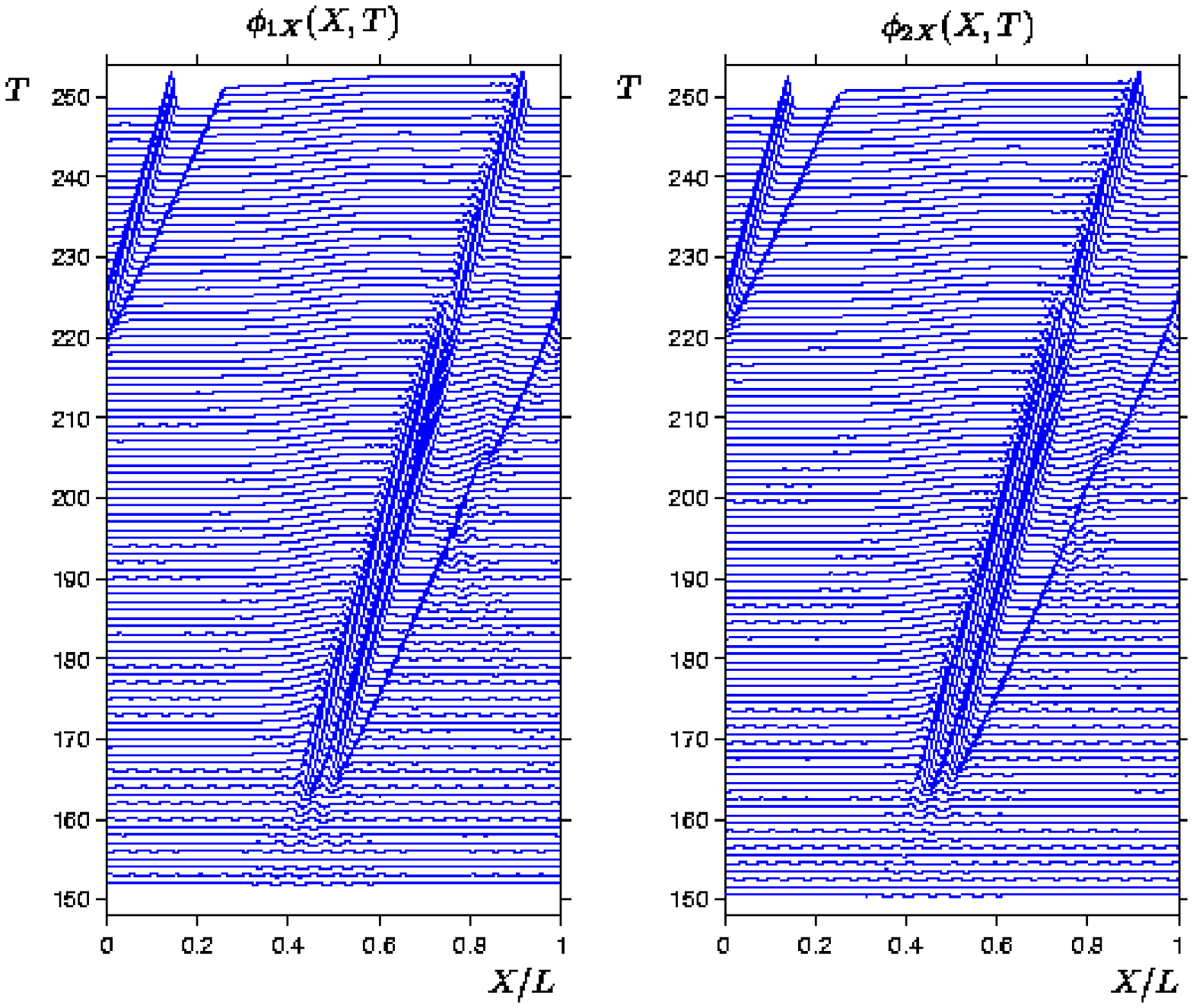}}
\centerline{\includegraphics[height=9cm]{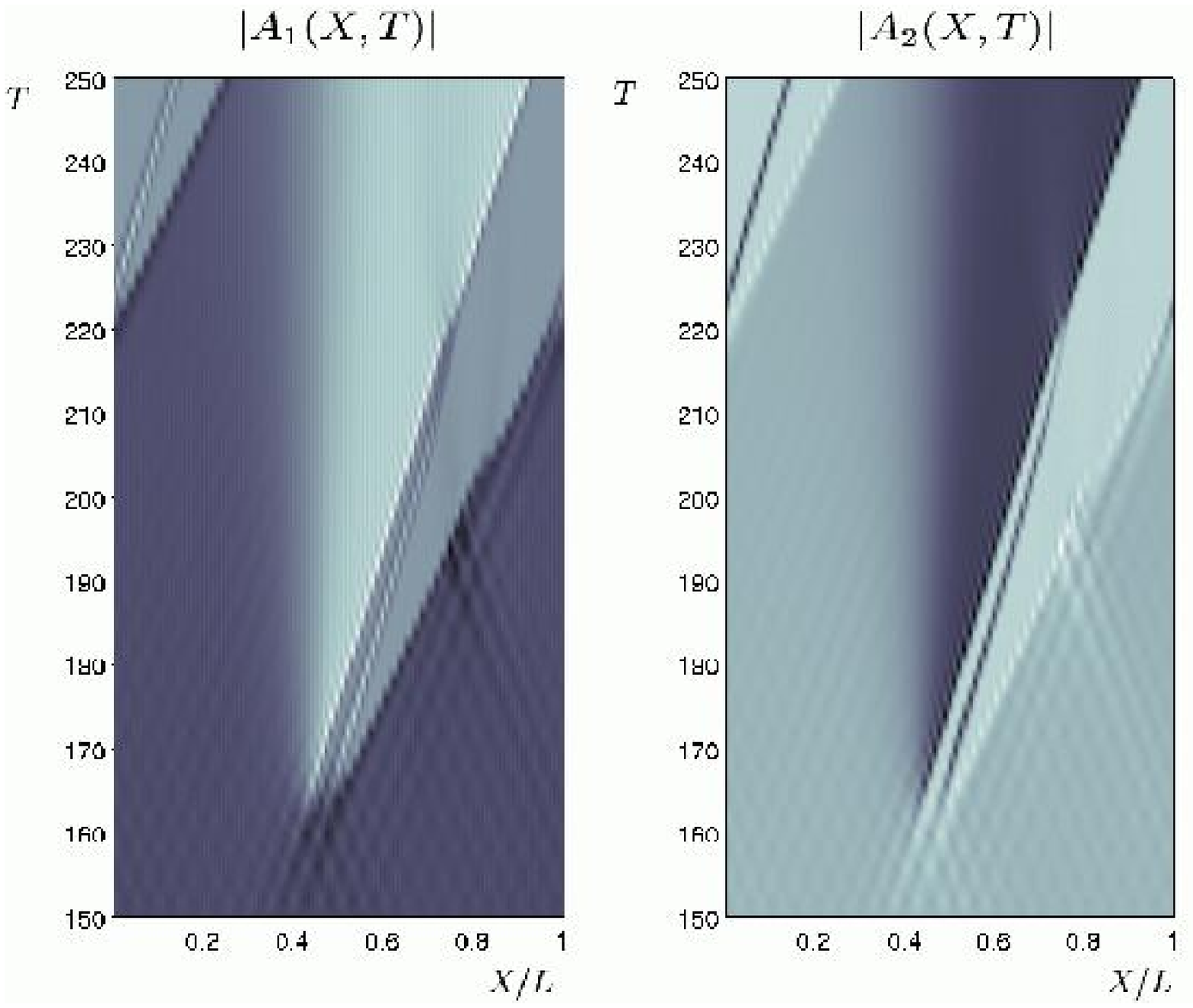}}
\caption{Space-time diagram for the magnitude and the phase-gradient
of  the amplitudes $A_1$  and $A_2$.
Parameters:  as in Fig.~\ref{fig:dgrm2n5withtw(a)}b but $\gamma=1.3$
and with $k=0.5$, $\mu=1.42$ and $L=250$. In  the gray-scale plots dark
 (light) stands for low (high) amplitudes. The temporal evolution of $A_2$ is also shown in
 movie [movie:fig.5.8.avi]; red and yellow lines show the real
and imaginary part of $A_2$, respectively, white its magnitude $|A|$,
and green the local wavenumber.}
\cierto{fig:spacetime7}
\end{figure}
%%%%%%%%%%%%%FIG%%%%%%%%%%%%%

%%%%%%%%%%%%%FIG%%%%%%%%%%%%%
\begin{figure}[htb]
\centerline{\includegraphics[height=18cm]{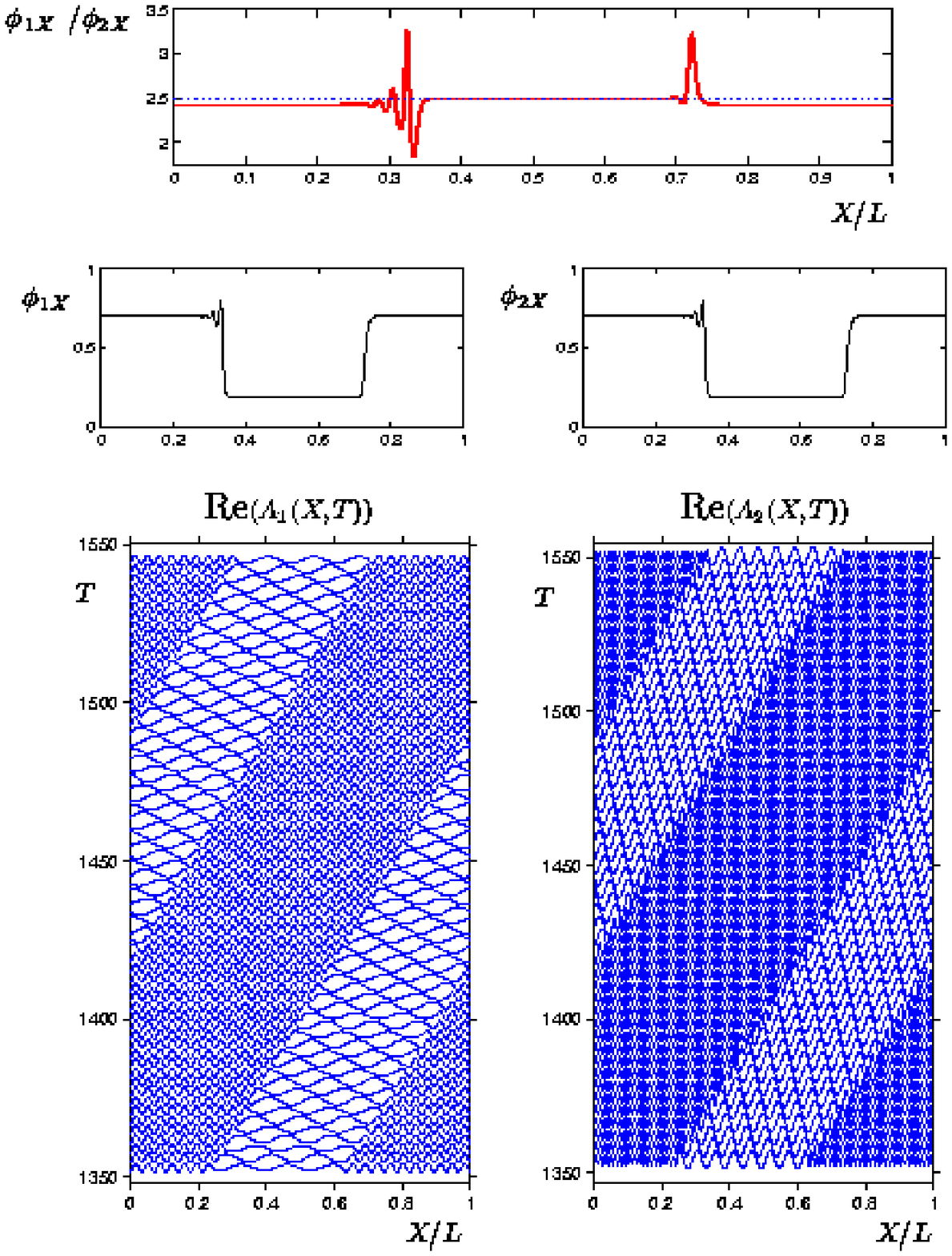}}
\caption{Final, localized drift wave reached after
the transient shown in Fig.~\ref{fig:spacetime7}. Dotted line in
plot of $\phi_{1X}/phi_{2X}$ indicates perfect $2:5$ resonance.}
\cierto{fig:spacetime8}
\end{figure}
%%%%%%%%%%%%%%%%%%%%%%%%%%%%%%%%

Finally, in order to determine the character of the bifurcations
 we have also integrated  Eqs.~(\ref{2},~\ref{3}) using the final states as
initial conditions and varying $\mu$ so as to take the system
back into the stable regions of mixed modes. The results
obtained indicate that the  instabilities discussed in this section are
subcritical (except for the cases shown in Fig.\ref{fig:puremixed}).
The solutions persist over a large interval of
the forcing giving rise to bi-stability between periodic and
quasi-periodic states. This behaviour has been found in all cases
we have considered. Therefore we expect that the coexistence of
periodic and quasi-periodic stable states may be regarded  as a
characteristic property of the system Eqs.~(\ref{2},~\ref{3}).

\clearpage

\section{Conclusions}

In this paper we have investigated a near-resonant steady $m:n$
mode interaction in one-dimensional dissipative systems with
reflection and translation symmetry using coupled
Ginzburg-Landau equations that describe the slow evolution of
the two relevant mode envelopes. One of the goals was to shed
some light on the competition between periodic and quasi-periodic
patterns by addressing in detail the stability of the periodic
solutions, emphasizing particular instabilities that can lead to
quasi-periodic patterns. 

The simplest stationary solutions of this system are
reflection-symmetric periodic patterns: the pure modes $S_{1,2}$ and
the mixed modes $S_\pm$.  Building on the results of Dangelmayr
\cite{Da86} we established the stability properties of these
solutions with respect to perturbations that preserve the periodicity
of the solution. Among the many possible choices of parameters we
restricted our attention to sets of values that produced two
supercritical primary bifurcations to pure modes followed by
secondary bifurcations to mixed modes - these mixed modes, in turn,
suffered symmetry-breaking bifurcations to traveling waves.

Side-band instabilities modify the stability regions of the pure and
mixed modes.  In particular, we find that these steady periodic patterns
can undergo longwave as well as shortwave instabilities.  For the
pure modes the side-band instabilities are always of
steady-state type.  The destabilizing longwave perturbations
correspond to the well-known Eckhaus instability and depend only on
the linear coefficients of Eqs.~(\ref{2},\ref{3}). In contrast, the
shortwave instabilities depend on nonlinear interactions and
 on the resonance considered (i.e., on $m$ and $n$). In particular, we
find:

1.  If $m>2$ and a pure mode, $S_1$ say, loses stability in a
shortwave steady-state bifurcation, the wavenumber of the
growing perturbation lies at the band center of the other mode
(i.e. of $S_2$)); this
instability is affected by the detuning parameter
$\hat\gamma=\gamma/2\delta$, which measures the mismatch between
the two critical wavenumbers.

2.  For wavenumber ratios of the form $2:n$ one of the resonant terms
enters into the stability calculation for the pure mode $S_1$.
Away from the band center this pure mode can experience
shortwave instabilities while close to the band
center it can lose stability only to homogeneous or longwave
perturbations. In the short-wave case, due to the resonant terms the
wavenumber selected by the instability is shifted with respect to the
band center of the other pure mode, $S_2$. As the band center is
approached the perturbation wavenumber of the short-wave
instability can go to 0. 
 
Close to their bifurcation from the pure modes, the mixed modes
inherit the stability properties of the pure modes with respect
to short-wave perturbations. In addition, they may 
undergo an oscillatory instability. It results
from the interaction of the translation and the
symmetry-breaking modes and can be strongly affected by the
wavenumber mismatch.

The nonlinear evolution ensuing from the instabilities was
investigated through direct numerical simulation of
Eqs.~(\ref{2},\ref{3}) with periodic boundary conditions for the
slowly varying envelopes. It was found that the transition from
the pure modes to the mixed mode can depend strongly on the
wavenumber mismatch. We have identified cases in which without a
wavenumber mismatch the instability of the pure mode leads to a
periodic mixed mode, while in the non-resonant case the
instability of the pure mode is determined by a side-band
instability resulting in a quasi-periodic pattern.
 
Side-band instabilities of the mixed modes  generically destroy
the spatial periodicity of the stationary, reflection-symmetric
states and produce quasi-periodic patterns. They are typically
characterized by a spatial modulation of the local wavenumber  with a
wavelength that is determined by the detuning
parameter (in a manner predicted by the linear analysis) and the
resonant terms (as indicated in Eq.(\ref{19})). The size of the
modulation depends strongly on the strength of the resonance
terms. For the weakly nonlinear 2:5 resonance discussed here the
effect is only noticeable for $S_-$ (cf.
Fig.\ref{fig:spacetime3}); for $S_+$ the resonant terms are too
small (cf. Fig.\ref{fig:spacetime2}). 

The oscillatory instability tends to lead to the formation of
propagating fronts separating reflection symmetric
patterns from drift waves with broken
reflection symmetry. The fronts may interact to form
stable localized drift waves. These may subsequently collide
with each other, destroying the initial periodic pattern
completely in some cases and only partially in others. In the
latter case the system relaxes to a very interesting pattern
composed of alternating localized stationary domains of periodic
and quasi-periodic states. We find other cases where all the
fronts travel in the same direction, eventually achieving the
same velocity and producing stable localized 
propagating domains of drift waves.  The
wavenumbers in these traveling domains are selected by the
resonance condition (i.e., they are in the ratio $m/n$), whereas
the wavenumbers associated with the surrounding steady periodic
pattern are no longer resonant.  Thus, the resulting patterns
can be described as localized periodic traveling wave structures
propagating through a stationary quasi-periodic background.

In this paper we have focused on the periodic solutions and
their stability. Complementary to this analysis would be a study
of the quasi-periodic solutions. Of course, the set of solutions
in which the two modes have non-resonant wavenumbers is
considerably larger than that of the periodic solutions and such
an analysis is beyond the scope of the present work. In
preliminary numerical simulations we have considered the
stability of certain quasi-periodic solutions and found that the
range in the wavenumber mismatch over which they are stable is
smaller than that of periodic solutions with nearby wavenumbers
\cite{HiRiunpub}. Thus, at least in the regime considered, the
resonant terms enhance the stability of the periodic
solutions. Another promising line of investigation would be the
reduction of the coupled Ginzburg-Landau equations (\ref{2},\ref{3})
to two coupled phase equations. This should be possible for long-wave
perturbations of the mixed-mode solution if the smallness of the
resonant terms is exploited and it is assumed that
such terms are of the same order as the gradients in the
magnitude of the amplitudes (cf. (\ref{6a},\ref{7a})).

We gratefully acknowledge discussions with J. Porter and P.
Umbanhowar. 

\appendix

\section{Stability of Pure and Mixed Modes}

In this appendix we provide some of the details of
the stability analysis of the pure and mixed modes, presented in
Sec.\ref{sec:side-band}.

\subsection{Pure modes}

The perturbed pure mode $S_1$ is written as
\begin{eqnarray}
&&A_1= R_1e^{ikX}\left(1 +a_1^+(T)e^{iQX}+a_1^-(T)e^{-iQX}\right)\cierto{ansta1S1} \\
&&A_2= e^{ik\frac {n} {m}X}\left(a_2^+(T)e^{iQX}+a_2^-(T)e^{-iQX}\right).\cierto{ansta2S1}
\end{eqnarray}
Inserting this ansatz in (\ref{2},\ref{3}) and linearizing in $a_1^\pm$ and $a_2^\pm$ we obtain
\begin{eqnarray}
\dot a_1^\pm& =&\left(\alpha - sR_1^2-\delta Q^2\mp (\gamma+2\delta k) Q\right)a_1^\pm-
sR_1^2a_1^\mp,\cierto{eqa1S1}\\
\dot a_2^\pm&=&\left(\beta -\rho' R_1^2-\delta' Q^2\mp 2\delta' Q(n k/m)\right)a_2^\pm
+\left\{\nu' R_1^n a_2^\mp\right\}_{m=2},\cierto{eqa2S1}
\end{eqnarray}
where the bracketed term with subscript $m=2$ is present only if $m=2$, and
$\alpha$ and  $\beta$ are defined by Eq.~(\ref{5b}).
The eigenvalues associated with Eq.~(\ref{eqa1S1}) are
\begin{equation}
\lambda_1^\pm =-(\alpha+\delta Q^2)\pm \sqrt{\alpha^2+Q^2(\gamma+2k\delta)^2},
\cierto{eigva1S1}
\end{equation}
and with  Eq.~(\ref{eqa2S1}) are
\begin{equation}
\lambda_2^\pm =(\beta-\rho'\frac {\alpha}{s}-\delta'Q^2)\pm \sqrt{4\delta'^2\left(k
\frac {n}{m}\right)^2 Q^2+\left\{\nu'^2\left(\frac 
{\alpha}{s}\right)\right\}_{m=2}}.
\cierto{eigva2S1} 
\end{equation}
Eqs.(\ref{eigva1S1}) and (\ref{eigva2S1}) are related to longwave and shortwave 
instabilities, respectively. Note that for $m\geq 3$ the perturbation amplitudes 
$a_2^+$ and $a_2^-$ decouple and the destabilizing mode is composed of just one 
wave of the form $(a_2^+, 0)$ or $(0 ,a_2^-)$. In contrast,  due to 
the resonant term when $m=2$, the eigevectors have (in principle) 
components in both directions and the destabilizing mode is composed of two 
waves with different wavenumbers.

For the pure mode $S_2$ we have 
\begin{eqnarray}
A_1&=& e^{ikX}\left(a_1^+(T)e^{iQX}+a_1^-(T)e^{-iQX}\right)\cierto{ansta1S3} \\
A_2&=&R_2 \,e^{ik\frac {n} {m}X}\left(1+a_2^+(T)e^{iQX}+a_2^-(T)e^{-iQX}\right),\cierto{ansta2S2}
\end{eqnarray}
where $a_1^\pm$ and $a_2^\pm$ satisfy
\begin{eqnarray}
\dot a_1^\pm& =&\left(\alpha - \rho R_2^2-\delta Q^2\mp (\gamma+2\delta k) Q\right)a_1^\pm,\cierto{eqa1S2}\\
\dot a_2^\pm&=&\left(\beta - 2 s' R_2^2-\delta' Q^2 \mp 2 \delta' (n k/m) Q\right)a_2^\pm
-s'R_2^2 a_2^\mp.\cierto{eqa2S2}
\end{eqnarray}
The stability of  $S_2$ is determined by two eigenvalues:
\begin{eqnarray}
\lambda_1^\pm & =&(\alpha -\rho \frac{\beta}{s'}-\delta Q^2 )\pm |Q||\gamma+2k\delta |,\nonumber \\[0.5cm]
\lambda_2^\pm &=&-(\beta+\delta' Q^2)\pm \sqrt{\beta^2+4\delta'^2\left(\frac{n}{m}k\right)^2Q^2},
\cierto{eigvS1}
\end{eqnarray}
which are associated with the shortwave and longwave instabilities, respectively.
  
\subsection{Mixed modes}

In this case we consider the system (\ref{6a})-(\ref{9a}) for the
evolution of the real amplitudes $R_j>0$ and phases $\phi_j$ (for
$j=1,2$). Thus (abusing notation) we write
  
\begin{eqnarray}
&&R_1(X,T)= R_1+r_1(X,T),\quad \phi_1(X,T)=kX+\varphi_1(X,T)\cierto{ansta1M }\\ 
&&R_2(X,T)= R_2+r_2(X,T),\quad \phi_2(X,T)=\frac{n}{m}kX+\varphi_2(X,T),\cierto{ansta2M} 
\end{eqnarray} 
where 

\begin{eqnarray*}
r_j &=& \left[r_j^+(T)-ir_j^-(T)\right]e^{iQX}+\left[r_j^+(T)+ir_j^-(T)\right]e^{-i
Q X } 
\\[0.5cm]
\varphi_j &=& \left[\varphi_j^+(T)-i\varphi_j^-(T)\right]e^{iQX}+\left[\varphi_j^+(
T) + i \varphi_ j^-( T)\right]e^{-iQX},
\end{eqnarray*}
for $j=1,2$. The 
linearized problem 
for the perturbations $r_j^\pm$ and $\varphi_j^\pm$ is given 
by 

\begin{eqnarray} 
\left( \begin{array}{c}
\dot {r_{1}^\pm}\\
\dot{\varphi _{1}^\mp}\\
\dot{r_{2}^\pm}\\
\dot{\varphi_{2}^\mp}
\end{array}\right)  =  \left( \begin{array}{cccc}
a_1 & -2\delta Q(k+\gamma) & a_2 &0\\
-2\delta Q (k+\gamma) & b_1 & 0 & b_2\\
c_2 & 0 & c_1 & -2\delta' \frac{n k}{m}\\
0 & d_2 & -2\delta'\frac{n k} {m} & d_1
\end{array}\right)
\left( \begin{array}{c}  
{r_{1}^\pm}\\ 
{\varphi _{1} ^\mp}\\ 
{r_{2}^\pm}\\ 
{\varphi_{2}^\mp} 
\end{array}\right), 
\cierto{a1} 
\end{eqnarray}
where 
\begin{eqnarray*}
a_1 &=& \mu-\delta k^2-\gamma k-\delta Q^2-(3s R_1^2+\rho R_2^2)+\nu (n-1) R_1^{n-2}R_2^m\cos(n\hat \phi_1-m\hat\phi_2),\\[0.5cm]
a_2 &=& -2\rho R_1 R_2+\nu m R_1^{n-1} R_2^{m-1},\\[0.5cm]
b_1&=&-\nu n R_1^{n-2}R_2^{m}-\delta Q^2,\quad b_2=\nu m R_1^{n-1}R_2^{m-1},\\[0.5cm]
c_1&=&\mu-\Delta\mu-\delta' (kn/m)^2-\delta' Q^2-(3s' R_2^2+\rho' R_1^2)+\nu' (m-1) R_2^{m-2}R_1^n\cos(n\hat \phi_1-m\hat\phi_2),\\[0.5cm]
c_2&=&-2\rho' R_1 R_2+\nu' n R_1^{n-1} R_2^{m-1},\\[0.5cm]
d_1&=&-\nu'mR_1^{n}R_2^{m-2}-\delta' Q^2,\quad d_2=\nu' n R_1^{n-1}R_2^{m-1}.
\end{eqnarray*}

\section{Derivation of (\ref{16a})}
 
To derive (\ref{16a}) we first consider the characteristic
equation in the form
\begin{align}
\lambda^4+G_3 \lambda^3+G_2\lambda^2 +G_1\lambda+G_0=0, 
\cierto{14a} 
\end{align} 
where $G_j$ ($j=0,\ldots,3$) are polynomials in $Q^2$ 
whose coefficients depend on the parameters directly and also through 
the amplitudes of the mixed modes. When the bifurcation is steady and of shortwave type
the location of $\Gamma_{\!+}$
and the wavenumber  $Q\neq 0$ 
are obtained as solutions of the system $G_0= \partial G_0/\partial
Q^2 =0$, which yields
\begin{equation} 
(g_1g_2)^2-4g_0g_2^3 - 4g_1^3-18g_0g_1g_2=0\cierto{14b} 
\end{equation} 
for the location of the lower part of  $\Gamma_{\!+}$ and 
\begin{equation}
Q^2=(g_1g_2-9g_0)/(6g_1-2g_2^2)\cierto{14c}
\end{equation}
for the perturbation wavenumber of the most unstable perturbation.
Here the funtions  $g_0$, $g_1$ and $g_2$ can be written as
\begin{eqnarray*}
g_0 &=&-(a_1c_1-a_2c_2)(\delta'b_1-\delta d_1)-\gamma ^2c_1d_1,\\
g_1 &=& \delta\delta' (a_1c_1-a_2c_2+a_1d_1+c_1b_1)+\delta^2c_1d_1+
{\delta'}^2a_1b_1+\gamma^2\delta'(c_1+d_1),\\
g_2 &=& -\delta^2\delta'(c_1+d_1)-{\delta'}^2\delta(a_1+b_1)-\gamma^2{\delta'}^2,
\end{eqnarray*}
where $a_j$, $b_j$, $c_j$ and $d_j$ ($j=1,2$) are defined in Appendix A.

To proceed, we consider the case $k=0$ 
and make use of the fact that for $\gamma=0$ the
stability limit $\Gamma^+$ coincides with $L_2$ and remains
close for $\gamma \ll 1$. We therefore expand the value of $\mu$
on the stability limit,
\begin{equation} 
\mu=\mu_c+\eta^2,\quad\text{with}\quad\eta=c_1\gamma+c_2\gamma^2+\ldots,
\quad |\gamma|\ll 1,
\cierto{15app} 
\end{equation} 
where $\mu_c=\rho\Delta\mu /(s'-\rho)$ is the value of $\mu$ on
the curve $L_2$ at $k=0$.  The small parameter $\eta$ is a
function of $\gamma$ and must vanish when $\gamma=0$.  Moreover
the amplitudes of the mixed mode $S_+$ on $\Gamma_{\!+}$ may be
expressed as \begin{align} &R_1=\eta  r_{11}+ \eta^2 r_{12}+
\cdots,\qquad R_2=r_{20}+ \eta^2 r_{22}+\eta^3
r_{23}\cdots.\cierto{15c} \end{align} where
$r_{20}=\sqrt{\Delta\mu/(s'-\rho)}$ is the amplitude of the pure
mode on the line $L_2$.  The coefficients $r_{ij}$ are
determined by substituting the expansions~(\ref{15c}) into
(\ref{6}) and solving at successive orders. The number of
coefficients $r_{ij}$ that must be kept depends on the
particular resonance.  In fact the expansion (\ref{15c}) must be
carried out to $O(\eta^{n-2})$ for an $m:n$ resonance. In order
to calculate the $c_j$ we substitute (\ref{15c}) into
(\ref{14b}) and  solve order by order. Finally, the wave number
$Q$ is obtained by substituting these results into (\ref{14c}).

We now sketch the necessary calculations for the case
$m:n=2:5$ at $k=0$.  Inserting (\ref{15app}) and (\ref{15c}) 
into (\ref{6}) yields 

\begin{equation} 
\left.\begin{array}{r} 
sr_{11}^2+2\rho r_{20} r_{22}=1,\\ 
2s'r_{20} r_{22}+\rho' r_{11}^2=1, 
\end{array}\right\} \quad \Longrightarrow \quad 
\begin{array}{l} 
r_{11}^2=(s'-\rho)/(ss'-\rho\rho'),\\ 
r_{22}=(s-\rho')/2r_{20}(ss'-\rho\rho') 
\end{array} 
\end{equation} 
at $O(\eta^2)$, while at $O(\eta^3)$ we have 
\begin{equation} 
\left.\begin{array}{r} 
2sr_{11}r_{12}+2\rho r_{20} r_{23}=-\nu r_{20}^mr_{11}^3,\\ 
  s' r_{20} r_{23}+\rho' r_{11}r_{12}=0 
\end{array}\right\} \quad \Longrightarrow \quad 
\begin{array}{l} 
r_{12}=-\nu s' r_{11}^2 r_{20}^m/ 2(ss'-\rho\rho'),\\ 
r_{23}=\nu\rho' r_{11}^3 r_{20}^{m-1}/ 2(ss'-\rho\rho'). 
\end{array} 
\end{equation} 
If $n>5$ we would have found  $r_{12}=r_{23}=0$ and it would be necessary  to 
consider higher order corrections.  Near the curve $L_2$ 
(i.e., near the point $(\mu,k)=(\mu_c,0)$) (see Fig.~\ref{fig:dgrm2n5withtw(a)}) 
the functions $g_j$, for $j=0,1,2$, may be written as 
\begin{align} 
g_0=\;&\eta^n \left(4 \delta' \nu (ss'-\rho\rho')r_{11}^n r_{20}^{m+2}\right)+ 
O(\eta^{n+1}),\nonumber\\ 
g_1=\;&\eta^2\left(4 
\delta\delta'(ss'-\rho\rho')r_{11}^2r_{20}^2\right)+\eta^{n-2}(2s'n+\Delta\mu(n-
1 ) r _ {20} ^{m})\delta\delta' \nu a_1^{n-2}\cierto{16}\\ 
&+\gamma^2 \delta'(-2s'r_{20}^2-\eta^2 4s'r_{20}r_{22}-\eta^2 
4s'r_{20}r_{23})+O(\eta^n+ \gamma^2\eta^{n-2})\nonumber\\ 
g_2=\;&\delta^2\delta'(2 s' r_{20}^2)+ O(\eta^2),\nonumber\end{align} where only  
those terms necessary to compute $Q$ have been kept in (\ref{16}). Finally  
composing the expansions~(\ref{16}) and (\ref{15app}) and substituting the result 
into (\ref{14c}) we find (after some algebra) the result (\ref{16a}).

%\bibliography{journal}

\end{document}